\documentclass[12pt,english]{article}
\pdfoutput=1

\usepackage[a4paper,lmargin=70pt,rmargin=70pt]{geometry}
\usepackage{setspace}
\usepackage{babel}
\usepackage{graphicx}
\usepackage{amsmath}
\usepackage{epsfig}
\usepackage{amssymb}
\usepackage{caption}
\usepackage{subcaption}
\usepackage[colorlinks=true,urlcolor=blue,anchorcolor=blue,citecolor=blue,filecolor=blue,linkcolor=blue,menucolor=blue,pagecolor=blue,linktocpage=true]{hyperref}

\usepackage[colorlinks=true,urlcolor=blue,anchorcolor=blue,citecolor=blue,filecolor=blue,linkcolor=blue,menucolor=blue,pagecolor=blue,linktocpage=true]{hyperref}

\newcommand{\bra}[1]{\left\langle #1 \right|}
\newcommand{\ket}[1]{\left|#1\right\rangle}

\newcommand{\Tr}{\text{Tr}} 

\begin{document}

\begin{flushleft} 
\today
\end{flushleft} 
\vspace{-1cm}
\begin{flushright} 
SU-ITP-18/01

YITP-18-15
\end{flushright} 

\vspace{0.1cm}

\begin{center}
	{\LARGE Onset of Random Matrix Behavior  \\ \vspace{.5em} in Scrambling Systems}
\end{center}
\vspace{0.1cm}
\begin{center}
	 Hrant G{\sc haribyan}$^{a}$,
  Masanori H{\sc anada}$^{abcd}$,
  Stephen  H. S{\sc henker}$^a$,
  
and  
  Masaki T{\sc ezuka}$^e$

\vspace{0.4cm}

$^a${\it Stanford Institute for Theoretical Physics,\\
Stanford University, Stanford, CA 94305, USA}

\vspace{0.2cm}

$^b${\it Yukawa Institute for Theoretical Physics, Kyoto University, Kyoto 606-8502, Japan}\\

\vspace{0.2cm}

$^c${\it The Hakubi Center for Advanced Research, Kyoto University, Kyoto 606-8501, Japan}\\

\vspace{0.2cm}
$^d$ {\it Department of Physics, University of Colorado, Boulder, Colorado 80309, USA}

\vspace{0.2cm}

$^e${\it Department of Physics, Kyoto University, Kyoto 606-8502, Japan}\\

\end{center}

\onehalfspacing

\begin{center}
  {\bf Abstract} 
  \end{center}
 The fine grained energy spectrum of quantum chaotic systems is widely believed to be described by random matrix statistics.  A basic scale in such a system is the energy range over which this behavior persists.   We define the corresponding time scale by the time at which the linearly growing ramp region in the spectral form factor begins. We call this time  $t_{\rm ramp}$.
 The purpose of this paper is to study this scale in many-body quantum systems that display strong chaos, sometimes called scrambling systems.   We focus on randomly coupled qubit systems, both local and $k$-local (all-to-all interactions) and the Sachdev--Ye--Kitaev (SYK) model. Using numerical results for Hamiltonian systems and analytic estimates for random quantum circuits we find the following results. For geometrically local systems with a conservation law we find $t_{\rm ramp}$ is determined by the {\it diffusion time} across the system, order $N^2$ for a 1D chain of $N$ qubits.  This is analogous to the behavior found for local one-body chaotic systems. For a $k$-local system with conservation law the time is order $\log N$ but with a different prefactor and  a different mechanism than the scrambling time.   In the absence of any conservation laws, as in a generic random quantum circuit,
 we find $t_{\rm ramp} \sim \log N$, independent of connectivity.
 
\newpage

\tableofcontents
\section{Introduction and Summary}

Dynamical aspects of many-body quantum chaos are of fundamental interest in various fields of physics,  
including condensed matter physics, quantum information theory and quantum gravity. 
Random Matrix Theory (RMT) provides us with a simple characterization of quantum chaos: 
if the level statistics at small energy separation agrees with RMT, one concludes the system under consideration is chaotic.   There is a rich theory of this  for few-body  semiclassical quantum systems ({for a general reference see \cite{Haake2006}).  For many-body systems there is a substantial amount of numerical evidence, but not  yet a fully developed theory (for recent analytic approaches see \cite{SerbynMoore2016, AltlandBagrets2017,Chan:2018dzt}).   

This work was motivated by recent numerical work we and others \cite{You:2016ldz,Garcia-Garcia:2016mno,Cotler:2016fpe}  performed on the Sachdev--Ye--Kitaev (SYK) model \cite{Sachdev:1992fk,Kitaev_talk,Maldacena:2016hyu} , which has become a reference model for the study of many-body quantum chaos.

Another diagnostic of strong chaos is sensitive dependence to initial conditions, classically diagnosed by the presence of a positive Lyapunov exponent.   The semiclassical  quantum analog of this in few-body chaos was discussed in the 1960s  \cite{Larkin}, and diagnosed by out-of-time-order correlation functions (OTOC).   Motivated by a desire to understand the quantum chaotic nature of black holes \cite{Hayden2007,Susskind2008}, these diagnostics were rediscovered in the general many-body context \cite{Almheiri:2013hfa} and have been studied extensively in recent years 
\cite{Kitaev_talk, kitaevprize,Lashkari2013, Shenker:2013pqa, Shenker:2013yza, Roberts:2014isa, Shenker:2014cwa, Maldacena2015}.   
A positive value of the analogous ``chaos exponent'' is a signature of strong many-body chaos.   Systems with such behavior are sometimes referred to as ``fast scramblers'' \cite{Susskind2008}.   There is a characteristic time scale for such systems, the scrambling time $t_{\rm scr}$,  that measures  when an initial perturbation affecting only a few degrees of freedom has chaotically spread to all the degrees of freedom in the system.   

Such systems also display a simpler signature of thermalization.  Two point functions of generic simple operators exponentially decay.  We call the decay time of such quantities the ``dissipation time"  $t_{\rm diss}$. In general this time is different from the inverse of the chaos exponent. 

There is also a natural time scale associated to RMT level statistics.  In the energy domain one can ask how far in energy space do random matrix statistics persist, that is how many energy levels apart can one go before the characteristic correlations of RMT are lost.   The inverse of this energy defines a time, which in the many-body context we will call the ``ramp time",  $t_{\rm ramp}$ for reasons that will be made clear below.  

The RMT behavior of the fine-grained energy spectrum is {\it universal} in quantum chaotic systems.\footnote{The type of random matrix ensemble just depends on the symmetry class of the Hamiltonian.}  The structure of the spectrum at time scales shorter than $t_{\rm ramp}$  contains nonuniversal information about the specific system.

In simpler, one-body, systems this time scale has been studied extensively \cite{altshuler1986repulsion, ErdosKnowles2015}.  For a single quantum particle propagating in a random medium, or equivalently for a banded random matrix,  this time scale is the time for a particle to diffuse across the system $t_{\rm diff}$.   The corresponding energy is called the ``Thouless energy" and we will sometimes refer to this time scale as the ``Thouless time."   

The goal of this paper is to study the relationship between these various time scales in strongly chaotic many-body quantum systems.\footnote{The first discussion of the Thouless (or ramp) time in the SYK model was presented in \cite{Garcia-Garcia:2018ruf}.  We discuss the limitations of their analysis in Section~\ref{sec:SYK-numerics}.}    These time scales are connected, via gauge/gravity duality, to phenomena in quantum black holes.  This connection provides an important motivation for our work.

To disentangle the various phenomena involved we study two different types of system: $k$-local systems, with degrees of freedom nonlocally coupled, but only $k$ degrees of freedom interacting at a time; and geometrically local systems, with only nearest neighbor couplings.  The $k$-local systems we study are the SYK model and a qubit model of pairs of qubits interacting with Gaussian random couplings which we call the ``2-local Randomly Coupled Qubits (RCQ)" model.   The geometrically local systems we study are a one dimensional nearest neighbor qubit model with Gaussian random couplings (the  ``Local RCQ") and, for reference, banded matrices.

Robust analytic techniques to study RMT phenomena in such many-body systems have not yet been developed.  This is a major goal for future research.   So of necessity one part of our work involves numerical investigation.   But powerful techniques have been developed to study related systems, ones in which the random couplings vary at each time step.   For discrete time steps these are referred to as random quantum circuits, in the continuous time limit these are called Brownian circuits.\footnote{Random quantum circuits and Brownian circuits have played an important role in a number of recent developments.   An early application in the black hole context was \cite{Banks:1983by}.   More recently these systems have been used in the quantum information community to provide rapid approximations to random unitary operators  \cite{Lloyd2005}.  An important advance was the use of Markov chain techniques to estimate convergence times \cite{Oliveira06}.  In particular these ideas have been applied to the study of approximate unitary $k$-designs \cite{Emerson2003, Gross2007, Arnaud2008, Low09, Harrow2009, Brown2010, Brown:2012gy, Harrow2015, Onorati2016, winter2016, Banchi2017}. These techniques were used to give the first logarithmic estimate of scrambling \cite{Lashkari2013}.  More recently they have been used to calculate OTOC for $k$-local \cite{Shenker:2014cwa} and geometrically local systems \cite{Nahum:2017yvy,  vonKeyserlingk:2017dyr, Khemani:2017nda}.  These models are currently being used as testbeds to establish ``quantum supremacy" \cite{Preskill2012, Martinis2017, Harrow2017, Vazirani2018}.}

As we will discuss various indicators of chaos are well captured by such systems.  But a major difference with time independent Hamiltonian systems is the absence of conservation of energy.  The fluctuating couplings pump energy into the system.   To study the effects of conservation laws we use a recently introduced random  quantum circuit that conserves spin \cite{Khemani:2017nda, rakovszky2017}.

We are not able to give a definitive analysis of $t_{\rm ramp}$ in time independent Hamiltonian systems, but are able to provide numerical results that are consistent with those obtained from  random quantum circuits with conservation laws.

\subsection{Summary of Results}

In Section~\ref{sec:RMT-review} we review the basics of RMT, including the density of states and the eigenvalue pair correlation function.  In Section \ref{diagnosticsofspectralrigidity} we introduce diagnostics of spectral rigidity, the hallmark of the spectrum of random matrices and quantum chaotic systems.   We first discuss the ``number variance" and point out artifacts that can affect this quantity in many-body systems.  We then define the spectral form factor which is the Fourier transform of the pair correlation function.   We discuss the robustness of this diagnostic in the presence of a varying density of states, and then point out the contamination of this signal due to sharp edges in the spectrum.  We then introduce a new quantity $YY^*$ that incorporates a Gaussian filter
into the spectral form factor that removes the effect of sharp edges.  We then discuss analogous quantities for random unitary matrices.

In section \ref{klocalhsystems} we study the ramp time $t_{\rm ramp}$ in various $k$-local scrambling systems numerically.  In the SYK model we use $YY^*$ to eliminate the ``slope" and show that $t_{\rm ramp}$ is very short, consistent with a $\log N$ dependence (although a slow power law cannot be ruled out).   A ramp time $t_{\rm ramp} \sim \log N$ is equivalent to spectral rigidity extending over $2^{N/2}/(\sqrt{N} \log N)$ eigenvalues, out of a total of $2^{N/2}$ eigenvalues.\footnote{Recall that the spacing between near-neighber eigenvalues is $\sim \sqrt{N} 2^{-N/2}$ .} 

We then study a $2$-local qubit system, the Randomly Coupled Qubit model (RCQ).   Here again we find a short ramp time, consistent with $t_{\rm ramp} \sim \log N$. 

In  order to understand the relation between this $\log N$ time and the scrambling time we turn to geometrically local systems in Section \ref{sec:local-Hamiltonian}.   We mainly focus on a one dimensional nearest neighbor version of the RCQ.   For such geometrically local systems the scrambling time is $t_{\rm scr} \sim N$.
Our numerical results 
indicate that $t_{\rm ramp} \sim N^2$ . Energy is conserved here and the diffusion time is $t_{\rm diff} \sim N^2$,  indicating that one must wait for all conserved quantities to diffuse across the system before  random matrix behavior is established.    This is one of our central findings and we give analytic arguments for it later.

In  Section \ref{sec:Random-and-Brownian-circuits} we turn to analytic methods.  In particular we study analogs of the spectral form factor for random and Brownian quantum circuits where the couplings are varied at each time step.   Powerful techniques using Markov chains exist to study such systems.    We study $k$-local and geometrically local RCQ random circuits  and find that the analog of the ramp time is $t_{\rm ramp} \sim \log N$, independent of connectivity. The scrambling time in the one dimensional geometrically local random circuit is $t_{\rm scr} \sim N$ so the mechanisms here are clearly different.  In fact we find the source of the logarithm is the exponential decay of $N$ two-point functions of single qubit operators.  So the coefficient of $\log N$ is set by the dissipation time, not the Lyapunov time.

Energy is not conserved in random circuits and hence its diffusion cannot be studied there. So in order to study diffusion  we examine in Section \ref{sec:spin-conserving-circuit} a random circuit that conserves spin, a version of the XXZ model.   Here we find that the ramp time depends on connectivity.   For the geometrically local case $t_{\rm ramp} \sim N^2$.
We are able to relate this to (many-body) diffusive effects.   For the $2$-local case we find that the diffusion time is order one, and so unimportant.  Here $t_{\rm ramp} \sim \log N$.  Again we are able to trace this to the decay of $N$ two point functions.   In the geometrically local case these decay diffusively  $\sim 1/t^{1/2}$ and in the $2$-local case they decay exponentially with a scale given by the dissipation time. 

Finally in Section \ref{sec:hamestimate} we discuss the connection between the ramp time and the relaxation time scale of correlation functions of simple operators in random quantum circuits. We expect the same connection to be true for Hamiltonian evolution, but point to an issue that makes it difficult to construct a precise argument. 

\subsubsection*{Other relevant work}
There has been other recent work aimed at  computing $t_{\rm ramp}$ in the SYK model
\cite{AltlandBagrets2017,Garcia-Garcia:2016mno, Garcia-Garcia:2018ruf}. Garc\'ia-Garc\'ia and Verbaarschot give a numerical study of the number variance.  They find that spectral rigidity extends over of order $N^2$ eigenvalues in the spectrum.  Translating their result to the time domain they predict $t_{\rm ramp} \sim 2^{N/2}/N^{5/2}$ \cite{Garcia-Garcia:2016mno, Garcia-Garcia:2018ruf}. This differs dramatically from our results.  We discuss a possible explanation for this discrepancy in Section \ref{numbervarsec}.  

Altland and Bagrets \cite{AltlandBagrets2017} use interesting techniques from few-body quantum chaos to argue that in the SYK model  $t_{\rm ramp} \sim \sqrt{N} \log N$.  This disagrees with the $\log N$ scaling we expect from conserved charge random quantum circuit intuitions.  In addition the $N=34$ numerical value for $t_{\rm ramp}$ they calculate disagrees with our numerical results presented below by roughly an order of magnitude.  We do not understand this discrepancy, but make a few remarks about it in Section \ref{closingremarks}.  Their techniques may well serve to demonstrate the ramp at later times.  

As this paper was in the final stages of editing the paper  appeared.  These authors \cite{Chan:2018dzt} introduce a Floquet model without local conservation law with large site dimension $q$ that can be treated analytically in the large $q$ limit.   This allows them to compute all moments of their unitary  (which in this case corresponds to time evolution) analytically, a significant advance.    For a one dimensional chain they find a ramp time of $\log N$ consistent with our results.  For higher dimensions they find $t_{\rm ramp} \sim 1$.  For our systems we would expect a higher dimensional model with $N$ sites to again have $t_{\rm ramp} \sim \log N$.  Perhaps the difference is due to the large $q$ limit.   

After submitting the first version of this paper we learned of a very interesting paper by Kos, Ljubotina, and Prosen \cite{Kos:2017zjh} that we had missed.  In this paper the authors discuss a clever model that injects just enough randomness to compute all moments of their Floquet unitary even with fixed site dimension.   They see the ramp as the sum of cyclically permuted ladders that give the result in RMT (as do the authors of \cite{Chan:2018dzt})  and make the connection to periodic orbit theory.   They find a ramp time of $\log N$ in a way that seems quite generic, in accordance with our results.   This technique looks quite powerful and seems amenable to generalization.

\section{Random Matrix Theory}\label{sec:RMT-review}
In this section we review a few basic properties of Random Matrix Theory (RMT) which are relevant for the following sections.  For simplicity we restrict ourselves to Gaussian random Hermitian matrices, the Gaussian Unitary Ensemble (GUE) \cite{dyson1962statistical}.  A standard reference is  \cite{mehta2004random}. 
The partition function for the GUE is given by 
\begin{eqnarray}\label{measure}
\mathcal Z_{\rm GUE}
=
\int \left(\prod_{i,j}dM_{ij}\right)e^{-\frac{L}{2}{\rm Tr}M^2}, 
\end{eqnarray}
where $M$ is an $L\times L$ Hermitian matrix.  We denote its eigenvalues   $E_i$, because $M$ is often taken to be the Hamiltonian of a quantum system.   We introduce the normalized eigenvalue density 
\begin{eqnarray}
\hat\rho(E) = \frac{1}{L} \sum_i^{L} \delta(E - E_i) ~.
\end{eqnarray}
At large $L$ the average density in this ensemble is given by the Wigner semicircle law
\begin{eqnarray}\label{rhosemicircle}
\langle \hat\rho (E) \rangle_{\rm GUE}=  \rho_{\rm sc}(E) =\frac{1}{2\pi} \sqrt{4-E^2}. 
\end{eqnarray}
We see that with these normalizations the width of the entire band is order one and so the typical spacing between nearest neighbor eigenvalues is $\sim 1/L$.   

The pair correlation function $R(E_1, E_2) = \langle \rho(E_1) \rho(E_2)\rangle $ is the probability of finding two eigenvalues at energies $E_1$ and $E_2$.  If their separation $|E_1 - E_2| \ll 1$ and their average ${\bar E} = (E_1+E_2)/2$ is not near the edge of the band  $R(E_1, E_2)$ is given by the sine kernel formula  \cite{mehta2004random} , most simply stated by introducing a scaled variable
\begin{eqnarray}\label{guepairdist}
x = \pi L (E_1 -E_2)\rho({\bar E}) ~.
\end{eqnarray}
Then we have
\begin{eqnarray}\label{sinekernel}
R(E_1, E_2) 
 &= \frac{ \rho_{\rm sc}(E_1)}{\pi L} \delta(E_1 -E_2) +\rho_{\rm sc}(E_1)\rho_{\rm sc}(E_2)\left(1- \frac{\sin^2 x}{x^2} \right) ~.
\end{eqnarray}

A sketch of this formula is given in Figure~\ref{energypaircorrelation}.   Note that $1/(L\rho_{\rm sc}({\bar E} ))$ is the spacing between nearest neighbor energy levels in the vicinity of ${\bar E}$.   In regions of small $\rho_{\rm sc}$ the eigenvalues are farther apart. The scale of the oscillations in \eqref{sinekernel} are set by this spacing.  RMT universality means that formula \eqref{sinekernel} holds after scaling by the local eigenvalue density.  When comparing different parts of the spectrum one must scale different regions differently.  This process is called ``unfolding," and becomes problematic when the fluctuations in the average density are large.  We will see some effects of this later.

\begin{figure}[t]
\center
\includegraphics[scale=0.5]{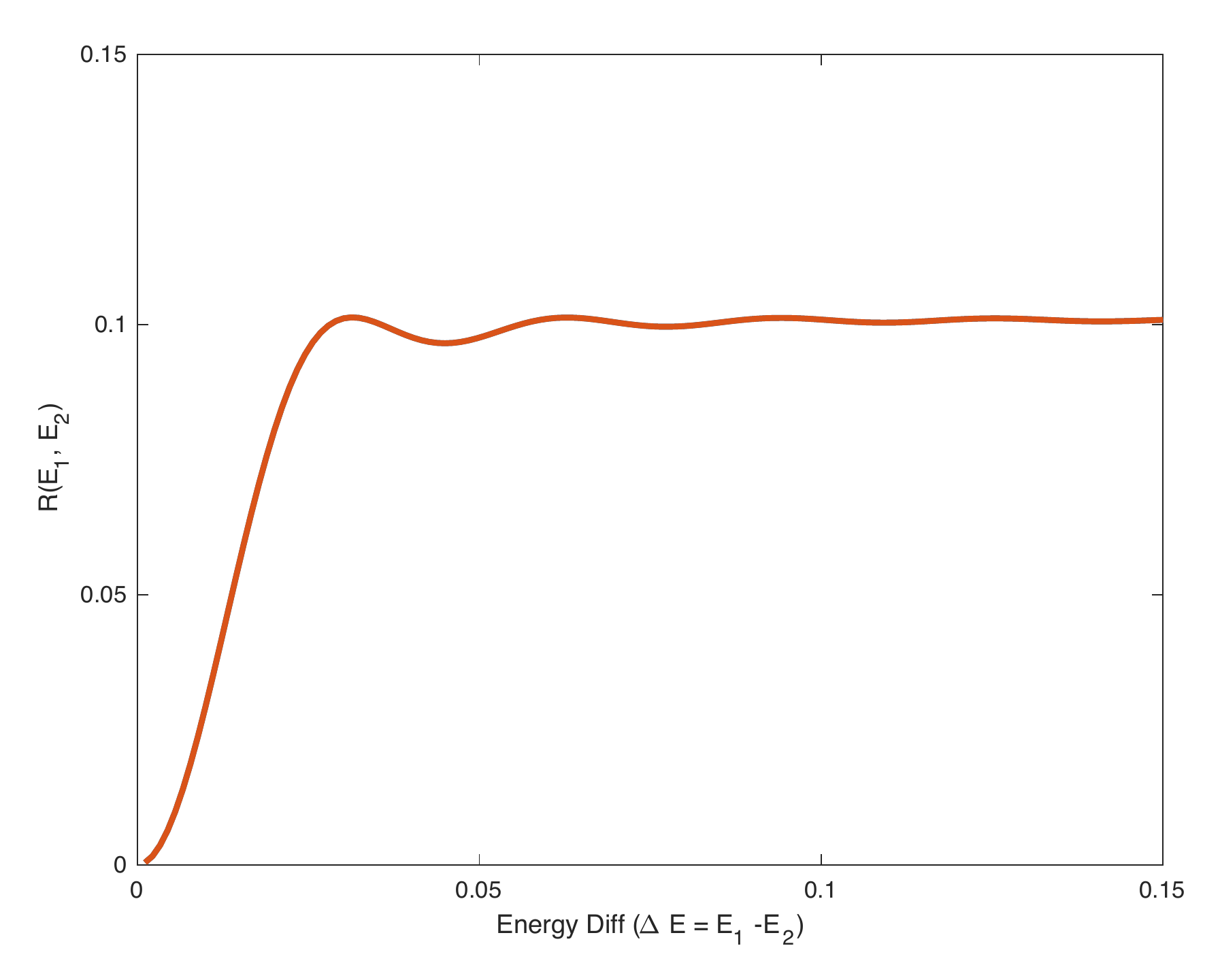}
\caption{Pair correlation function $R(E_1, E_2)$ plotted when $E_2=0$ and $\Delta E =E_1$ is varied. For simplicity, we have used $L=100$ and have suppressed the delta function in equation~\eqref{sinekernel}.  The oscillations here are at the scale of the nearest neighbor eigenvalue spacing.}
\label{energypaircorrelation}
\end{figure}

Note that when $|E_1 -E_2| \gg 1/L$, the nonconstant part of the pair correlation behaves like $\frac{1}{x^2}$.   This shows that long range fluctuations in the eigenvalue ``gas" are suppressed,  a phenomenon referred to as ``spectral rigidity."   We can now give a more precise definition of the Thouless energy: it is the energy difference where the RMT form of spectral rigidity ceases to be valid.

In the following sections we will often use ensembles of unitary matrices.   The Lie group of $L\times L$ unitary matrices has a unique measure that is invariant under left and right multiplication,  Haar measure. In random matrix theory, the unitary group U($L$) with  Haar probability measure is called the Circular Unitary Ensemble (CUE). The eigenvalues of each matrix are of the form $\lambda_k  = e^{i \theta_k}$, where the eigenphases are defined in the range  $0\leq \theta_k \leq 2\pi$, and the probability density is \cite{mehta2004random} 
\begin{eqnarray}
\rho(\theta_1, \theta_2, ..., \theta_L) 
= \frac{1}{\mathcal Z_{\rm CUE}} \prod_{1\leq k< j \leq L} |e^{i \theta_k} - e^{i \theta_j} |^2 ~.
\end{eqnarray}
where $\mathcal Z_{\rm CUE} =  L!\cdot (2\pi)^L$ denotes the partition function. Note that the single eigenvalue density $\rho(\theta)$ is uniform on the unit circle.  There are no edges, unlike the semicircle.   Again, we see that with these normalizations the width of the entire band is order one and so the typical spacing between nearest  neighbor eigenphases $(\theta_k)$ $\sim 1/L$.  

The pair correlation function $R(\theta_1, \theta_2) = \langle \rho(\theta_1) \rho(\theta_2)\rangle $ of eigenvalues is given by the sine kernel formula \cite{mehta2004random},
\begin{eqnarray}\label{sinekernelhaar}
R(\theta_1, \theta_2) 
 &= \frac{1}{ 2 \pi L} \delta(\theta_1 -\theta_2) + \frac{1}{(2 \pi)^2}\left(1- \frac{\sin^2 L(\theta_1-\theta_2)/2}{ L^2\sin^2((\theta_1-\theta_2)/2)} \right) ~.
\end{eqnarray}

Note that, unlike the GUE ensemble, this formula is exact for all values of eigenvalue differences and not just for a small window.   When the eigenvalues are close, expanding the denominator yields the same answer as GUE, as expected, because the same logarithmic repulsion acts in both cases.

\section{Diagnostics of Spectral Rigidity}\label{diagnosticsofspectralrigidity}
\subsection{Number Variance}\label{numbervarsec}

A standard diagnostic of spectral rigidity in random matrix theory  and  quantum chaos is the number variance.  This was first applied to the SYK model in \cite{Garcia-Garcia:2016mno}. The number variance is defined as 
\begin{eqnarray}\label{sigmadef}
\Sigma^2(K) = \langle n^2(E, K) \rangle -  \langle n(E, K) \rangle^2
\end{eqnarray}
where $n(E, K)$ denotes the count of levels in the interval $[E, E+K {\bar \Delta}]$ where ${\bar \Delta}$ is the average nearest neighbor level spacing and the computation is done after unfolding  the spectrum. If RMT universality is valid we expect this quantity to be essentially independent of $E$.   By the definition of ${\bar \Delta}$ we have $ \langle n(E, K) \rangle = K$ and in an interval of length $K$ the range of the expected number of energy levels is of order
\begin{eqnarray}
K \pm \sqrt{\Sigma^2(K)}  ~.
\end{eqnarray}
The number variance can be expressed in terms of the pair correlation function (see for example \cite{Guhr:1997ve}).  The result for GUE is
\begin{eqnarray}
\Sigma^2(K) = \frac{1}{\pi^2}  \left( \log (2\pi K) +\gamma +1 \right) + O\left(\frac{1}{L^2}\right), 
\end{eqnarray}
where $\gamma=0.57721...$ is Euler's constant.  The crucial thing here is the very slow $\log K$ behavior which indicates a very rigid spectrum.  Poisson distributed eigenvalues would give $\Sigma^2(K) \sim K$.
This quantity has proved a very useful numerical diagnostic in various studies of few-body quantum chaotic systems \cite{Guhr:1997ve}.  But in many-body quantum chaotic systems, like SYK, there is an additional contribution to $\Sigma^2(K)$ that obscures the RMT signal.   This complication was first pointed out  in the study of embedded Gaussian ensembles \cite{Flores2001}.   The basic issue is that the fluctuations in the coarse grained eigenvalue density in these systems are only suppressed by powers of $1/N$.  In contrast  these fluctuations in random matrix theory and few-body chaos are suppressed by powers of $1/L$.  Here $L$ is the dimension of the Hilbert space and in many-body systems $ L \sim e^{N}$.  So the density fluctuations are much larger in many-body systems.   

The main effect of this on $\Sigma^2(K)$ is easy to understand.  Assume that the result of each sample is described by random matrix statistics rescaled by the eigenvalue density realized in that sample, {\it not} the mean eigenvalue density.  Because the typical eigenvalue spacing $\Delta$ in each sample depends on this density it varies a lot from sample to sample.  This means that  $\Sigma^2(K)$ is actually of the following form (see formula 1 in \cite{Flores2001})
\begin{eqnarray}\label{sigmacorr}
\Sigma^2(K) \sim K^2 \frac{\langle( \Delta - {\bar \Delta})^2 \rangle}{{\bar \Delta}^2} + {\cal O} (\log K) + \ldots
\end{eqnarray}
For a system with zero mean energy ($\langle \Tr H \rangle =0$) and small variance a rough estimate is given by\footnote{This does not account for more complicated fluctuations in the shape of the energy spectrum.}
\begin{eqnarray}\label{deltavar}
\frac{\langle( \Delta - {\bar \Delta})^2 \rangle}{{\bar \Delta}^2} \sim \frac{\langle  (\Tr H^2)^2 \rangle - (\langle \Tr H^2 \rangle)^2}{(\langle \Tr H^2 \rangle)^2}
\end{eqnarray}

For GUE \eqref{deltavar} is of order $1/L^2$ so the correction in \eqref{sigmacorr} is of order $K^2/L^2$ which is small until $K$ covers a finite fraction of the whole energy band.   In contrast in SYK ($q=4$)  the correction is of order $K^2/N^4$ (see for example \cite{Cotler:2016fpe} appendix F), which becomes large at $K \sim N^2$. This is the same answer found by \cite{Altland:2017eao} and according to \cite{Garcia-Garcia:2018ruf} is consistent with numerics.   In SYK the spectrum consists of  $2^{N/2}$ eigenvalues and so this result would correspond to an exponentially small fraction of the spectrum.    Note that this $K^2/N^4$ behavior  has nothing to do with spectral rigidity and in fact will swamp the logarithmic signal for larger energy scales.    We will argue later that spectral rigidity in this system extends over $2^{N/2}/(\sqrt{N} \log N)$ eigenvalues, corresponding to a Thouless time of $t \sim \log N$.

We have made a preliminary check that this artifact is responsible for the results in \cite{Garcia-Garcia:2016mno,Garcia-Garcia:2018ruf} by rescaling each sample to have the same variance.  This increases the range of logarithmic behavior by roughly a factor of $2$ for the largest $N$ studied.  These results are discussed in detail in Appendix~\ref{sec:NR-NV}.  
It is not clear to us whether there is even an in principle way to fully correct for these large fluctuations in this observable.\footnote{Note that the variance in \eqref{deltavar} goes like $1/N^q$ for general $q$ SYK.  So studying large $q$ will improve this statistic.  But we must take $q^2 < N$ to preserve $k$-locality (see \cite{erdHos2014phase,Cotler:2016fpe}).  This is not sufficient to make the error in \eqref{sigmacorr} smaller than the desired signal.}

\subsection{Spectral Form Factor}\label{sec:sff-gue-cue}

We now discuss another diagnostic that is less sensitive to the kind of errors discussed above, the spectral form factor.  This diagnostic also has a long history in random matrix theory and quantum chaos; see e.g. \cite{mehta2004random}.\footnote{For elegant analytic formulas in RMT see \cite{Brezin-Hikami-1997} and \cite{Brezin:1997zz}.}
 It was first used to study the SYK model in \cite{Cotler:2016fpe}, motivated by a finite temperature extension discussed in the context of black hole physics in \cite{Papadodimas:2015xma}. 
Here we only consider problems with finite dimensional Hilbert spaces and focus on generic behavior in the band so we restrict ourselves to infinite temperature.  
 
 The spectral form factor is defined as follows:

\begin{eqnarray}\label{sffdef}
|Z(t)|^2=\left|{\rm Tr}e^{-i\hat{H}t}\right|^2=\left|\sum_j e^{-itE_j}\right|^2 = \sum_{ij} e^{i(E_i -E_j)t} 
\end{eqnarray}
Using the eigenvalue correlator we can write 
\begin{eqnarray}
\langle |Z(t)|^2 \rangle = \int dE_1 dE_2 L^2 R(E_1, E_2) e^{i(E_1 -E_2)t}. 
\end{eqnarray}
We will often discuss the normalized spectral form factor 
\begin{eqnarray}
g(t) = \langle |Z(t)|^2 \rangle/L^2
\end{eqnarray}
defined so that  $g(0) = 1$.  Here again $L$ is the dimension of the Hilbert space.  

To get oriented let us look at this quantity for the GUE ensemble.  As a first approximation we can replace $\rho_{\rm sc}(E)$ in \eqref{rhosemicircle} by $1$, so the eigenvalue spacing is uniform and of order $1/L$ all across the band.\footnote{We approximate further by extending the integration ranges to $(-\infty, \infty)$}   Then we can compute the Fourier transform of \eqref{sinekernel} giving the result 

\begin{eqnarray}\label{rampplateau}
\langle |Z(t)|^2 \rangle \sim 
\begin{cases}
t/(2 \pi), \quad t < 2L\\
L/\pi, \quad t \geq 2L
\end{cases}
\end{eqnarray}

\begin{figure}
\includegraphics[scale=0.65]{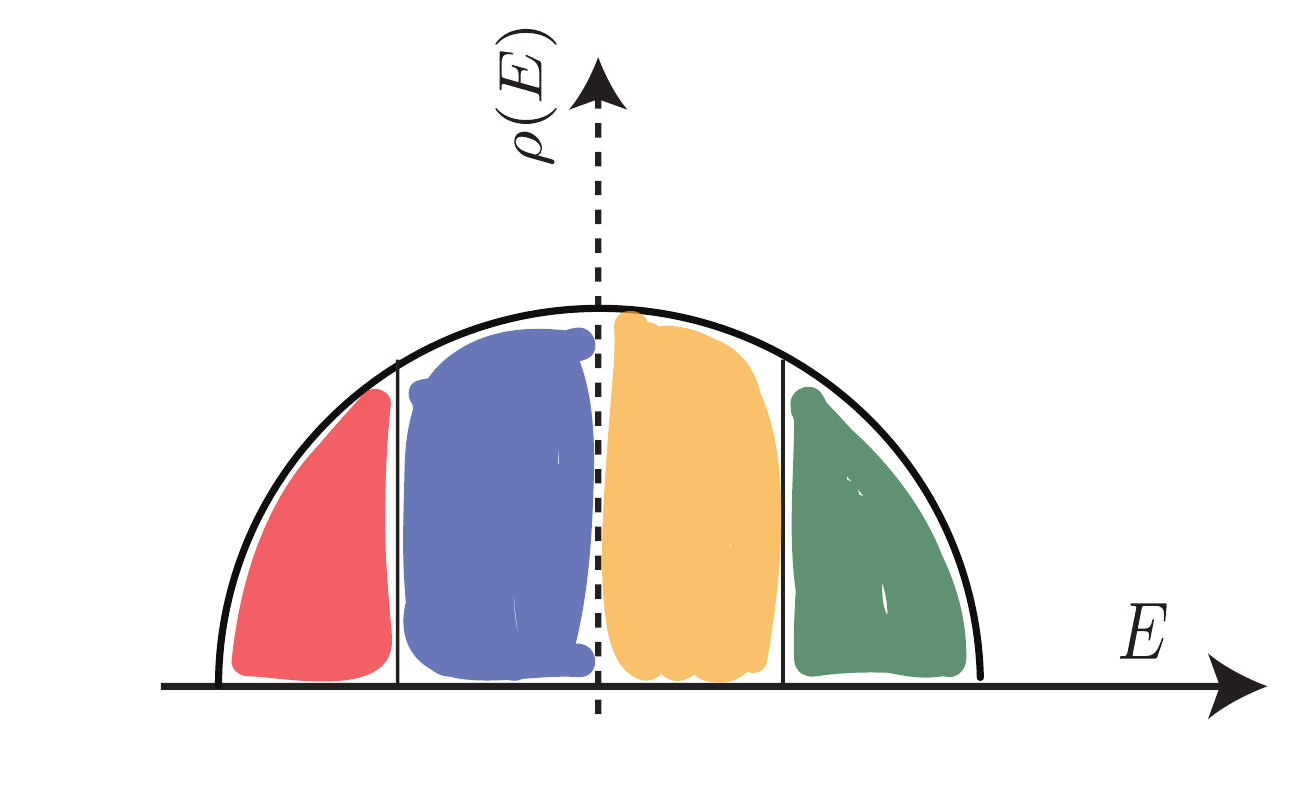}
\includegraphics[scale=0.6]{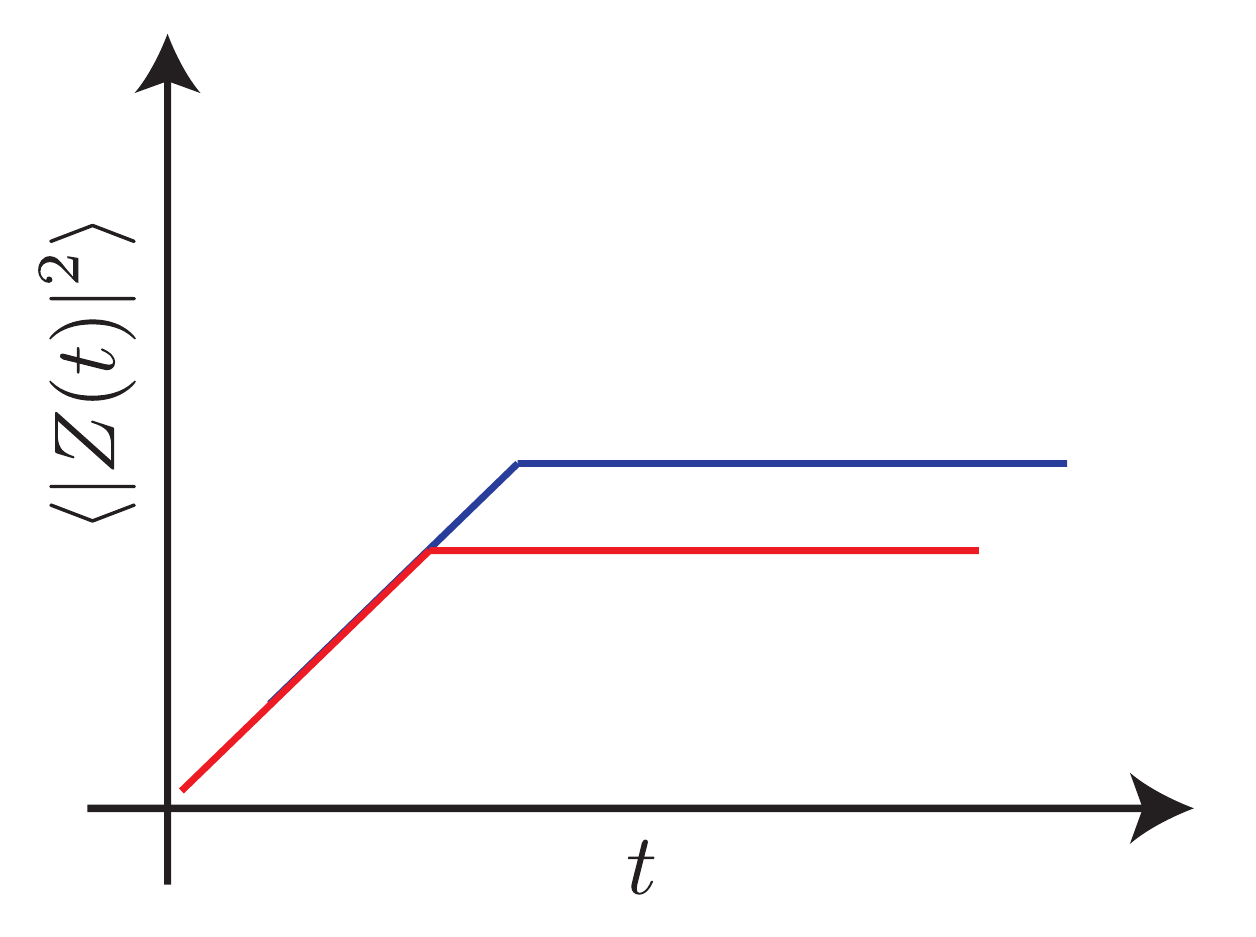}
\caption{[Left] Equal width slices of the energy spectrum marked with different colors. [Right] Ramp and plateau coming from two slices marked with the corresponding colors. }
\label{multipleramps}
\end{figure}

This is plotted in the right panel of Fig.~\ref{multipleramps} for two different energy slices with different values of $\rho$.  We called the linear $t$ behavior  at early time the ramp.  It is the time domain reflection of spectral rigidity.   The Thouless energy determines the earliest time for which the linear ramp behavior holds -- the ramp time, $t_{\rm ramp}$.    In GUE this time is order one, set by the width of the band.

 We call the constant region the ``plateau" and the time at which it begins the ``plateau time, " which is of order $L$.\footnote{Note that from the definition \eqref{sffdef} the presence of a plateau of height exactly $L$ at sufficiently large $t$ is manifest.  The factor of $1/\pi$ here is an artifact of the approximations involving $\rho$.}

We now discuss the consequence of having an eigenvalue density $\rho({\bar E})$ that varies across the band.   We divide the entire energy band up into a large number of regions of equal energy width narrow enough so that  $\rho$ is roughly constant in each.  Changing the constant value $\rho(\bar E)$ in \eqref{sinekernel} does not change the overall form of \eqref{rampplateau}.  It only changes the plateau time and the height of the plateau (see Fig.~\ref{multipleramps}).   This means that summing up the contributions from each region to compute  the total spectral form factor is very simple.  Along the ramp one just adds up linear functions with equal coefficients.    The final plateau height is just $L$, the total number of eigenvalues.  The only significant change is a rounding of the sharp transition to the plateau, because different regions of the spectrum have different plateau times.   The size of this effect depends on the shape of $\rho(E)$ and depends on the system being studied.  

This insensitivity of the ramp means that fluctuations from sample to sample of the eigenvalue density do not affect the ramp structure, making it a robust diagnostic of the Thouless time even without unfolding.    This is in contrast to the number variance, discussed above.

There is a problem, though,  with using the spectral form factor to diagnose  spectral rigidity when it is very long ranged because the ramp time will be very short and the height of the ramp will be very small compared to the plateau.   So small effects can hide the ramp.  The GUE ensemble offers a simple example of this effect when the variation of the spectral density is included. The leading contribution at relatively early times to the spectral form factor is just given by the Fourier transform of $\rho_{\rm sc}(E)$,
\begin{eqnarray}
\langle |Z(t)|^2\rangle \sim L^2 \left|\int dE \rho_{\rm sc}(E) e^{-iEt}\right|^2 = L^2 \Big(\frac{ J_1(2t)}{t}\Big)^2 \sim L^2 \frac{\cos^2(2t)}{t^3}
\end{eqnarray}

The $\frac{L^2}{t^3}$ behavior for $t \gg1$ is due to the sharp square root edge of the semicircle distribution and dominates until the ``dip time" $t_{\rm dip}$ when it becomes smaller than the ramp, i.e., at a time when $L^2/t_{\rm dip}^3 \sim t_{\rm dip}$, $t_{\rm dip} \sim L^{1/2}$.\footnote{Many plots illustrating these phenomena are displayed in \cite{ Cotler:2016fpe}. Another approach was discussed in \cite{Gaikwad:2017odv}, where authors use a non-Gaussian potential for RMT ensemble to change the edge.
}    This  early time ``slope" obscures the ramp until after $t_{\rm dip}$, but the ramp time is typically much earlier than this.  The necessity of following the ramp underneath the slope in SYK was already emphasized in  \cite{Cotler:2016fpe}.

There are strategies to deal with this.  In GUE one can just subtract off the disconnected part of the spectral form factor, computing
\begin{eqnarray}
\langle|Z(t)|^2\rangle_c = \langle|Z(t)|^2\rangle - |\langle Z(t) \rangle|^2 ~.
\end{eqnarray}
But this does not work in the SYK model, as explained in \cite{Cotler:2016fpe}.   This model has a square root edge in its eigenvalue density
 \cite{Cotler:2016fpe, Bagrets:2016cdf, Stanford:2017thb} but fluctuations in the edge position from sample to sample are only suppressed by a power of $N$, the number of fermions, rather than $L$ (equal to $2^{N/2}$ in SYK) as is the case in GUE.  This fluctuation converts the power law slope into a Gaussian falloff in $|\langle Z(t) \rangle|^2 $.  But fluctuations in the edge position cancel out in 
 $\langle|Z(t)|^2\rangle$, leaving the $1/t^3$ slope.    
 
In this paper we employ the alternate strategy, suggested to us by Douglas Stanford, of using a type of microcanonical ensemble with a Gaussian window to suppress the contribution of the edge.  
To see the onset of the ramp we introduce a variant of the spectral form factor 
\begin{eqnarray}\label{YY*definition}
|Y(\alpha,t)|^2=\left|\sum_i e^{-\alpha E_i^2 -itE_i}\right|^2. 
\end{eqnarray}
In other words, we replace the energy spectrum $\rho(E)$ with $e^{-\alpha E^2}\rho(E)$. This is in the spirit of a microcanonical density of states centered around $E=0$ but with smooth edges to the energy window that will not create spurious signals in the time domain.\footnote{  Of course we could choose to center this window around any energy we shoose, but to maximize the number of eigenvalues we choose to center it around $E=0$.  We expect the same universal features to be present as long as the center of the window is not extremely close to the edge of the spectrum.}
Although this changes the global shape of the energy spectrum, 
locally it is just a multiplication by a constant factor, and hence it does not spoil the RMT universality of the energy correlations.

When $\alpha$ is sufficiently large, 
any sharp edge of the spectrum is replaced with a Gaussian.
Then $Y$, which is essentially the Fourier transform of this `Gaussian-filtered energy spectrum', 
decays as a Gaussian at early time.   
Hence the non-universal part can be strongly suppressed, while preserving the universal part.

We also introduce the normalized quantity $h(\alpha,t)\equiv |Y(\alpha,t)|^2/|Y(\alpha,t=0)|^2$.  Note $h(0, t) = g(t)$.

In the GUE case  at early time
\begin{eqnarray}
|Y(\alpha, t)|^2 \sim L^2 e^{-t^2/\alpha^2} ~.
\end{eqnarray}
This becomes of order one at $t \sim \sqrt{\log L}$.  So spectral rigidity can be probed out to this scale.  In fact $t_{\rm ramp} \sim 1$ here, so we cannot fully diagnose this system using this technique.  In the many-body systems we will study we expect $t_{\rm ramp}$ to be large enough that, in principle at least, it can be fully revealed by this measure.

In our analysis we will also discuss the analog of the spectral form factor for the Haar random CUE ensemble.   Here the relevant quantity is the ($k$-th moment) of the trace 
\begin{eqnarray}
\langle |\Tr(U^k)|^2\rangle_{\rm CUE} = \Big\langle \sum_{ij} e^{i(\theta_i -\theta_j)k} \Big\rangle_{\rm CUE} =  \int d\theta_1 d\theta_2 L^2 R(\theta_1, \theta_2) e^{i(\theta_1 -\theta_2)k}
\end{eqnarray}
Using the exact formula for the pair correlation function \eqref{sinekernelhaar} and doing the integral we get \cite{1999chao.dyn..6024H, Diaconis-Evans}
\begin{eqnarray}
\langle |\Tr(U^k)|^2\rangle_{\rm CUE} = 
\begin{cases}
k, \quad k < L\\
L, \quad k \geq L
\end{cases}
\end{eqnarray}
Note that the ramp and plateau are exact for the CUE, even at finite $L$.  There is no slope because there is no eigenvalue edge.  No unfolding is necessary because the eigenvalue density is  uniform. 

With these preliminaries out of the way we now turn to a systematic analysis of the models.   We first discuss numerical data on time independent Hamiltonian systems, beginning with $k$-local systems.

\section{Time Independent Hamiltonian Systems}\label{timeindependenthsystems}
\subsection{$k$-local Hamiltonian Systems}\label{klocalhsystems}
$k$-local systems have been studied as a good model for scrambling in black holes \cite{Hayden2007,Susskind2008}.   
These models thermalize very quickly, with a dissipation time of order one and a scrambling time of order $\log N$.  Diffusive effects will also be very rapid.   Here we will see numerically that $t_{\rm ramp}$ is also short, with a weak $N$ dependence.   In Section \ref{sec:hamestimate} we will give an analytic argument that $t_{\rm ramp}$ also scales as $\log N$ but the physical mechanism is different from scrambling.

\subsubsection{SYK Model}\label{sec:SYK-numerics}

Let us start with the SYK model \cite{Sachdev:1992fk,Kitaev_talk,Kitaev2017awl,Maldacena:2016hyu}. 
This is a quantum mechanical model of $N$ Majorana fermions satisfying the anticommutation relation 
\begin{eqnarray}
\{\hat{\psi}_a,\hat{\psi}_b\}=\delta_{ab}. 
\end{eqnarray}
The dimension of the Hilbert space is $L = 2^{N/2}$.
The Hamiltonian is given by 
\begin{eqnarray}
H
=
\frac{1}{4!}\sum_{a,b,c,d=1}^NJ_{abcd}\hat{\psi}_a\hat{\psi}_b\hat{\psi}_c\hat{\psi}_d, 
\end{eqnarray}
where $J_{abcd}$ is a Gaussian random number with a mean zero and width $\sqrt{\frac{6}{N^3}}J$, 
and totally antisymmetric with respect to $a,b,c,d$. 
The disorder average (the average with respect to $J_{abcd}$) is taken after calculating observables.

In the first and second panels of Fig.~\ref{fig:Y-original-SYK-spectrum}, 
$h$ with $N=32$ has been plotted for various values of $\alpha$. 
 As mentioned above  the spectral form factor of the SYK model has a $\frac{1}{t^3}$  `slope' which hides the onset of the ramp.   This is visible in the $\alpha = 0$ curves where $h(t) = g(t)$ in Fig.~\ref{fig:Y-original-SYK-spectrum}.   
In Fig.~\ref{fig:Z-Y-comparison} (using the optimal value of $\alpha$)  we can see that the ramp continues all the way down to  the Gaussian envelope, and may well continue past it.  This intersection time, which we call $t_\mathrm{min}$, gives an upper bound to the ramp time $t_\mathrm{ramp}$.
\begin{figure}[t]
\includegraphics{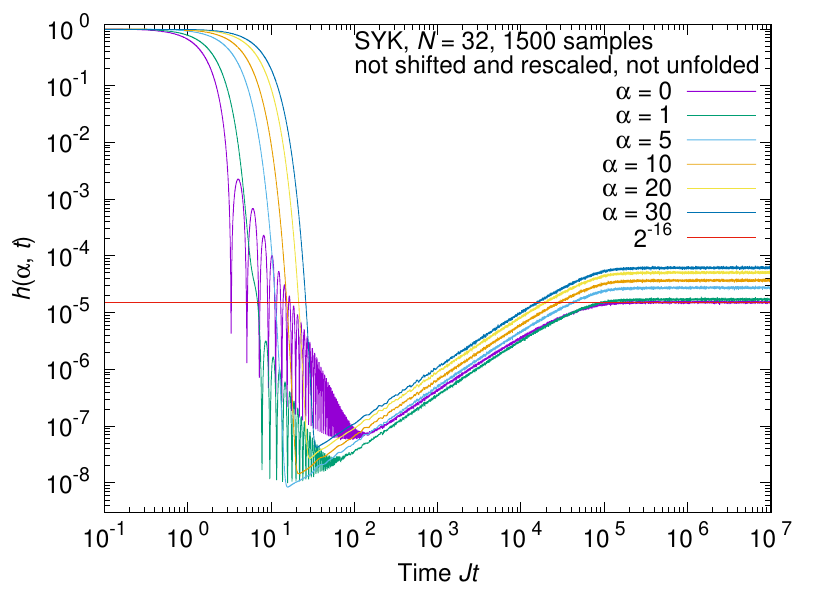}
\includegraphics{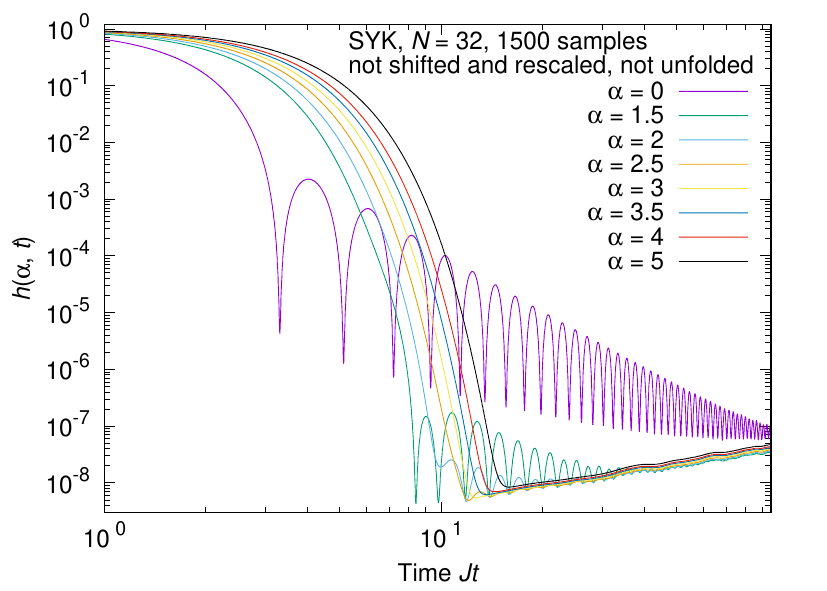}
\caption{$h(\alpha,t)\equiv |Y(\alpha,t)|^2/|Y(\alpha,t=0)|^2$ at $N=32$, 
with various values of $\alpha$. Note that $g(t) = h(\alpha =0, t)$.}\label{fig:Y-original-SYK-spectrum}
\end{figure}
\begin{figure}[ht]
\begin{center}
\includegraphics[width=0.6\textwidth, clip]{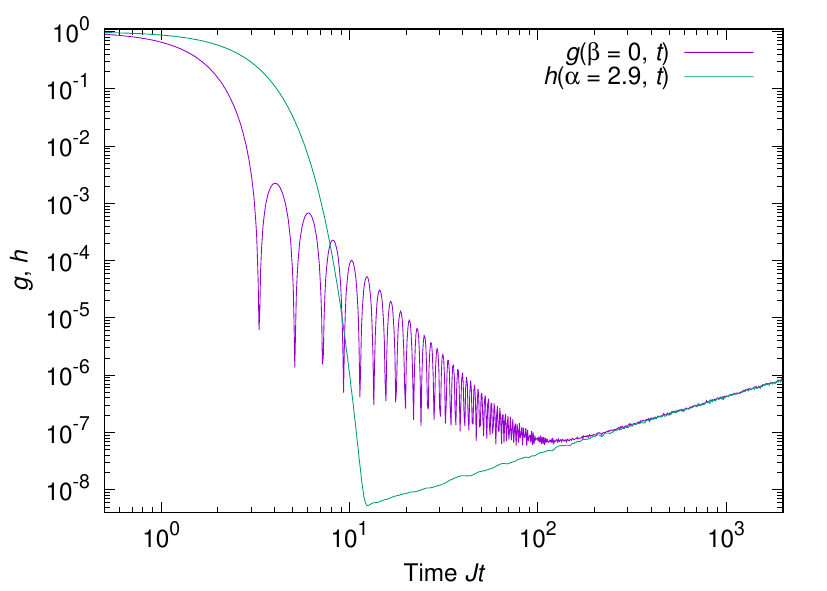}
\end{center}
\caption{
$g(t)=|Z(t)|^2/|Z(0)|^2$  and
$h(\alpha,t)=|Y(\alpha,t)|^2/|Y(\alpha,t=0)|^2$ with $\alpha=2.9$
for $N=32$, 1500 samples. }\label{fig:Z-Y-comparison}
\end{figure}
In the data displayed the ramp extends down to a time $t_{\mathrm{min}}$ of order 10. The plateau time is of order $10^5$ here, so this represents a ratio of $10^4$ in time or energy scales.  Note that there are only $2^{16} \sim 66\ 000$ energy levels\footnote{After removing the two fold fermion parity degeneracy which is the only degeneracy present for $N=32$ \cite{Cotler:2017jue}.} in the spectrum so rigidity extends across an appreciable fraction of the entire spectrum!\footnote{Of course a precise comparison would require evaluating various numerical factors of order one.} 
 We should emphasize that as expected $t_\mathrm{min}$ is far less, over an order of magnitude here, than $t_{\rm dip}$.
   
It seems from the graph that the ramp probably extends somewhat further but is just being masked by the Gaussian envelope.  
For this finite value of $N$ it is impossible to do better because an envelope that decays faster in time corresponds to a Gaussian filter that is broader in energy and extends closer to the edge of the spectrum, allowing more contamination from the sharp edge.   But as $N$ grows this effect goes away because the edges of the spectrum are at energies $\pm cN$ (even though the standard deviation of the energy is of order $\sqrt{N}$).   The Gaussian filter lets in an edge signal in $\langle |Y(\alpha, t)|^2\rangle$ of order $e^{- 2\alpha (c N)^2 +2 s_0 N}/t^3$ where $s_0 N$ is the zero temperature entropy.   The early time ramp signal is of order one. So choosing $\alpha$ as small as $\frac{s_0}{c^2 N}$ is sufficient to mask the slope contribution from the edge.   The Fourier transform of the Gaussian envelope decays  $\sim e^{2 s_{\infty} N - \frac{t^2}{4\alpha}}$ where $s_{\infty}N$ is the infinite temperature entropy.  So an $\alpha \sim 1/N$ is small enough for the  Gaussian envelope to become order one in a time of order one, allowing the study of $t_{\rm ramp}$ as short as order one.    Thus this quantity provides a reliable definition of the Thouless time in SYK and other many-body systems with similar properties.  The required $N$ values may well be computationally prohibitive though.

We discuss our detailed algorithm for determining  $t_\mathrm{min}$ in Appendix~\ref{sec:t_min}. The oscillations in the data make this determination quite noisy. The values listed in  Table~\ref{tbl:tmin} in the Appendix provide upper bounds for $t_\mathrm{ramp}$.    The best we can say about the $N$ dependence is that it is weak. Power laws faster than $N^1$  and exponentials,  with order one coefficients, are disfavored.   A $\log{N}$ behavior would certainly be consistent.

To gain more experience we now examine a $k$- local model without a sharp edge in its density of states. 

\subsubsection{2-local Randomly Coupled Qubits (RCQ)}\label{sec:2-local-RCQ-Hamiltonian}

We now consider  the 2-local-RCQ model.\footnote{This model is often referred to as the ``Quantum Spin Glass" \cite{erdHos2014phase} . 
We see no evidence of a spin glass phase in our system so we prefer a different name.} This is a model of $N$ spin-$(1/2)$ particles on a complete graph that interact with couplings of random strength.  It has the same  kind of $k$-local connectivity as the SYK model and hence the same pattern of sparseness in its Hamiltonian matrix.   We therefore expect similar scaling properties of its random matrix statistics.
 The Hamiltonian is of the form 
\begin{align}
H = \frac{1}{\sqrt{8(N-1)}} \sum_{i<j}^{N} \sum_{\alpha, \beta =0}^3 J_{ij}^{\alpha \beta} \sigma_{i}^\alpha \otimes  \sigma_{j}^\beta
\end{align}
where $\sigma^\alpha_i$ represents an element of the set $\{\sigma^0,\sigma^1,\sigma^2,\sigma^3\}=\{ I, X, Y, Z \}$ acting at the $i$-th site and couplings $J_{ij}^{\alpha \beta}$ are randomly drawn from a normal distribution. This choice of normalization is analogous to the typical SYK normalization discussed in Section \ref{sec:SYK-numerics}, such that 
\begin{eqnarray}
\langle \overline{\Tr} H^2 \rangle_{J}  = \frac{1}{L} \langle \Tr H^2 \rangle_{J} = N
\end{eqnarray}
where $L = 2^N$ is the dimension of the Hilbert space, $\langle \cdot \rangle_{J}$ denotes  ensemble averaging over Gaussian couplings $J_{ij}^{\alpha \beta}$ and $\overline{\Tr}$ denotes the normalized trace. 

We plot the eigenvalue density in Fig.~\ref{Fig:QSG-var-dos-1}.\footnote{We give details about this figure and present other data about this model, including nearest neighbor eigenvalue histograms and the number variance in Appendix~\ref{sec:2-local-RCQ}.}
We see that
the energy distribution function $\rho(E)$ has a Gaussian tail for sufficiently large $N$ $(N \gg 4)$ (see \cite{erdHos2014phase} for discussion of the spectral density).
Therefore the slope behaves like $e^{-Nt^2}$ which decays very quickly, unlike SYK.

\begin{figure}
\includegraphics{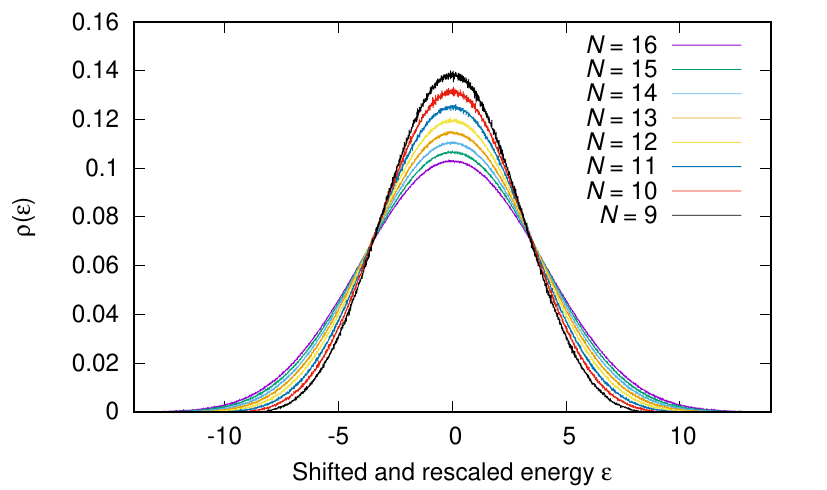}
\includegraphics{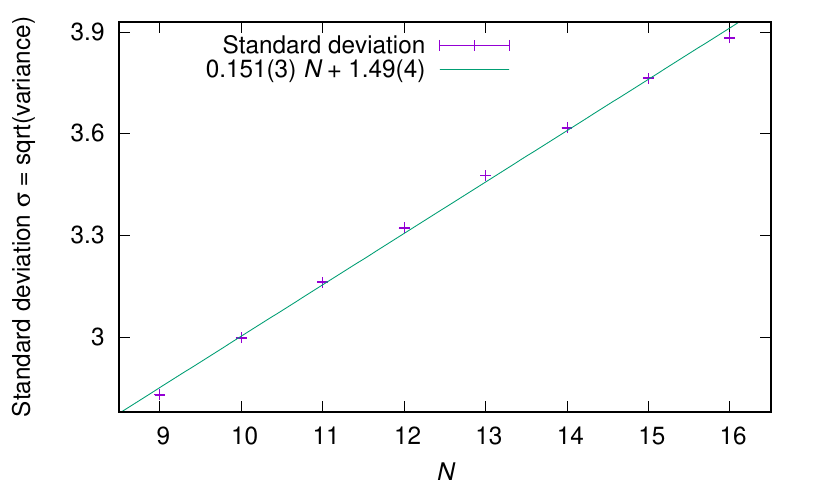}
\caption{
The density of shifted and renormalized eigenvalues for the 2-local RCQ model for $N=9,10,\ldots,16$,
and the averaged variance of eigenvalues with a straight line fit.
$(2^{16-N}\times 100)$ samples have been used for each $N$.
}
\label{Fig:QSG-var-dos-1}
\end{figure}
%
\subsubsection*{Spectral Form Factor}\label{sec:sff-RCQ}
\hspace{0.51cm}

The spectral form factor for the $2$-local RCQ system with $N$ from $8$ to $16$ is shown in Fig.~\ref{fig:sff-k-QSG}. 
There are several noteworthy features of this graph.  First, as just discussed, the slope decays very quickly. 
Then there is a bump.  This is followed by a short intermediate region.   Then a ramp begins, at quite early times. 
\begin{figure}[!ht]
  \centering
    \includegraphics[width=0.6\textwidth, clip]{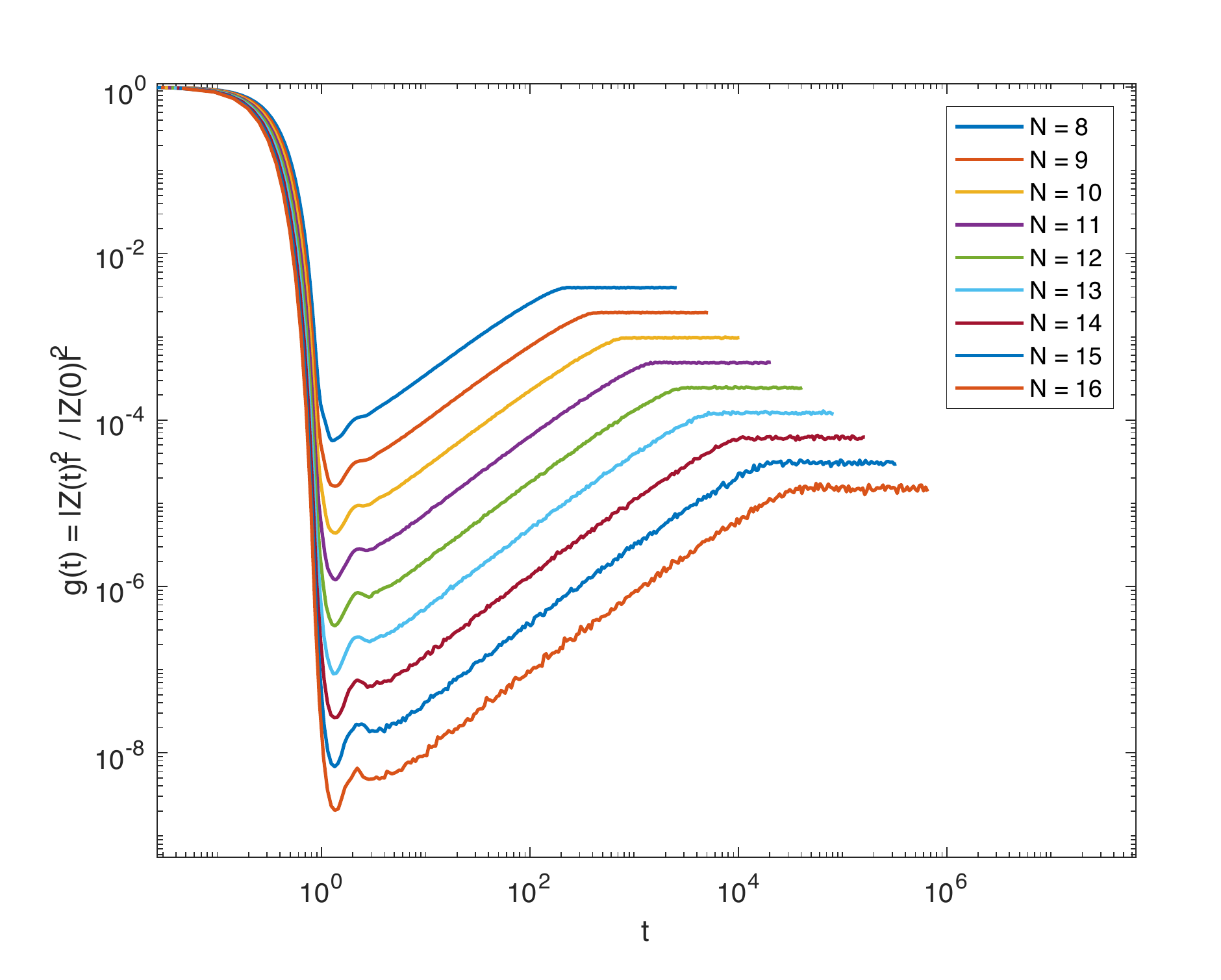}
      \caption{The spectral form factor for the $k$-local randomly coupled qubits (RCQ) with 8 to 16 qubits. We observe the Gaussian decay at the early times, follow by a `bump', ramp and plateau.  }\label{fig:sff-k-QSG}
\end{figure}
To study the bump we have experimented with using $YY^*$. We were unable to isolate a sizable transition region to the ramp while excluding the bump.   We do not understand its origin, unfortunately.

 Nonetheless it is clear that the ramp begins early, at times between 2 and 5, increasing slowly with $N$.    For $N =16$ (see Fig. \ref{fig:sff-k-QSG-fractional-error}) the plateau time is about $10\ 000$ times greater than the ramp time.   The entire spectrum contains $2^{16} \sim 66\ 000$ eigenvalues.  So as in the SYK model we see rigidity extending across an appreciable part of the entire spectrum.   

We now attempt a more quantitative study of the $N$ dependence of $t_\mathrm{ramp}$. In the absence of a detailed theory of the early time behavior we proceed phenomenologically.
\begin{figure}[!ht]
  \centering
    \includegraphics[width=0.6\textwidth, clip]{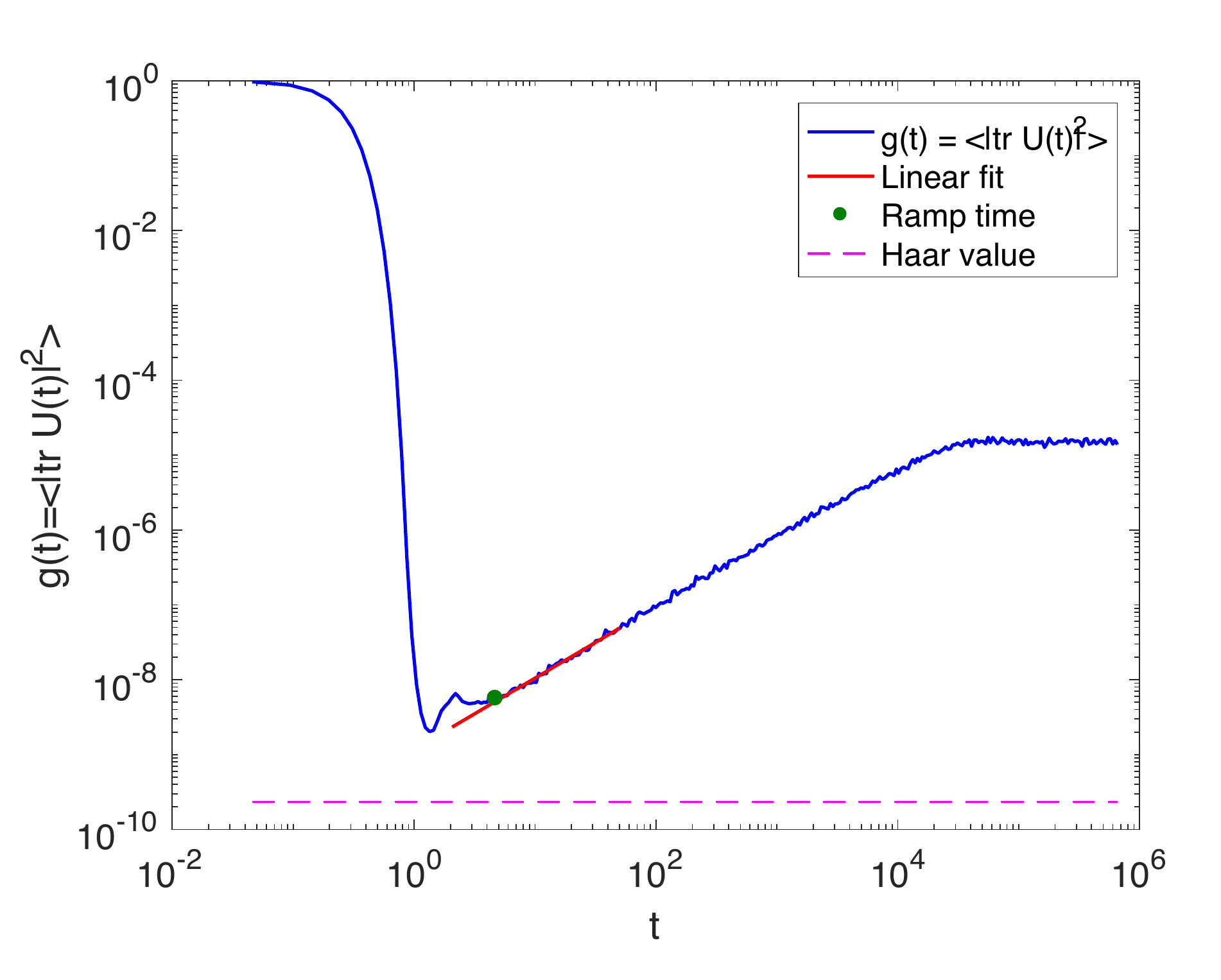}
     \caption{The spectral form factor for the $k$-local randomly coupled qubits (RCQ) for $N=16$ qubits is given by the blue curve. We fit a line to the ramp (red line) and determine the time where the fractional error from the linear fit $\epsilon(t)$ equal to $5~\%, 10~\%, 20~\%$ or $30~\%$. In the plot the green dot represents the ramp time estimate when fractional error is $10~\%$. }\label{fig:sff-k-QSG-fractional-error}
\end{figure}
\begin{figure}[!ht]
  \centering
    \includegraphics[width=0.6\textwidth, clip]{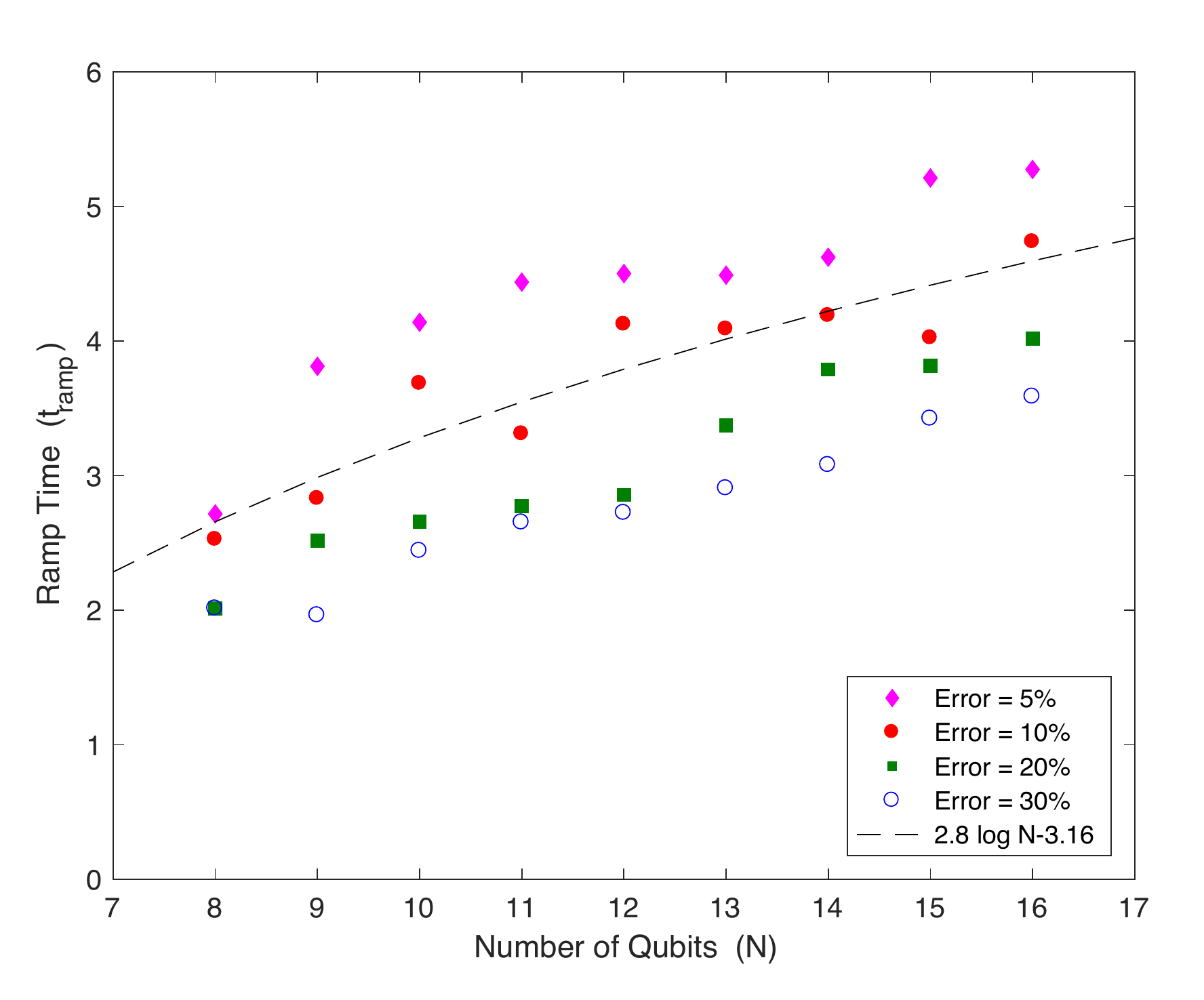}
      \caption{We plot the estimate of $t_\mathrm{ramp}$ for different $N$ and different choices of fraction errors. The dashed line is the $t(N) = A \log N +B$ fit to the red dots. We find that the best fit for ($\epsilon = 10~\%$) is given by $t_\mathrm{ramp} = 2.8 \log N - 3.16$. }\label{fig:sff-k-QSG-ramp-time}
\end{figure}
We define the time scale where the bump ends and the ramp starts as the ramp time,  $t_\mathrm{ramp}$.
In order to define the ramp time systematically, 
we quantify the distance of the spectral form factor from the linear fit to the ramp, by using the fractional error 
\begin{eqnarray}
\epsilon(t) = \frac{g(t) - g_\mathrm{ramp}(t)}{g_\mathrm{ramp}(t)}.  \label{fractional-error}
\end{eqnarray}
Here we use $g_\mathrm{ramp}(t)$ to denote the linear fit to the ramp. We consider four different estimates for the ramp time, picking the points where $\epsilon(t)$ equal to $5~\%, 10~\%, 20~\%$ and $30~\%$. 
See Fig.~\ref{fig:sff-k-QSG-fractional-error} for an example of an actual fit. 
We perform this procedure for all the values of $N$ ranging from 8 to 16 and plot the estimates of the ramp time with respect to $N$. In Fig.~\ref{fig:sff-k-QSG-ramp-time}, we plot an estimate of the ramp time for four different values of the fractional error. Different values of the error systematically shift the curve. 

As mentioned previously our theoretical expectation is that $t_\mathrm{ramp}$ scales like $\log{N}$ although for reasons unrelated to the logarithmic behavior of the scrambling time. 
So in Fig.~\ref{fig:sff-k-QSG-ramp-time}, we fit a function of the form $A \log N +B$ to the data with $\epsilon =10~\%$.
(For other values of the fractional error the fit will have the same $A$, but different values of the offset $B$.)  
$t_{\rm ramp}$ is consistent with $A \log N +B$, although it is far from conclusive.  A power law like $N^{1/2}$ fits equally well.    A more rapid power, faster than $N^1$ or so, or an exponential dependence, with order one coefficients is disfavored.  Larger $N$ would be required to distinguish these functional forms. 

To help disentangle the physical mechanisms at work here we now turn to a geometrically local system.

\subsection{Local Hamiltonian Systems}\label{sec:local-Hamiltonian}
We will consider two examples of one dimensional local Hamiltonian systems, a nearest neighbor coupled RCQ model and the XXZ chain with random magnetic field.   
\subsubsection{Local Randomly Coupled Qubits (Local-RCQ)}\label{sec:LRCQ}

In geometrically local systems (which we will refer to as ``local", as opposed to $k$-local) the various interesting time scales become distinct from one another and so it is easier to diagnose the effect of different physical mechanisms. 
\begin{figure}[h]
  \centering
    \includegraphics[width=0.6\textwidth, clip]{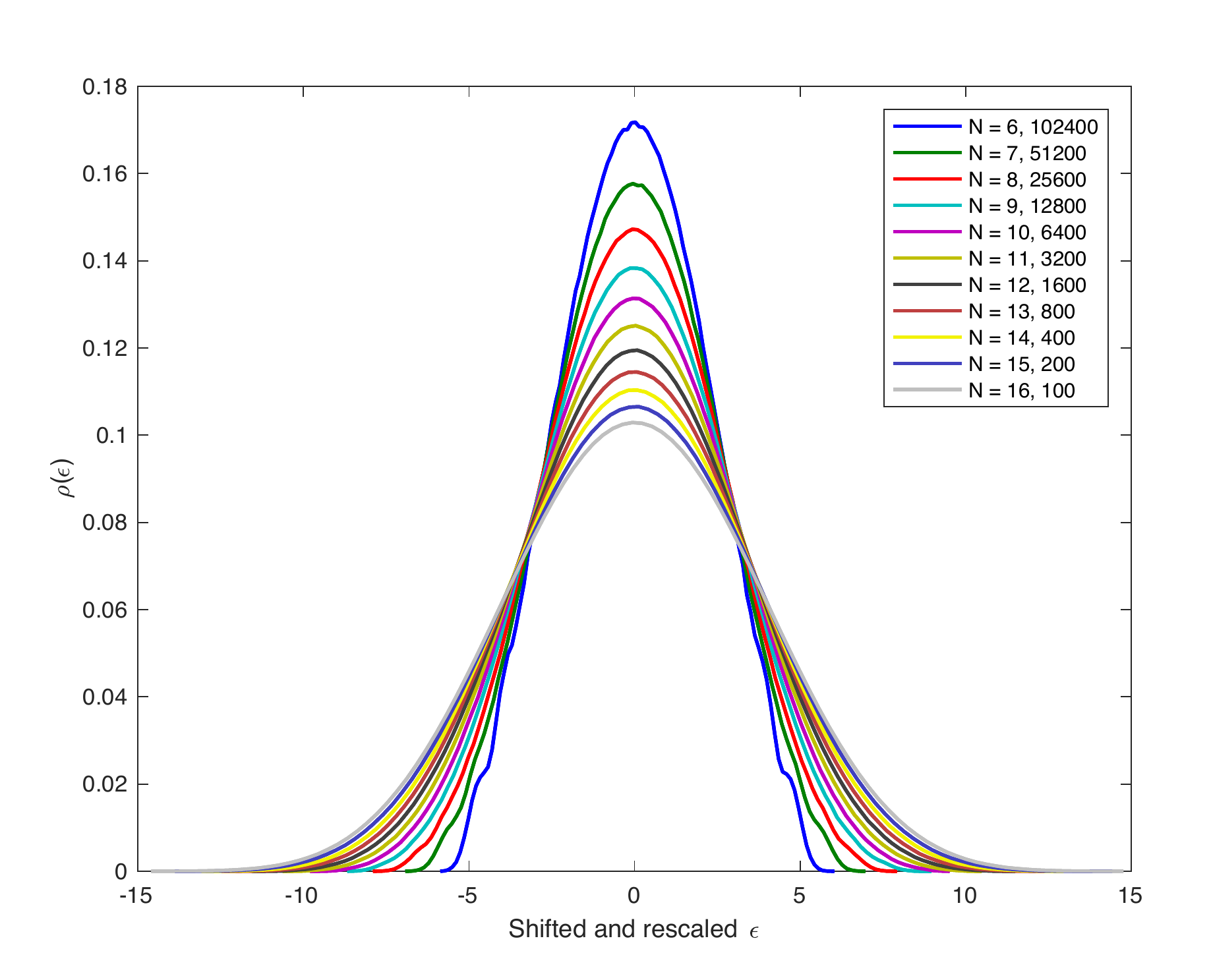}
      \caption{Spectrum of the local spin chain (Local-RCQ) for $N=6$ to 16, as well as number of samples used for each $N.$ Spectrum appears close to Gaussian as low as $N=6$, where it starts developing edges characteristic to semi-circle distribution. }\label{fig:SpectrumN6-16-LQSG}
\end{figure}
 In a one dimensional local spin system we expect the dissipation time $t_{\rm diss} \sim 1$, the scrambling time $t_{\rm scr} \sim N$ and the diffusion time $t_\mathrm{diff} \sim N^2$.   
These time scales vary more quickly with $N$ than in the $k$-local case making them easier to study numerically.  In this system we find that $t_{\rm ramp} \sim N^2$ and hence conjecture that it is related to diffusion, not scrambling.

Let us start with a one-dimensional local randomly coupled qubit model with nearest-neighbor random interactions, 
\begin{align}
H = \frac{1}{4} \sum_{i=1}^{N-1} \sum_{\alpha, \beta =0}^3 J_{i}^{\alpha \beta} \sigma_{i}^\alpha \otimes \sigma_{i+1}^\beta, 
\end{align}
where parameters $J_{i}^{\alpha \beta}$ are normal random variables with mean zero and variance one, 
so that $\overline{\Tr} H=0$ and  $\overline{\Tr} H^2=N-1$. 
This is analogous to SYK normalization.\footnote{ The Trotterization of the Hamiltonian with this scaling will correspond to $N$ parallel gates acting on the system in an order one time.  This is the normalization we will use in our random circuit discussion.}
We will explore the properties of this system numerically. 

\begin{figure}[!h]
  \centering
    \includegraphics[width=0.6\textwidth, clip]{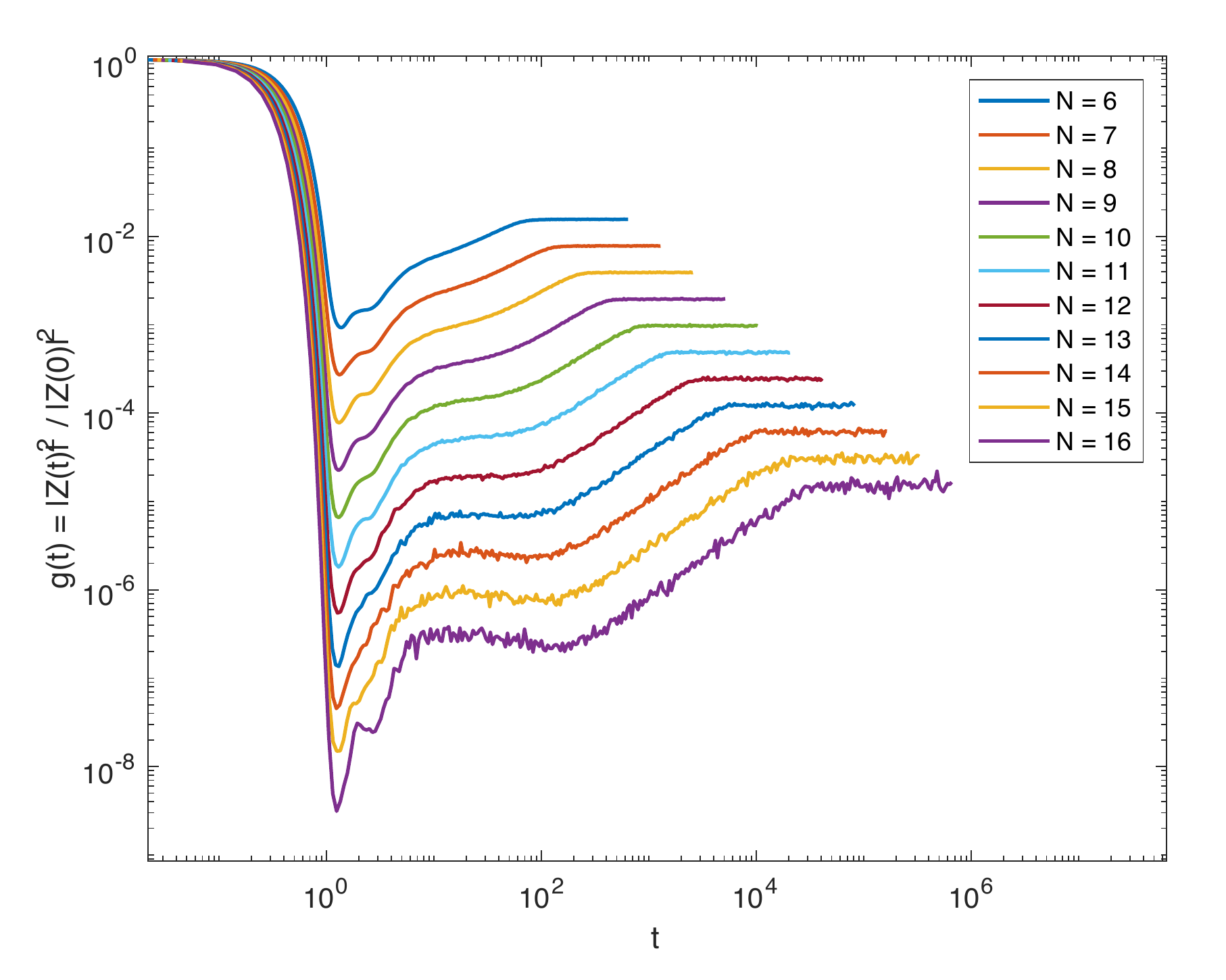}
      \caption{The spectral form factor for the local randomly coupled qubits (Local-RCQ) with 8 to 16 qubits. We observe universal behavior of  $g(t)$; Gaussian decay at the early times, follow by an intermediate region, then a ramp and plateau. The time scale where the intermediate region ends and the ramp starts we will call the `ramp time'. }\label{fig:LQSG-SFF}
\end{figure}

\subsubsection*{Energy Spectrum}
\hspace{0.51cm}
The spectrum of this Hamiltonian was studied numerically \cite{Keating-Linden-Wells-1, Keating-Linden-Wells-2}, 
and shown to have a Gaussian shape. 
We repeated the same calculation as a sanity check. 
The energy spectra for $N=6,7,8,\cdots,16$ are shown in Fig.~\ref{fig:SpectrumN6-16-LQSG}. 
The spectrum becomes closer to Gaussian as $N$ becomes larger. 
We have also looked at nearest-neighbor level spacing distribution and do not see any hints of a MBL phase anywhere in the spectrum.

\subsubsection*{Spectral Form Factor}\label{sec:sff-LRCQ-Loc}
\hspace{0.51cm}
The spectral form factor is shown in Fig.~\ref{fig:LQSG-SFF}. 
We can see that an intermediate region emerges as $N$ grows. Note that this region lasts much longer than than in the $k$-local spin chain.  We do not have a theoretical understanding of this region and regard it as  an interesting topic for future research.  Here we proceed phenomenologically and define the ramp time to be where this region ends and the ramp starts.

In order to define the ramp time systematically, we adopted two methods (see Fig.~\ref{fig:LQSG-tramp-fit}):
\begin{itemize}
\item
Pick up the points where the fractional error $\epsilon(t)$ defined by \eqref{fractional-error} decreases to certain values, as we have done for the $k$-local spin chain. 
We have used $\epsilon(t)$ equal to $5~\%, 10~\%, 20~\%$ and $30~\%$.

\item
Fit the intermediate region and ramp by powers of $t$ (straight line in a log-log plot) and identify the intersection of two fitting lines with $t_\mathrm{ramp}$. 

\end{itemize}
\begin{figure}[t]
  \centering
    \includegraphics[width=0.6\textwidth, clip]{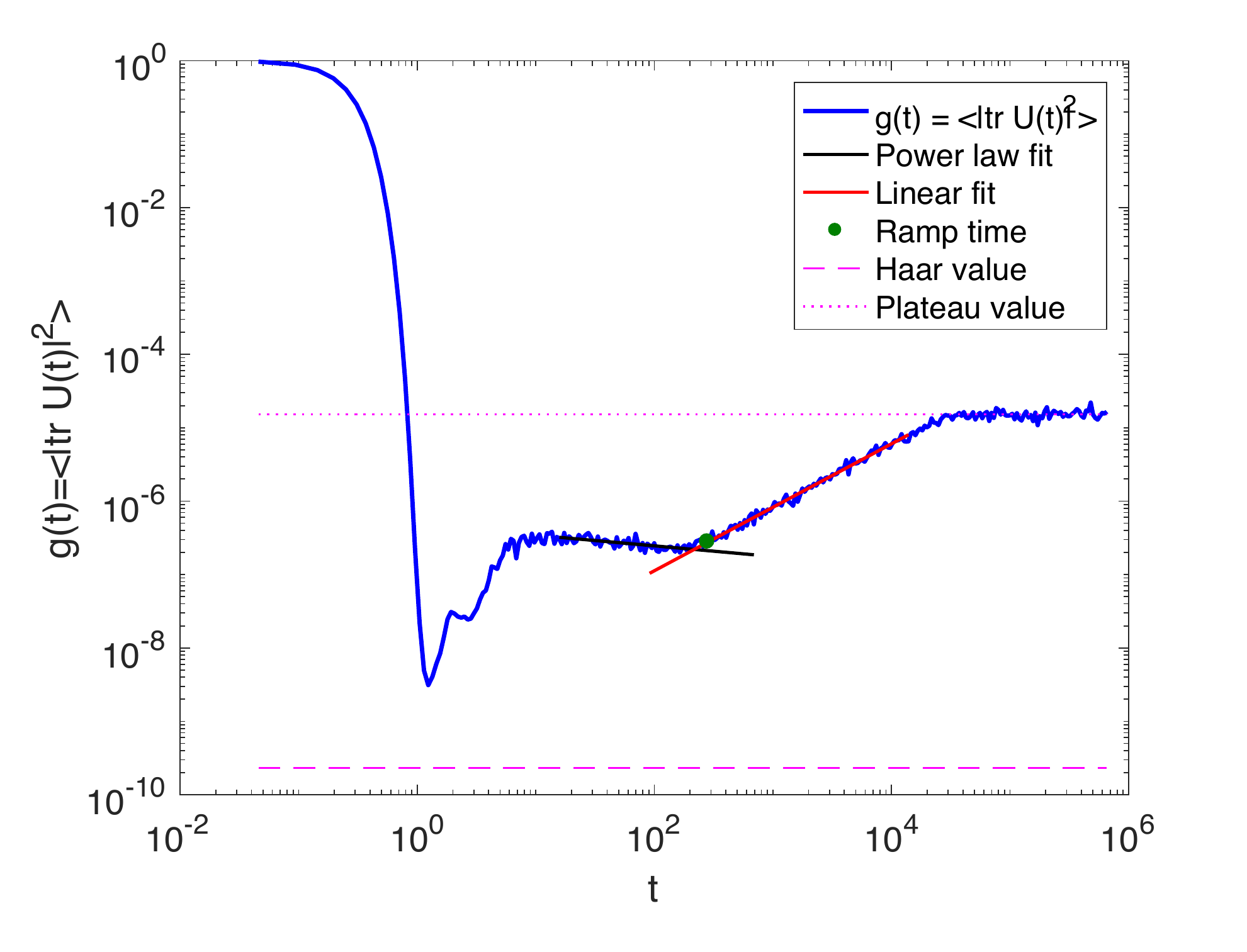}
     \caption{The spectral form factor for the local randomly coupled qubits (Local-RCQ) for $N=16$ qubits is given by the blue curve. We fit two different lines in the log-log plot to the bump and the ramp (two red lines) and determine, the first, time scale where two red lines intersect and, the second, where fractional error of the blue curve from the best fit of the ramp is $\epsilon$. In the plot the green dot represents the ramp time estimate with the second method, when fractional error is $5~\%$.}\label{fig:LQSG-tramp-fit}
\end{figure}
We do this procedure for all the values of $N$ ranging from 8 to 16 and plot the estimates of the ramp time with respect to $N$. 
In Fig.~\ref{fig:tramp-vs-N-local-QSG}, we plot estimates of the ramp time determined in these ways. 
The ramp time can be fit reasonably well by an ansatz
\begin{eqnarray}
t_{\mathrm{ramp}}(N) = A N^2 +B,  
\end{eqnarray}
as shown in Fig.~\ref{fig:tramp-vs-N-local-QSG}. 
Here we used $t_{\mathrm{ramp}}$ determined from the intersection of the power law fit to the intermediate region  and a linear fit to the ramp; 
other choices lead to more or less the same $A$, but different values of the offset $B$. The best fit has $A = 0.85 \pm 0.09$ and $B=-14.57 \pm 6.28$. A linear fit to the graph is significantly worse:
a parabolic fit has smaller mean square error and the coefficients are of order one. 
Hence we conjecture that $t_{\mathrm{ramp}}$ should be related to $t_{\mathrm{diff}}$ rather than $t_{\mathrm{scr}}$.  We will give analytic evidence for this below.
\begin{figure}[!ht]
  \centering
    \includegraphics[width=0.6\textwidth, clip]{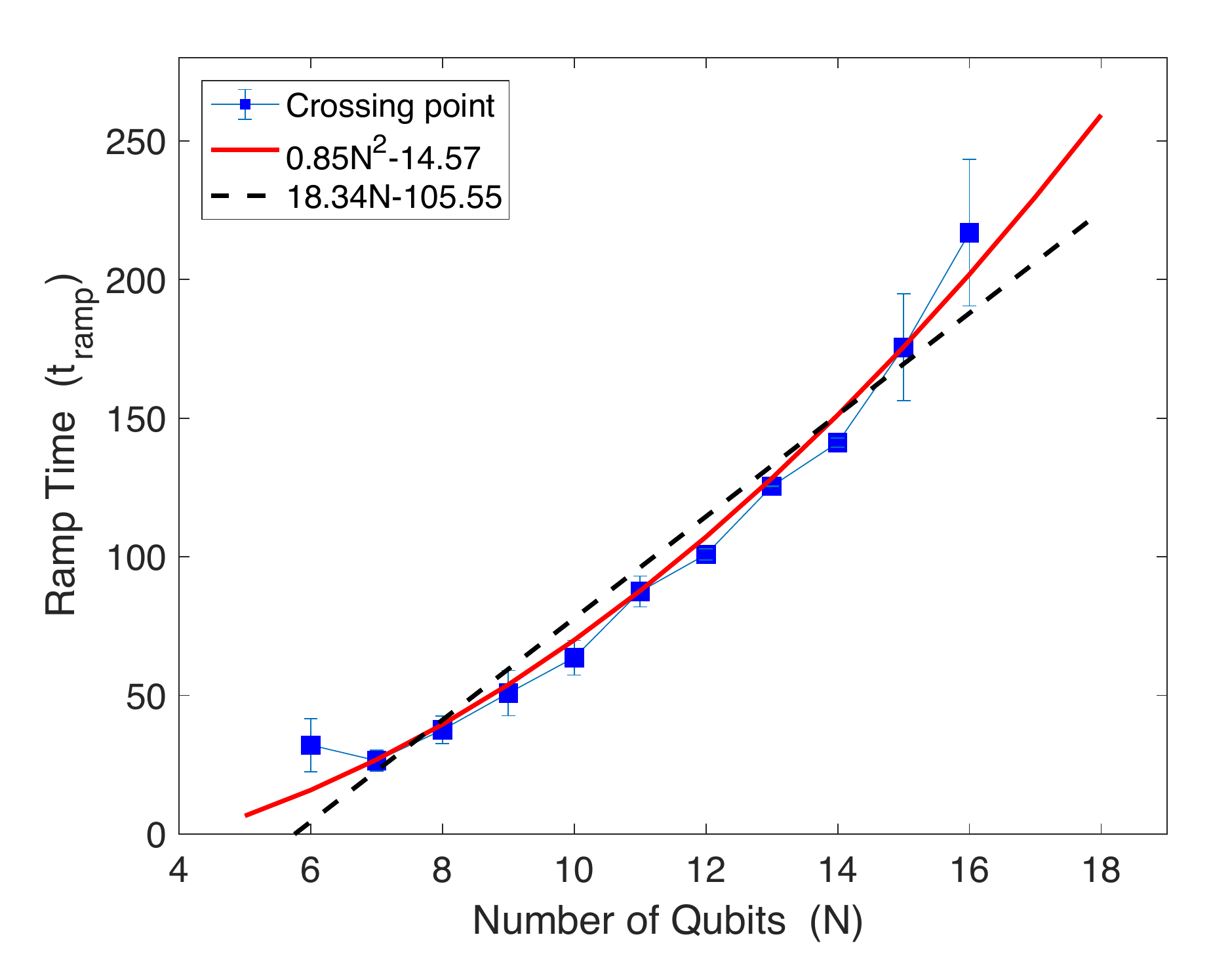}
      \caption{The blue dots represent the estimates of the ramp time from the crossing point of the intermediate region power law fit and the linear fit to the ramp. The error bars illustrate the difference between this estimate and the ramp estimate using an error of $\epsilon=20\%$. The best fit is given by the red parabola and the alternative, linear fit is show as a black dashed line.}
      \label{fig:tramp-vs-N-local-QSG}
\end{figure}

\subsubsection{XXZ Model} \label{sec:XXZ_Hamiltonian}

Next let us consider the one-dimensional XXZ Hamiltonian system with random magnetic field.   This model has many interesting features, including a very well studied MBL phase \cite{HusePal2010, Luitz2015}.   We are interested in the model because it has another conservation law besides energy -- the $z$ component of spin.    This conservation law can be preserved (unlike energy) in a random quantum circuit as we will explore in Sec.~\ref{sec:spin-conserving-circuit}.  The random circuit analysis allows us to make the connection between $t_{\rm ramp}$ and $t_{\rm diff}$ manifest.   Here we present some preliminary numerical results for the time independent Hamiltonian system.  The intermediate region here grows as one approaches the MBL phase.  We are unsure of the significance of this result.

\begin{figure}[t]
  \center
    \includegraphics[width=0.55\textwidth, clip]{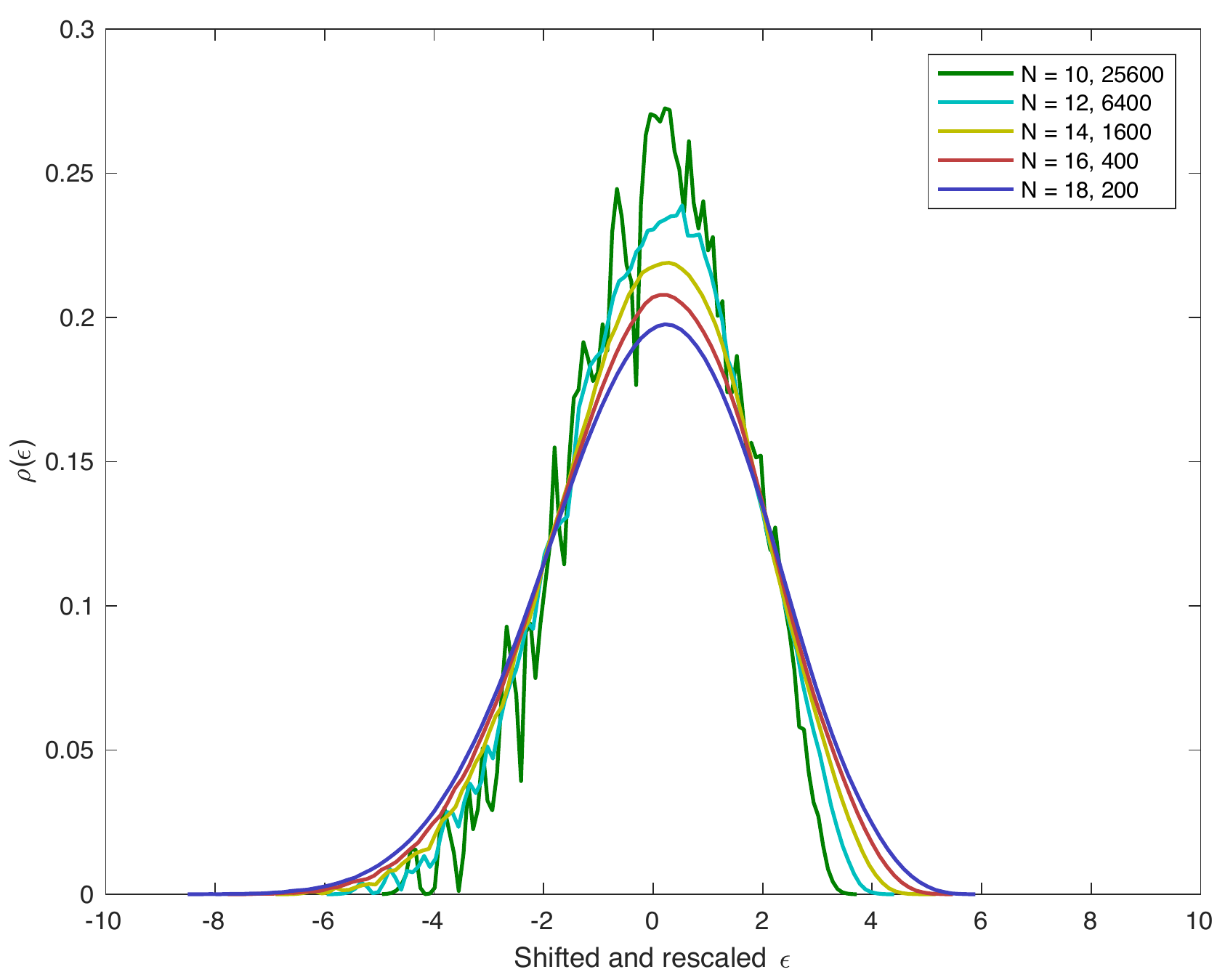}
      \caption{The spectrum of the XXZ Hamiltonian for $N=10, 12, 14, 16, 18$, $W=0.5$ and spin zero sector.  }\label{fig:XXZ-spectrum}
\end{figure}
The Hamiltonian we consider is 
\begin{align}
H =  \sum_{i=1}^{N} \Big( \frac{1}{4} \vec{\sigma}_i \cdot  \vec{\sigma}_{i+1} +  \frac{\omega_i}{2} \sigma^3_{i} \Big)
\label{Hamiltonian-XXZ}
\end{align}
where $\vec{\sigma}_i=(\sigma_i^1,\sigma_i^2,\sigma_i^3)$ are Pauli matrices and $\omega_i$,  a random magnetic field, 
is a random number selected from the uniform distribution on the interval $[-W, +W]$. 
Here we use periodic boundary conditions, so the sums are modulo $N$.
It has been argued that this model is in the MBL phase for $W \geq 2.75$ \cite{Luitz2015}. 

This Hamiltonian commutes with the total spin along the $z$-direction, $S_z^{\rm tot}=\frac{1}{2}\sum_i\sigma_i^3$. 
Thus we will need to look into specific spin sectors to detect nearest-neighbor Wigner--Dyson statistics. Unless otherwise specified we will consider the total spin zero sector for an even number of spins.\footnote{
If we do not restrict the spin, we expect a  Poisson distribution. }

\subsubsection*{Energy Spectrum}\label{sec:number-variance-XXZ}
\hspace{0.51cm}

We are interested in the diffusive regime, where the Hamiltonian exhibits chaotic properties and Griffiths effects, contributions of rare many-body localized regions, are not yet significant \cite{agarwal2017}.
So we will investigate this Hamiltonian for $W=0.5$ which is in the heart of diffusive regime.
The energy spectrum is shown in Fig.~\ref{fig:XXZ-spectrum}. 
We have checked that the nearest-neighbor level correlations are consistent with the GOE ensemble
when we look into each conserved spin sector separately.

\subsubsection*{Spectral Form Factor}\label{sec:sff-RCQ-Loc}
\hspace{0.51cm}

In this section, we study $|Z|^2$ for the XXZ model. 
Again we take $W=0.5$, which is in the heart of the diffusive phase. 
The qualitative behavior of $|Z|^2$ is similar to the local RCQ, however it is more sensitive to finite $N$ effects in the edge of the spectrum. 

\begin{figure}[t]
  \centering
    \includegraphics[width=0.6\textwidth, clip]{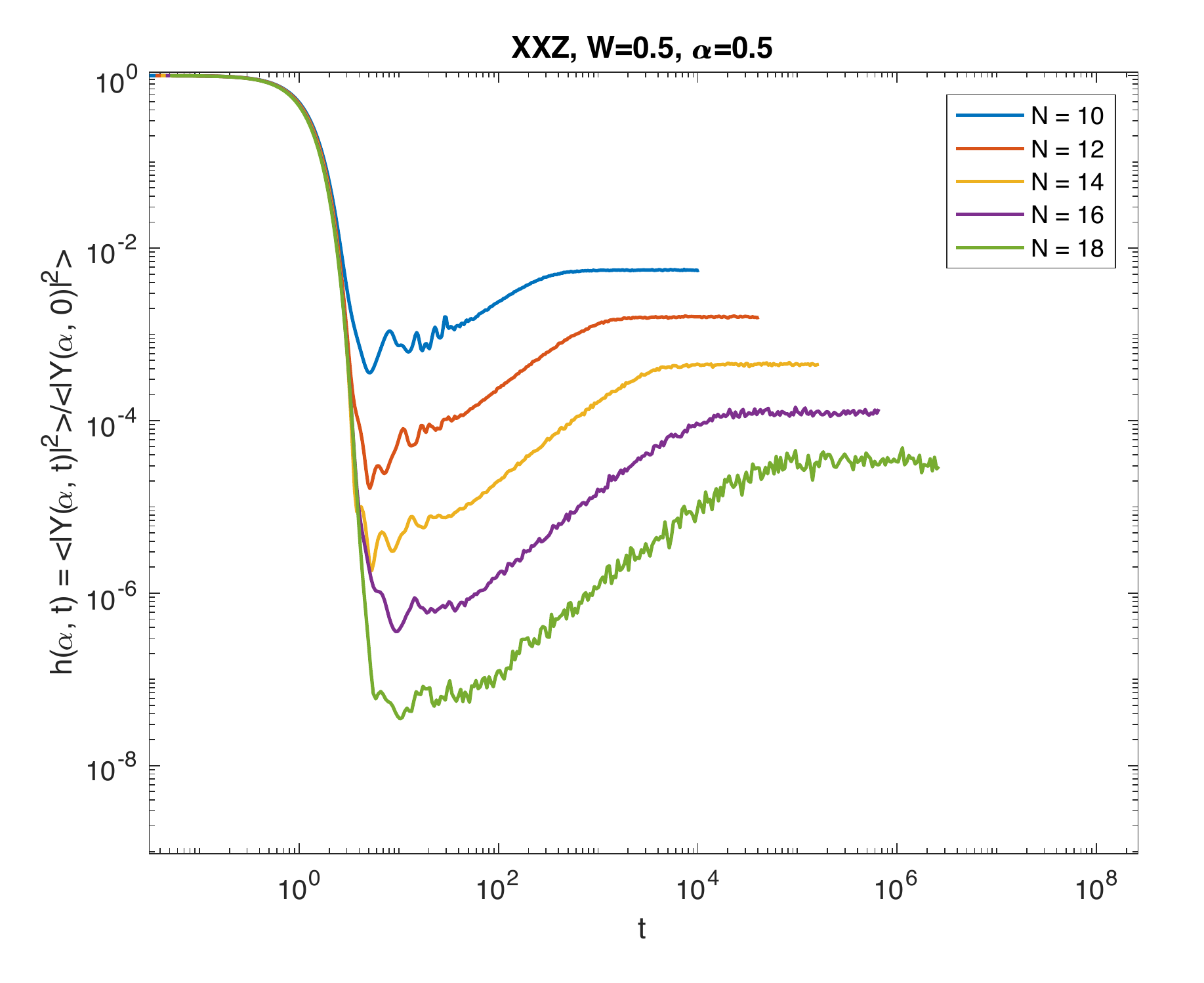}
      \caption{The spectral form factor for the XXZ Hamiltonian with 10 to 18 qubits for the S=0 sector and $W=0.5$. We observe universal behavior for the $h(\alpha, t)$, Gaussian decay at the early times, flat region, ramp and plateau. The time scale where the `bump' ends and the ramp starts we will again call ramp time. }\
\end{figure}

\begin{figure}[h]
\includegraphics[width=0.5\textwidth, clip]{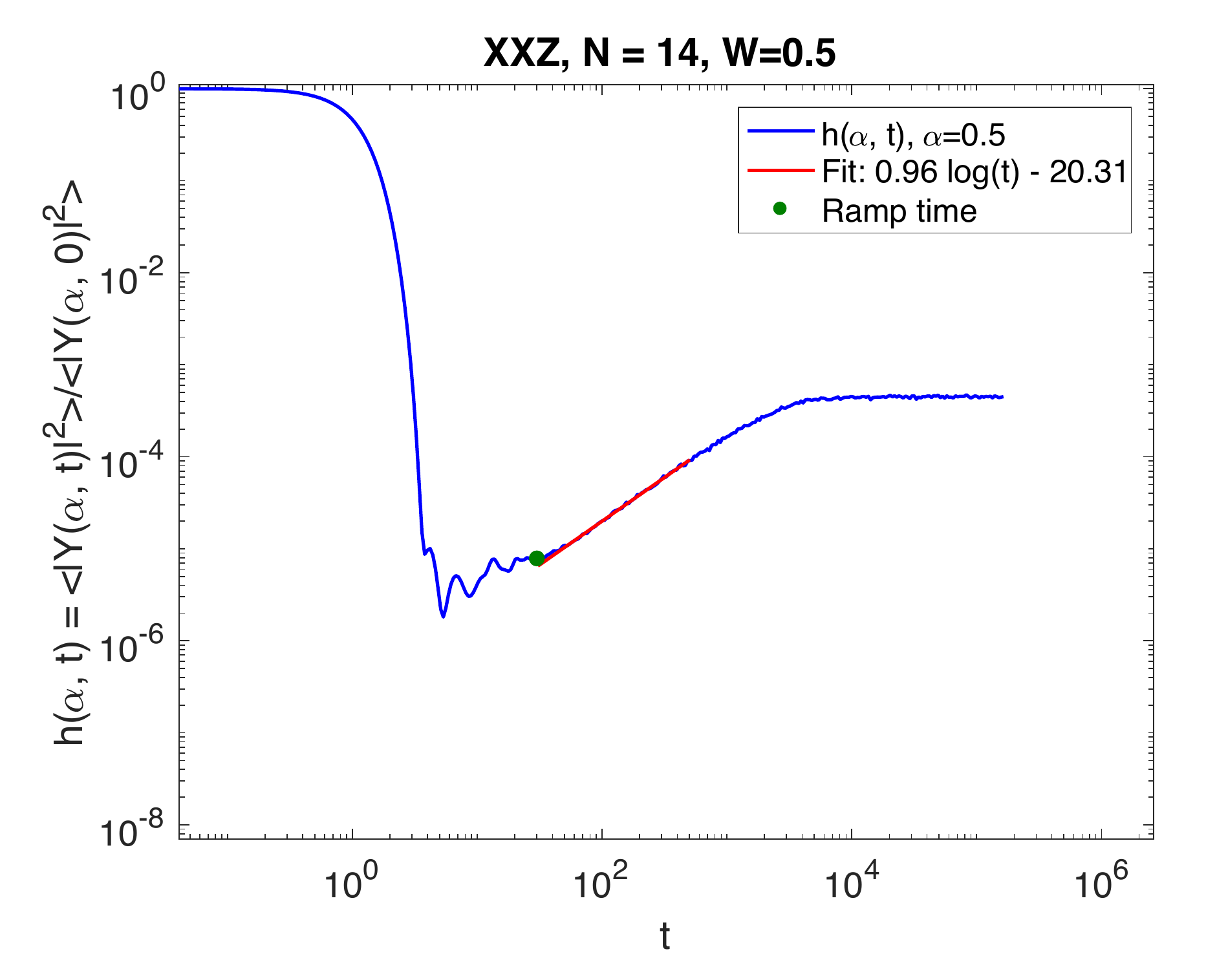}
\includegraphics[width=0.5\textwidth, clip]{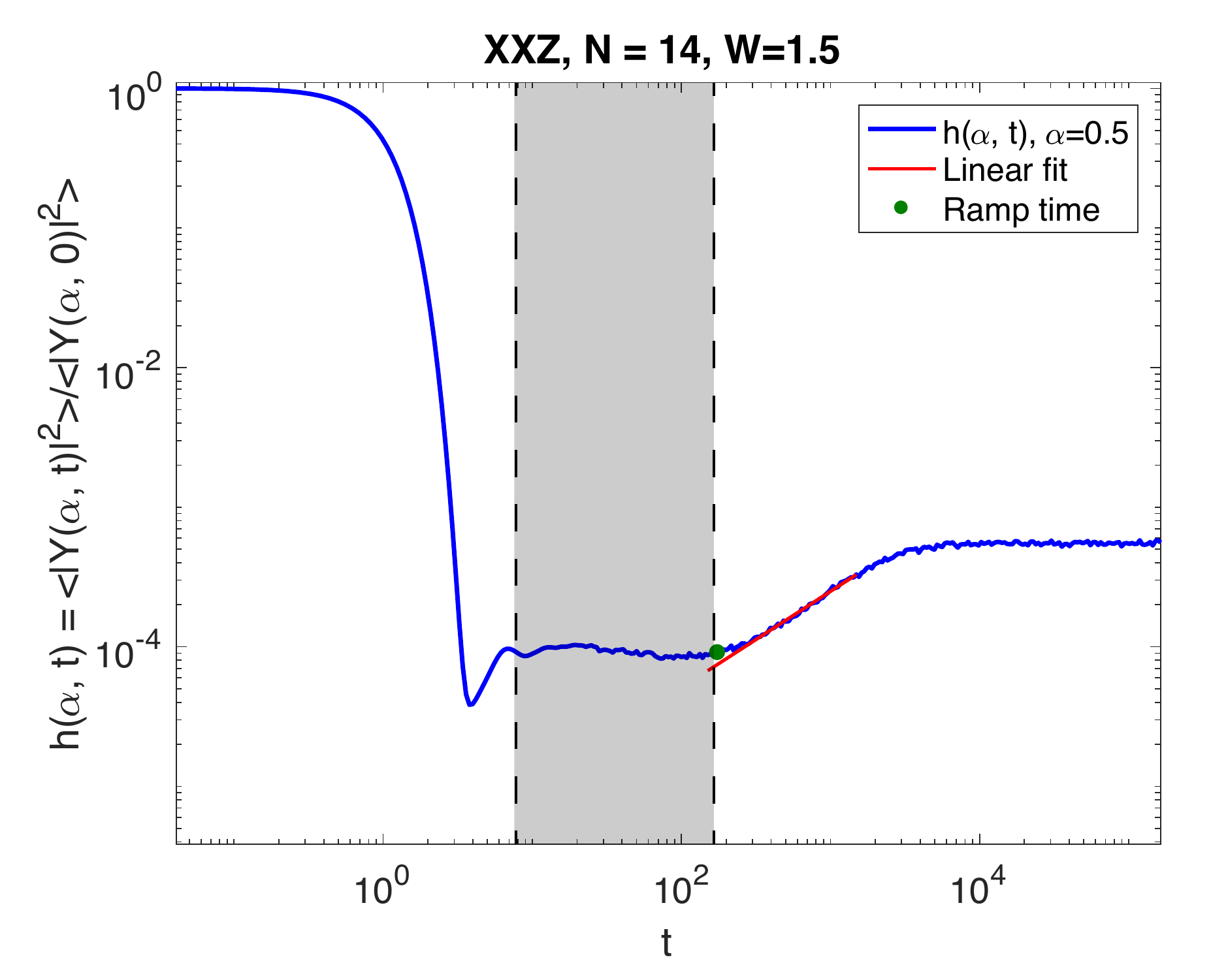} 
\caption{[Left] $h(\alpha =0.5, t)$ for the XXZ Hamiltonian in the ergodic regime $(W=0.5)$ for $N=14$ qubits and total spin zero sector. [Right] The $h(\alpha =0.5, t)$ for the XXZ Hamiltonian in the subdiffusive regime $(W=1.5)$ for $N=14$ qubits. The shaded region highlights an almost constant region of the spectral form factor whose origin we would like to explore. }
\label{Fig:XXZ-SFF}
\end{figure}

As in  the local RCQ, the intermediate region lasts longer than in the  $k$-local spin chain. 
We determine $t_{\mathrm{ramp}}$ in exactly the same way as in Sec.~\ref{sec:LRCQ}. However, for this system the  numerical analysis is more subtle. For the small values of $W$ that are in the diffusive regime there are some approximate conservation laws that produce oscillating behavior at  early time and
make it difficult to determine the ramp time accurately so we do not present results.

If one increases  $W$ the system moves toward the MBL phase and a rather long intermediate region  appears in the spectral form factor (see Fig.~\ref{Fig:XXZ-SFF}). At this point, we do not have a good theory to explain the shaded region for values of the $W\geq 1$. Perhaps it is related to the subdiffusive regime between ergodic and MBL phases. 

These numerical results are inconclusive, 
but as we will see in Sec.~\ref{sec:spin-conserving-circuit}, the random circuit analogue of this model provides an ideal setup to test our proposal that 
$t_{\rm ramp}$ and $t_{\rm diff}$ are the same. 

\section{Random/Brownian Quantum Circuit Models}\label{sec:Random-and-Brownian-circuits}

We now discuss analytic methods.   In contrast to few-body systems analytic approaches  for time independent Hamiltonian many-body systems have not yet been fully developed.   Instead we turn to a related class of systems that display similar thermalization behavior and are analytically tractable.   These are random quantum circuit and Brownian quantum circuit models.  

In  Brownian circuit models, the Hamiltonians are chosen randomly from certain ensembles, 
independently at each small time step $dt$. After $M$ steps, the unitary evolution
will be given by
\begin{eqnarray}
U(t) = e^{-iH_M dt} e^{-iH_{M-1} dt}\cdots e^{-iH_{1} dt} 
\end{eqnarray} 
where $t = M dt$ and each of the operators $H_{i}$ is independently drawn from a given ensemble.  A random quantum circuit is a discrete version of this where each $e^{-iH_i dt}$ is replaced by a unitary drawn from a suitable ensemble of unitary ``gates."   If  the $H_i$ are chaotic we expect that $U(t)$ performs a kind of random walk on the unitary group U$(L)$, eventually converging to the Haar random ensemble.   The analogous question concerning the onset of random matrix behavior becomes the question about how quickly the spectrum of $U(t)$ starts displaying CUE statistics.   A number of recent papers have studied thermalization in various correlations functions using these techniques \cite{Lashkari2013,Shenker:2014cwa,Nahum:2017yvy, vonKeyserlingk:2017dyr, Khemani:2017nda}.

The conclusion of all this work is that appropriate  Brownian/random quantum circuits display known thermalizing behaviors --- two point function decay (dissipation); chaotic/scrambling behavior as diagnosed by OTOCs and subsystem density matrices; and diffusion --- with mechanisms and time scales that are very much analogous to fixed Hamiltonian systems.    
We will assume for now that
  Brownian/random quantum circuits will also display the onset of random matrix behavior via analogous mechanisms and with the same time scales as fixed Hamlitonian systems with the same locality properties and conservation laws.    Because of their more efficient randomization we expect that at the least the time scales in these models are a lower bound for those in the coresponding Hamiltonian systems.
  
All the results we obtain are consistent with the Hamiltonian numerics discussed above. In addition these results are consistent with a heuristic framework for an analytic argument for Hamiltonian systems we discuss in Section \ref{sec:hamestimate}. In any case we find this assumption to be a very plausible one, and now turn to working out its consequences.

 Because of the Brownian nature of these systems the
naive analog of the spectral form factor $\left\langle|{\rm Tr}U(t)|^2\right\rangle$ decays to the  value given by the average using Haar measure, which we call the Haar value, and stays there. 
We call the time when this value is attained the `Haar time' $t_\mathrm{Haar}$.

But this does not probe the full range of correlations in the spectrum of $U(t)$.  To do this imagine running the quantum circuit for a time $t$, 
and after that  repeating the same time evolution.\footnote{In condensed matter physics this is called a ``Floquet system."}  Then the time evolution from $t=0$ 
to $\tilde{t}=k t$ ($k=1,2,\cdots$) is described by $U(\tilde{t})=(U(t))^k$. The analog of the spectral form factor for time $\tilde{t}$ is then
 \begin{eqnarray}\label{utk}
 u(k, t)\equiv \left\langle|{\rm Tr}(U(t))^k|^2\right\rangle.\label{SpectralFormK}
 \end{eqnarray} 
Note that $u(k, t)=k$ for $k \le L$ when $U(t)$ is drawn from the Haar (CUE) ensemble (see discussions in Section \ref{sec:RMT-review} and Appendix~\ref{appendix-analytics-Haar}). This is just a discrete version of the ramp.  The time $t$ at which the initial moments approach this value, $t_\mathrm{Haar}$, is thus the analog of $t_{\mathrm{ramp}}$ in the Hamiltonian systems discussed in previous sections.  
Hopefully  the physical meaning of $t_\mathrm{ramp}$ can be understood if we understand the meaning of $t_\mathrm{Haar}$. 

An important question is whether $t_\mathrm{Haar}$ depends on $k$.    We are able to explore $k=1, 2$ analytically.   We then appeal to random matrix theory intuition that suggests the finer grain structure of the eigenvalue distribution is easier to establish in a chaotic system than the coarse grained structure.  Since higher $k$ probes finer structure we expect, at least for large enough $k$, that $t_\mathrm{Haar}$ will, at the least,  not increase as $k$ increases.  Our numerics, discussed below, are consistent with this and seem to indicate that $k=2$ is already in the stable range.\footnote{These results are consistent with analytic results on the SYK brownian circuit \cite{saad2018}.}

 In this section we consider  Brownian circuit versions of the RCQ and XXZ models, both local and $2$-local.\footnote{Some results on random banded matrices are presented in Appendix~\ref{sec:band-matrix}.}
In earlier sections we suggested that for geometrically local systems $t_\mathrm{ramp}$ is the diffusion time. 
An essential difference between the random/Brownian quantum circuit and time-independent Hamiltonian system is that the former does not conserve energy and hence in general there is no diffusion.  To study the role of diffusion in this context we use a random circuit where spin, not energy, is conserved.  In particular we study
 a random circuit version of the XXZ model introduced by \cite{Khemani:2017nda, rakovszky2017}
in Sec.~\ref{sec:spin-conserving-circuit}.\footnote{We are grateful to David Huse for introducing us to this model and to him and Vedika Khemani for discussions about it.} 
There we will see it is possible to relate $t_\mathrm{Haar}$ and hence $t_\mathrm{ramp}$ to the diffusion time, distinguishing it from the scrambling time.   In the $2$-local case we show that diffusion happens in order one time, and is not important for setting $t_\mathrm{\mathrm{ramp}}$.  

\subsection{Local and $k$-local RCQ Random/Brownian Circuit}\label{localandk-localRCQrandomcircuit}

\subsubsection{Analytic Estimate}
Here we provide a calculation of the spectral form factor for random circuits.  Specifically, we are interested in the behavior of $u(k,t)$ defined by \eqref{SpectralFormK}. 
We first describe the $k=1$ case, which is conceptually quite simple.

\begin{figure}
  \centering
    \includegraphics[width=0.4\textwidth]{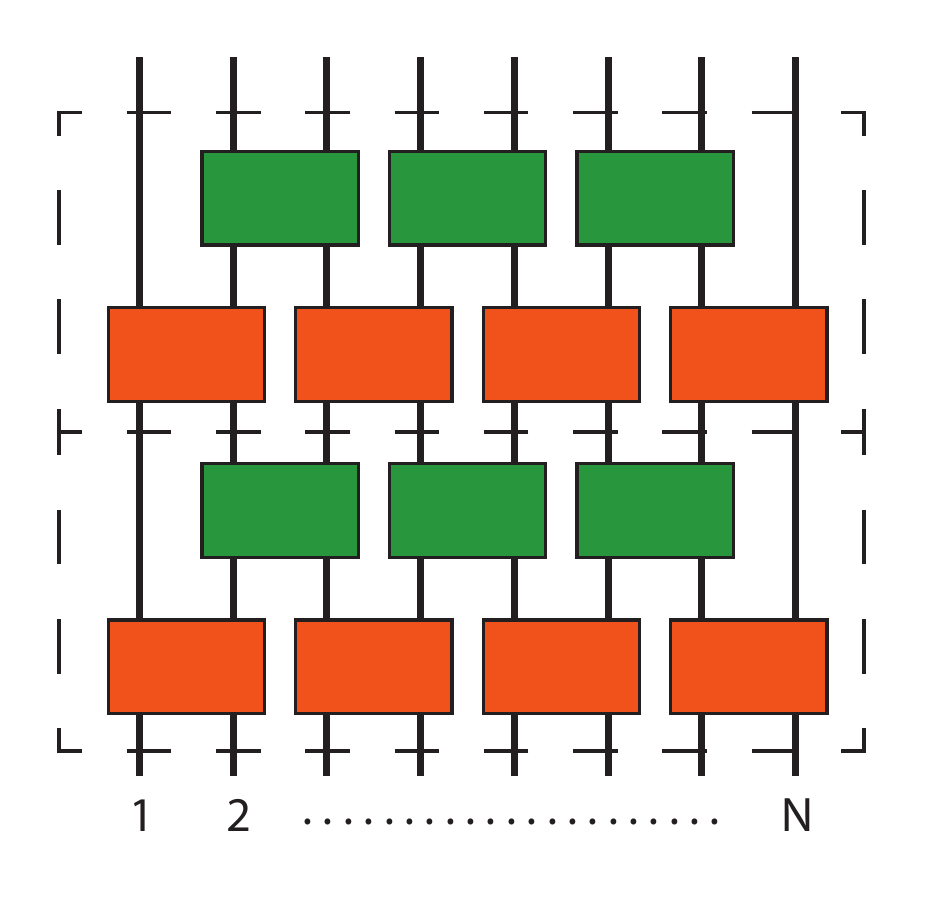}
    \includegraphics[width=0.4\textwidth]{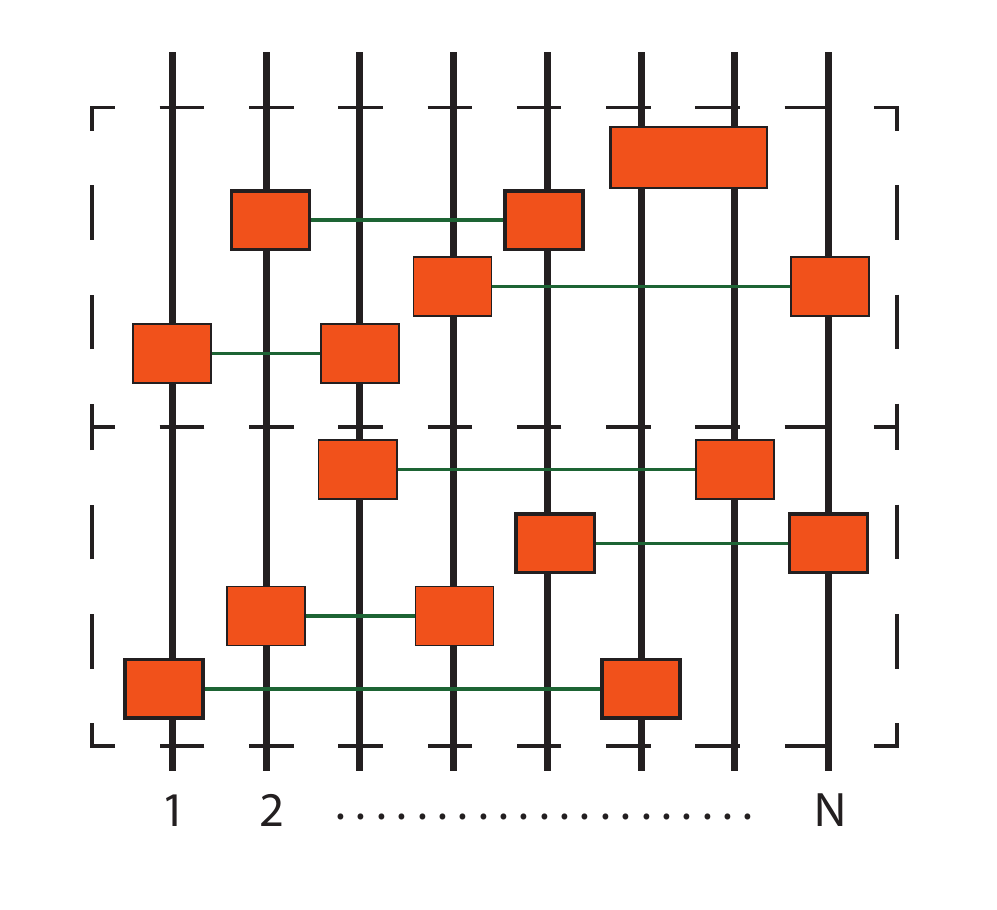}
\caption{[Left] Local random circuit evolution on $N$ qubits, each gate is randomly selected from a  gateset $\Gamma \in U(4)$. [Right] 2-local random circuit on $N$ qubits, in each step qubits are randomly paired and random gate is applied from a gateset $\Gamma$.}
\label{fig:local-random-cicuit}
\end{figure}

We use techniques presented in Ref.~\cite{Harrow2009}. We will investigate the parallelized, local random circuit model shown in the left part of Fig.~\ref{fig:local-random-cicuit}, where each step consists of two layers acting on pairs of qubits and each gate is selected at random from some 2 qubit  universal gate set  $\Gamma \subseteq {\rm U}(4)$. 
For general $k$ we can express $\Tr U^k$ as a sum of monomials, 
\begin{align}
\Tr U^k = \sum_{i_1, i_2, ..,i_k = 1}^L  U_{i_1 i_2} U_{i_2 i_3}...U_{i_k i_1}, 
\end{align}
where $L=2^N$ is the dimension of the Hilbert space. 
Therefore, \eqref{SpectralFormK} can be written as 
\begin{align}
u(k,t) = \left\langle \left|\Tr U^k \right|^2\right\rangle =  \sum_{i, j} \left\langle  U_{i_1 i_2} U_{i_2 i_3}...U_{i_k i_1} U^*_{j_1 j_2} U^*_{j_2 j_3}...U^*_{j_k j_1}\right\rangle.  
\end{align}

\subsubsection*{Calculating $\left\langle\left| \Tr \ U(t) \right|^2\right\rangle$}
We want to study 
\begin{eqnarray} 
u(1,t) = \left\langle |\Tr U(t)|^2\right\rangle = \sum_{i, j = 1}^L \left\langle U_{ii} U^*_{jj} \right\rangle.  
\end{eqnarray}
There are two types of terms we need to study: $i=j$ and $i\neq j$. Without loss of generality we can set $i=1$.

First let us consider $U_{11} U^*_{11}$. 
In a natural $0$ (up spin) and $1$ (down spin) basis, $i=1$ is $\ket{0,0,\cdots,0}=\ket{0}^{\otimes N}$, 
and 
\begin{eqnarray}
U_{11}(t)=\bra{0,0,\cdots,0}\hat{U}(t)\ket{0,0,\cdots,0}.  
\end{eqnarray}
By using a density matrix $\hat{\rho}(t)=\hat{U}(t)\hat{\rho}(0)\hat{U}^\dagger(t)$ 
with the initial condition 
\begin{eqnarray}
\hat{\rho}(0) 
=
\ket{0,0,\cdots,0}\bra{0,0,\cdots,0}
=
\left(\ket{0}\bra{0}\right)^{\otimes N} 
\end{eqnarray}
at $t=0$, we can write it as 
\begin{eqnarray}
U_{11}(t) U^*_{11}(t) = \Tr \Big[\hat{\rho}(t)\hat{\rho}(0) \Big]. 
\end{eqnarray}
The density matrix can be expanded in tensor products of strings of Pauli matrices $\sigma_p$, where $p$ is a label specifying 
a Pauli string,\footnote{
More precisely, $p$ is a set of $N$ numbers $(p_1,p_2,\cdots,p_N)$, where $p_i=0,1,2$ or $3$, and 
$\sigma_p=\sigma^{p_1}\otimes\sigma^{p_2}\otimes\cdots\otimes\sigma^{p_N}$.
Here $\sigma^0$ is the $2\times 2$ identity matrix and $\sigma^{1,2,3}$ are Pauli matrices $\sigma^{x,y,z}$.  
} as 
\begin{eqnarray}
\rho(t) = 2^{-N/2} \sum_{p} \gamma_t(p) \sigma_p. 
\end{eqnarray}
Note that $ \Tr (\rho(t)) = 1$ holds for any $t$ due to the unitarity of time evolution. 
We use $p=0$ to denote the identity matrix, and so $\gamma_t(0) = 2^{-N/2}$ for any $t$. 
By using these expansion coefficients $\gamma_t(p)$, we obtain 
\begin{eqnarray}
 U_{11} U^*_{11}  &=& 2^{-N} \Tr \Big[ \sum_{p} \gamma_t(p) \sigma_p \times \sum_{q} \gamma_0(q) \sigma_q \Big]  \\
&=& \sum_{p, q} \gamma_t(p) \gamma_0(q) \delta_{pq}  \\
&=& 2^{-N} +\sum_{p\neq 0} \gamma_t(p) \gamma_0(p). 
\end{eqnarray}

We can easily determine $\gamma_0(p)$, by noticing that 
\begin{eqnarray}
\hat{\rho}(0) 
=
\left(\ket{0}\bra{0}\right)^{\otimes N} = \left(\frac{1}{2}(I +Z)\right)^{\otimes N},    
\end{eqnarray}
where $I=\sigma^0$ and $Z=\sigma^3$. 
Namely $\gamma_0(p)$ is nonzero 
only when $\sigma_p$ is a Pauli string containing only  $I$ and $Z$'s
(equivalently, when $p$ is a set consisting only from $0$ and $3$),  
and all nonzero elements are $2^{-N/2}$.  
Therefore, 
\begin{eqnarray}
\left\langle U_{11} U^*_{11} \right\rangle
= 
2^{-N} + 2^{-N/2}\sum_{p'} \left\langle\gamma_t(p')\right\rangle , 
\end{eqnarray}
where $p'$ runs through all possible sets of $0$ and $3$ but $(0,0,\cdots,0)$. 

For a given choice of circuit the Pauli chain of $I$'s and $Z$'s evolves into other chains with positive and negative weights.  On averaging these tend to cancel and $\langle\gamma_t(p')\rangle$ decays exponentially.\footnote{In fact for gates chosen randomly from the Haar distribution on $U(4)$ the average $\langle\gamma_t(p')\rangle$ drops to zero in one step.}
A pair $II$ is left unchanged by the evolution.   So the only invariant chain is all $I$'s.  
The rate of decay of other chains is proportional to the number of $Z$'s in the chain, which we will call $d$. 
Lemma 4.1 (Section 4.1) of Ref.~\cite{Harrow2009} bounds the one and two norms of $\gamma_t(p)$ for the random circuit. 
Their bound is 
\begin{eqnarray}
\sum_{p'} \left\langle\left|\gamma_t(p')\right|\right\rangle 
\le 
\sum_{d=1}^Ne^{-\Delta t d} \sum_{d(p') = d} |\gamma_0(p')|.   
\end{eqnarray}
where $d(p')$ is a number of Pauli-$Z$'s in $\sigma_{p'}$
and $\Delta$ is the `gap parameter' which depends on the choice of the gate set $\Gamma$. 
Thus, we can bound the sum as
\begin{eqnarray}
\sum_{p'} \left\langle\left|\gamma_t(p')\right|\right\rangle 
\le 
\sum_{d=1}^Ne^{-\Delta t d} \sum_{d(p') = d} |\gamma_0(p')|.   
\end{eqnarray}

In our normalization, for both local and $k$-local RCQ, $\Delta$ should be of order 1, independent of the system size $N$.  In our case $\sum_{d(p') = d} |\gamma_0(p')|=2^{-N/2} {N \choose d}$, and hence 
\begin{eqnarray}
\left\langle U_{11} U^*_{11} \right\rangle
\le
2^{-N} + 2^{-N}\sum_{d=1}^Ne^{-\Delta t d}  {N \choose d}
=
2^{-N} + \left(2^{-N}\left(1+e^{-\Delta t}\right)^N-1\right). 
\label{U11U11*}
\end{eqnarray}

Next we study the behavior of $U_{11}U^*_{jj}$. 
The $j$-th state is represented by a sequence of $0$'s and $1$'s. As we will see, the number of $1$'s, which we denote by $q$, 
controls the decay rate. Hence let us denote the $j$-th state by $\ket{1^q 0^{N-q}} $, neglecting the ordering of $0$'s and $1$'s. 
We consider the time evolution of an operator $\hat{O}(t)$, whose initial condition is specified by 
\begin{eqnarray}
\hat{O}(0) = \ket{1}\bra{j} = \ket{0^N} \bra{1^q 0^{N-q}}.  
\end{eqnarray}
By definition, 
\begin{eqnarray}
\hat{O}(t) = \hat{U}(t) \hat{O}(0) \hat{U}^\dagger(t), 
\end{eqnarray}
and hence 
\begin{eqnarray}
U_{11} U^*_{jj}
=
\bra{0^N} \hat{U}(t) \ket{0^N} \bra{1^q 0^{N-q}}  \hat{U}^\dagger(t) \ket{1^q 0^{N-q}} 
= 
\Tr \Big[  \hat{O}(t) \hat{O}^\dagger(0) \Big]. 
\end{eqnarray}
Noticing that  
\begin{eqnarray}
\ket{0}\bra{1} = \frac{1}{2}(\sigma^1 +i \sigma^2) =  \frac{1}{2}(X +i Y),  
\end{eqnarray}
we can write $\hat{O}(0)$ as\footnote{
We have shown the expression for $\ket{j}=\ket{1,1,\cdots,1,0,0,\cdots,0}$ for simplicity. 
Expressions for other orderings can easily be obtained by changing the locations of $\sigma^{1,2}$. 
} 
\begin{eqnarray}
\hat{O}(0) 
=
2^{-N/2} \sum_{\alpha_1, .., \alpha_q =1}^2 \sum_{d=0}^N \sum_{i_1 \leq..\leq i_d} 2^{-N/2} (+i)^{f(\vec \alpha)} \sigma_1^{\alpha_1} \otimes ..\otimes \sigma_q^{\alpha_q}\otimes \sigma^3_{i_1} \otimes \sigma^3_{i_2} \otimes ...\otimes \sigma^3_{i_d}  
\nonumber\\
\label{O-zero-expansion}
\end{eqnarray}
where $f(\vec \alpha) = \sum_{i=1}^q (\alpha_i -1)$ counts the number of $\sigma^2$'s. 
By expanding $\hat{O}$ as we have done for $\hat{\rho}$ as 
\begin{eqnarray}
\hat{O}(t) = 2^{-N/2} \sum_{p} \gamma_t(p) \sigma_p,  
\end{eqnarray}
we obtain $\gamma_0(p) = 2^{-N/2} (+i)^{f(\vec \alpha)}$. 
Hence
\begin{eqnarray}
\left|
  U_{11} U^*_{jj} 
  \right|
=  
\left|
\sum_{p'}  \gamma_t(p') \gamma_0^*(p') 
\right|
\le 
\sum_{p'}|\gamma_t(p')|\cdot  |\gamma_0(p')|
\end{eqnarray}
where we sum over all the Pauli strings of the form $XYY..XIIZZ..Z$, that have $X, Y$ elements for the first $q$ sites and $N-q$ $I, Z$ elements for the remaining sites.
The coefficients of these terms will decay as
\begin{eqnarray}
\left\langle |\gamma_t(p')|\right\rangle \leq e^{- \Delta t (q+d(p'))} |\gamma_0(p')|  
\end{eqnarray}
except for $q+d(p')=0$, 
because the `conversion' can take place anywhere an $X, Y$ or $Z$ is located. 
(Note that this expression, and hence the following estimate, hold for all choices of the ordering of $0$'s and $1$'s in the $j$-th state.) 
Thus, for $q>0$ (i.e. $j\neq 1$), 
\begin{eqnarray}
\left\langle \left| U_{11} U^*_{jj}\right| \right\rangle
&\le&
 \sum_{p' } \left\langle  \ |\gamma_0(p)|^2 \right\rangle e^{-\Delta t (d(p') +q)} 
\nonumber\\
&=&
 2^{-N} e^{- \Delta t q} \sum_{d=0}^{N-q} e^{- \Delta t d} {N-q \choose d} 
\nonumber\\
 &=&
  2^{-N}e^{- \Delta t q} (1+e^{- \Delta t })^{N-q}.  
\end{eqnarray}

Now we can get a good bound for $u(1,t)$,   
\begin{eqnarray}
\left\langle  |\Tr U(t)|^2  \right\rangle
&=&
 \sum_{i = 1}^{2^N}  \sum_{j=1}^{2^N} \left\langle U_{ii} U^*_{jj} \right\rangle= 2^N \sum_{j=1}^{2^N}  \left\langle U_{11} U^*_{jj} \right\rangle
 \nonumber\\
&\leq&  
2^N 
\sum_{q=0}^N  2^{-N}  {N \choose q} e^{- \Delta t q} (1+e^{- \Delta t })^{N-q} 
\nonumber\\
&=&  
 (1+e^{- \Delta t })^{N} \sum_{q=0}^N   {N \choose q} \Big(\frac{e^{- \Delta t}}{1+e^{-\Delta t }}\Big)^q
\nonumber\\
&=&
(1+2 e^{- \Delta t })^{N}. 
\end{eqnarray}
Hence, for each fixed $N$, at sufficiently late time, 
\begin{eqnarray}
\left\langle  |\Tr U(t)|^2  \right\rangle - 1 
\lesssim
2N e^{- \Delta t }. 
\end{eqnarray}
As we have mentioned above, generic gate sets have  $\Delta$ that is $O(1)$ . 
If this inequality is almost saturated (numerical results explained below suggest this is actually the case), then 
if we plot $\frac{\left\langle  |\Tr U(t)|^2  \right\rangle - 1}{N}\sim 2e^{- \Delta t }$ as a function of $t$ for various values of $N$, 
they should lie on top of each other when $N$ is sufficiently large. 
The time scale for $\left\langle  |\Tr U(t)|^2  \right\rangle-1$ to decay as small as $\epsilon$ is 
\begin{eqnarray}
t_{\epsilon} \simeq \frac{1}{\Delta} \log\left(\frac{2N}{\epsilon}\right)
\sim 
\log N.  
\label{t_eps_bound}
\end{eqnarray}
It is clear that this estimate $t_{\rm Haar}$  does not depend on the connectivity of the circuit.

We should note that we can relate this estimate to the decay of two point functions.   As we show in \eqref{trQtrUrewrite} of Section \ref{sec:hamestimate} we can rewrite 
\begin{eqnarray}\label{kequalonetwopt}
\left\langle |\Tr U(t)|^2\right\rangle &= 1+  \sum\limits_{j=2}^{L^2}  \frac{1}{L} \Big\langle \Tr \Big( U^{\dagger}(t) P_j U(t) P_j \Big) \Big\rangle \\
&=  1+  \sum\limits_{j=2}^{L^2}  \frac{1}{L} \Big\langle \Tr \Big( P_j(t) P_j \Big) \Big\rangle
\end{eqnarray}
where the $P_j$ are a complete basis of Pauli operators acting on the $N$ qubit Hilbert space.   The decay of $u(1,t)$ is thus related to the decay of various two point functions.  The content of the random circuit analysis is that this sum is dominated by the decay of the $P_j$ containing just one Pauli.   This two point function decays like $e^{- \Delta t}$ and there are $N$ such operators.  This give the result \eqref{t_eps_bound}.

\subsubsection*{Sketch of the calculation  for $u(2,t) = \left\langle\left| \Tr \ U^2(t) \right|^2\right\rangle$}

In the above section we have estimated the convergence of the first moment $u(1,t)$ and shown the Haar time is $\log N$ for both local and 2-local circuits.   To fully understand the approach to random matrix statistics it is important to examine the Haar time for the $k$-th moment $u(k,t)$.  This is related to the study of approximate $k$ designs.  The techniques used above become quite challenging for general $k$ and have only been carefully examined for $k =2$.   In the Appendix \ref{app:k=2RCQcalc} we carry out an analysis for $u(2,t)$ using the techniques of Ref.~\cite{Harrow2009} for a particularly simple gate set, $U(4)$ random two qubit gates.  For this gate set $u(1,t)$ converges to Haar in one time step while $u(2,t)$ converges in $\log N$ time.    Here we give a summary of the argument.

The analysis for $u(2,t)$ boils down to a Markov process on two copies of a string of $N$  Paulis.   When the strings are unequal their weight vanishes in one time step.   For a more general gate set the $k=1$ rate would control their decay.   When the strings are equal, a different Markov process involving transitions between pairs of Paulis controls the convergence.  Again the strings with very few $Z$'s have the slowest convergence, and cause a deviation from the Haar value of $2$ of the form
\begin{eqnarray}
u(2,t) \sim 2 + Ne^{-\tilde{\Delta}t} ~.
\end{eqnarray}
So $u(2,t)$ becomes $\epsilon$ close to Haar in a time $\frac{1}{\tilde{\Delta}} \log \frac{N}{\epsilon}$.  Again we see a $\log N$ dependence but here $\tilde{\Delta}$ is different from the  $\Delta$ discussed above.  It is determined by the Markov chain on 
pairs of Paulis.   We note that the gap of the full Markov chain does not control the convergence of this quantity, as explained in Appendix  \ref{app:k=2RCQcalc}.

We can again interpret these results in terms of two point function decay.   The quantity described in \eqref{kequalonetwopt} vanishes in one step for $U(4)$ gates.  But we can define  ``two step" correlators using equation \eqref{trQtrUrewrite} of Section \ref{sec:hamestimate} to rewrite 
\begin{eqnarray}
\left\langle |\Tr U^2(t)|^2\right\rangle &= 1+  \sum\limits_{j=2}^{L^2}  \frac{1}{L} \Big\langle \Tr \Big( (U^{\dagger}(t))^2 P_j U^2(t) P_j \Big) \Big\rangle \\
\end{eqnarray}
These  ``two step"  two point functions will decay exponentially and the slowest ones will be the operators with a single Pauli.  The analysis in Appendix \ref{app:k=2RCQcalc} shows that these decay like $e^{-\tilde \Delta t} =(1/15)^t$. 

\subsubsection{Numerical Results for Brownian Circuit}

\subsubsection*{2-local RCQ Brownian circuit}
We start with the Brownian circuit made from the 2-local RCQ Hamiltonian introduced in Sec.~\ref{sec:2-local-RCQ-Hamiltonian}. 
In the following, $3000$, $20000$ and $10^5$ samples have been  used for $N=11, 10$ and $N\leq 9$ cases, respectively.
In Fig.~\ref{fig:SFF-QSG-circuit-various-k-2}, 
$u(k,t)$ minus the Haar value, $\langle\vert\Tr U^k(t)\vert^2\rangle-k$, is plotted
for $N=9$ and $dt=0.2$.\footnote{
As long as $dt$ is not too large, the specific value of $dt$ does not affect the result. 
For details regarding this point, especially in what sense $dt$-dependence becomes trivial,  
see Appendix~\ref{sec:Brownian-circuit-band-matrix}. 
} We can see clear exponential decays at early time and late time. 

\begin{figure}
  \centering
    \includegraphics[width=0.55\textwidth]{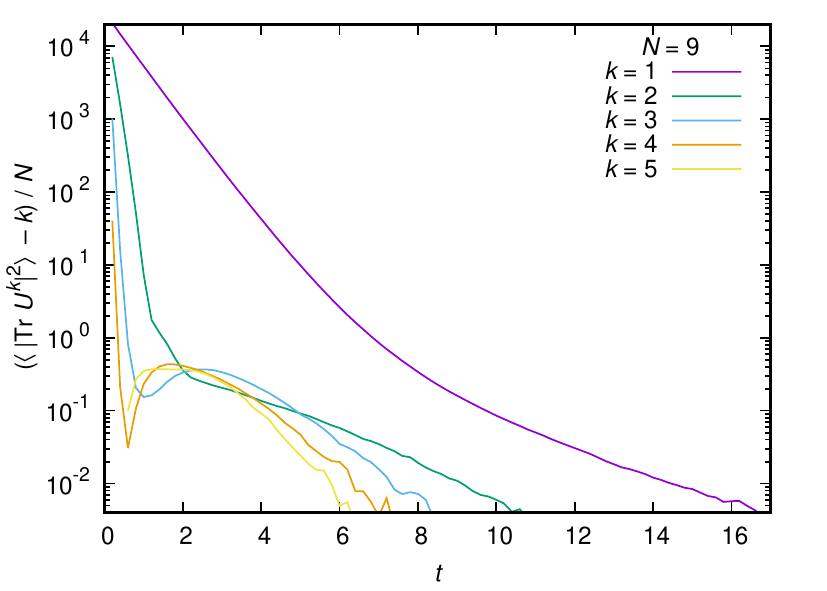}
\caption{Plot of $(\langle\vert\Tr U^k\vert^2\rangle-k)/N$ for $k=1, 2, 3, 4, 5$ against $t$ for the 2-local RCQ chain with $dt = 0.2$. $N = 9$, $10^5$ samples.
These exhibit exponential decays with rates that do not decrease as $k$ increases.}
\label{fig:SFF-QSG-circuit-various-k-2}
\end{figure}

As shown in the left panel of Fig.~\ref{fig:SFF-QSG-circuit-2}, 
the early time decay for $k=1$  can be fitted by an exponential function of the time
$\exp(-C N t)$, in which $C\simeq 0.189$ is a constant independent of $N$.
Therefore, the early-time decay to $\langle\vert\Tr U(t)\vert^2\rangle-1\sim O(1)$ takes place within order one time. 

For late time, 
$\left\langle  |\Tr U(t)|^2  \right\rangle - 1 
\sim
N e^{- \Delta t }$, with $\Delta=O(N^0)$, is expected.  
As shown in the right panel of Fig.~\ref{fig:SFF-QSG-circuit-2},
the late time decays are consistent with $\Delta\simeq 0.5$,
when $N$ is sufficiently large. 

\begin{figure}
  \centering
\includegraphics[width=0.47\textwidth]{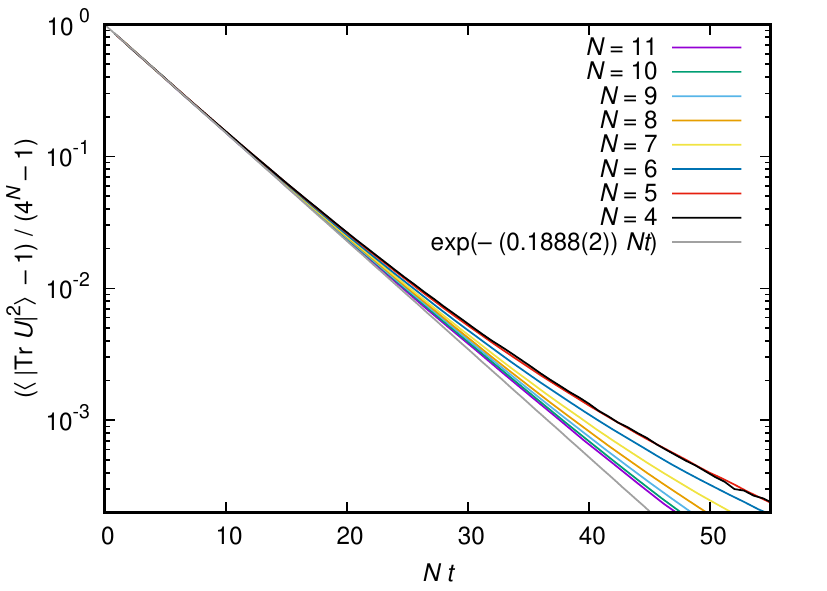}
\includegraphics[width=0.47\textwidth]{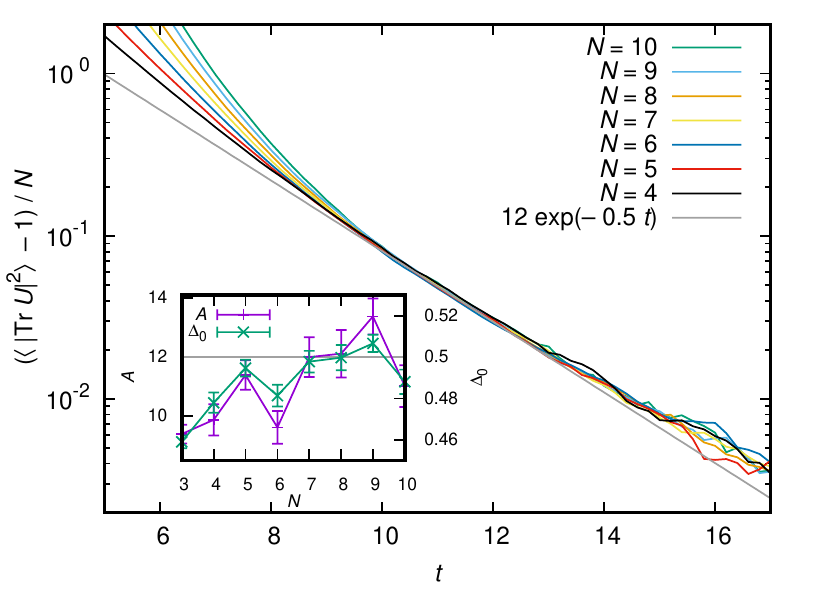}
      \caption{[Left] Plot of $(\langle\vert\Tr U\vert^2\rangle-1) / (4^N-1)$ against $Nt$ for the RCQ chain with $dt = 0.2$ for $N = 11, 10, 9, 8, 7, 6, 5, 4$ qubits. The result of a single parameter fit by an exponential function of $Nt$ is also shown.
[Right] Plot of $(\langle\vert\Tr U\vert^2\rangle-1) / N$ against $t$ for the RCQ chain with $dt = 0.2$ for $N = 10, 9, 8, 7, 6, 5, 4$ qubits. The inset shows the result of fit for $t\in[10:14]$ with $A e^{-\Delta_0 t}$ by taking $A$ and $\Delta_0$ as fitting parameters. The line corresponding to $(A, \Delta_0) = (12, 0.5)$ is also shown in the main plot.
}\label{fig:SFF-QSG-circuit-2}
\end{figure}

In Fig.~\ref{fig:SFF-QSG-circuit-various-k}, we used the same fitting ansatz for $k>1$, 
as $(\langle\vert\Tr U^k(t)\vert^2\rangle-k)/N\sim e^{-\Delta_k t}$. 
For each $k$, we can see  convergence to  exponential decay at large $N$. 
The exponent $\Delta_k$ increases with $k$ (see also Fig.~\ref{fig:SFF-QSG-circuit-various-k-2})\footnote{$\Delta_1$ actually is approximately the same as $\Delta_2$.},
which means the convergence to the Haar values at larger $k$ happens earlier.  
Therefore, after $t_{\rm Haar}$ is determined from $k=1$, higher moments have already converged.

\begin{figure}
  \centering
    \includegraphics[width=0.95\textwidth]{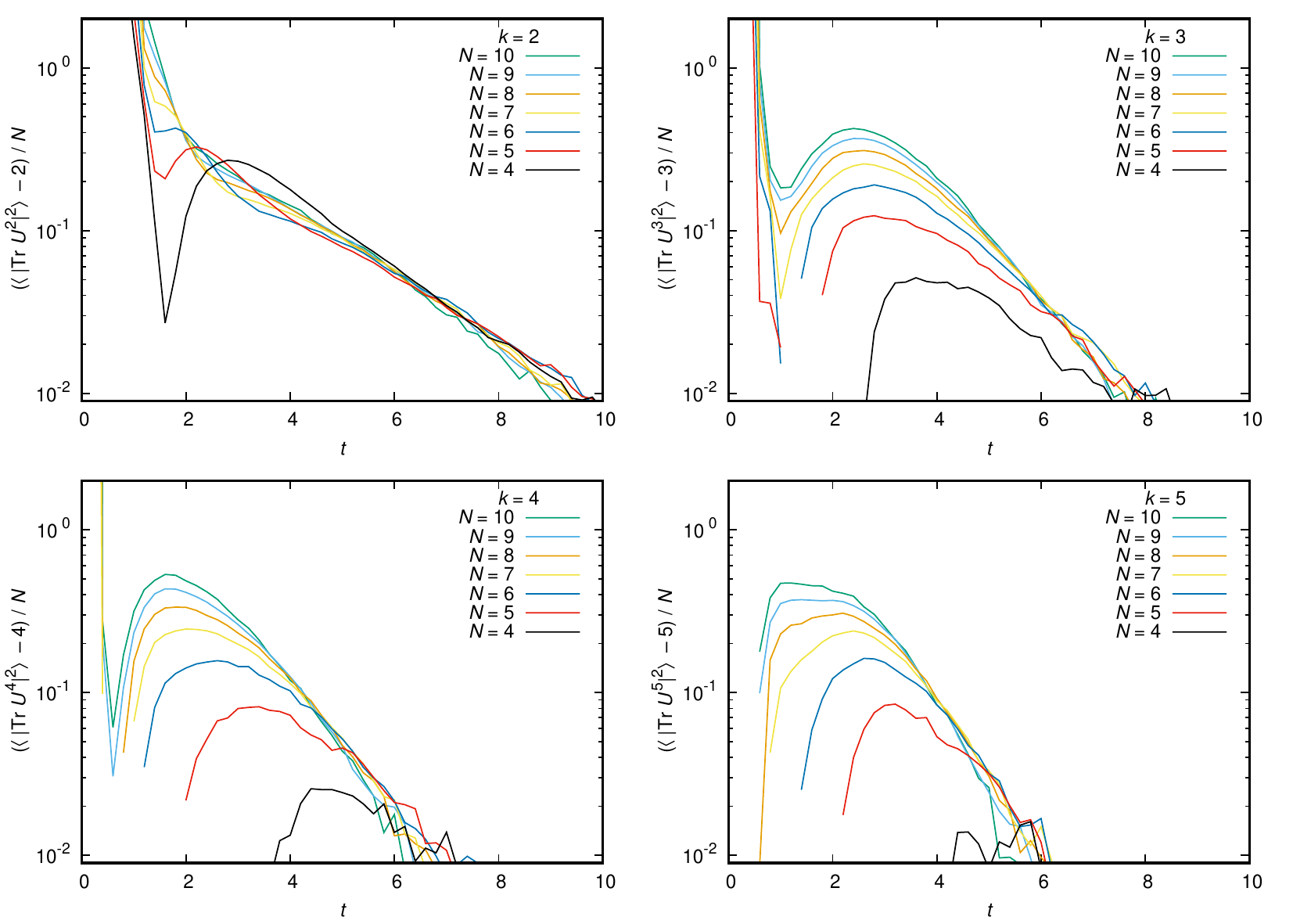}
\caption{Plot of $(\langle\vert\Tr U^k(\tilde{t})\vert^2\rangle-k)/N$ against $t$ for various values of $N$, $k=2,3,4,5$.
}\label{fig:SFF-QSG-circuit-various-k}
\end{figure}

\subsubsection*{Local RCQ Brownian circuit}
Next we consider the Brownian circuit made from the local RCQ Hamiltonian introduced in Sec.~\ref{sec:LRCQ}. 
We expect the same decay pattern as in the $2$-local RCQ. We have used the same number of samples as in the 2-local RCQ Brownian circuit case.

As shown in the left panel of Fig.~\ref{fig:SFF-LQSG-circuit}, the early-time decay can be fit by 
$(\langle\vert\Tr U(t)\vert^2\rangle-1) \simeq (4^N-1)\cdot e^{-c(N-1)t}$ where $c\simeq 0.187$. 
Therefore, the early-time decay to $\langle\vert\Tr U(t)\vert^2\rangle-1\sim O(1)$ takes place within order one time. 
As shown in the right panel of Fig.~\ref{fig:SFF-LQSG-circuit} and in Fig.~\ref{fig:SFF-LQSG-circuit-various-k}, 
the late time decay appears to be consistent with $\tilde{c}Ne^{-\Delta_k t}$ for $k=1,2,3,4$ and $5$, where $\tilde{c}$ is an order one constant. 
These observations are the same as the case of the 2-local Brownian circuit.   Note that here $\Delta_1$ is larger than $\Delta_2$.  This is not surprising given the analysis above which shows they result from somewhat different mechanisms.   But for $k>2$, $\Delta_k > \Delta_2$ indicating that $k=2$ is already in the stable range for determining the approach to random matrix behavior.

\begin{figure}
  \centering
\includegraphics[width=0.55\textwidth]{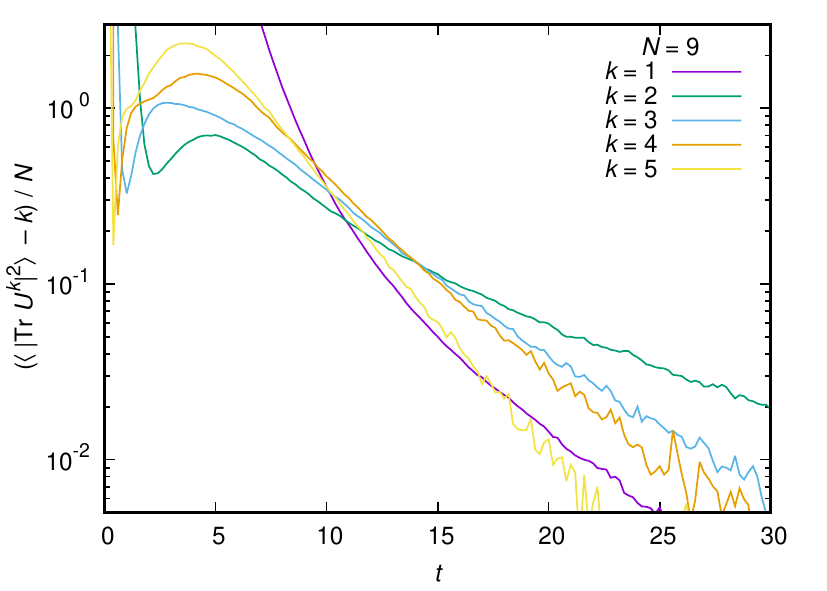}
\caption{Plot of $\langle\vert\Tr U^k\vert^2\rangle-k$ for $k=1,2,3,4,5$ for the local RCQ chain with $dt = 0.2$ for $N = 9$ qubits.
The decay time scale appears to grow with $k$ for $k \ge 2$.
\label{fig:SFF-LQSG-circuit-various-k-2}
}
\end{figure}

\begin{figure}
  \centering
\includegraphics[width=0.47\textwidth]{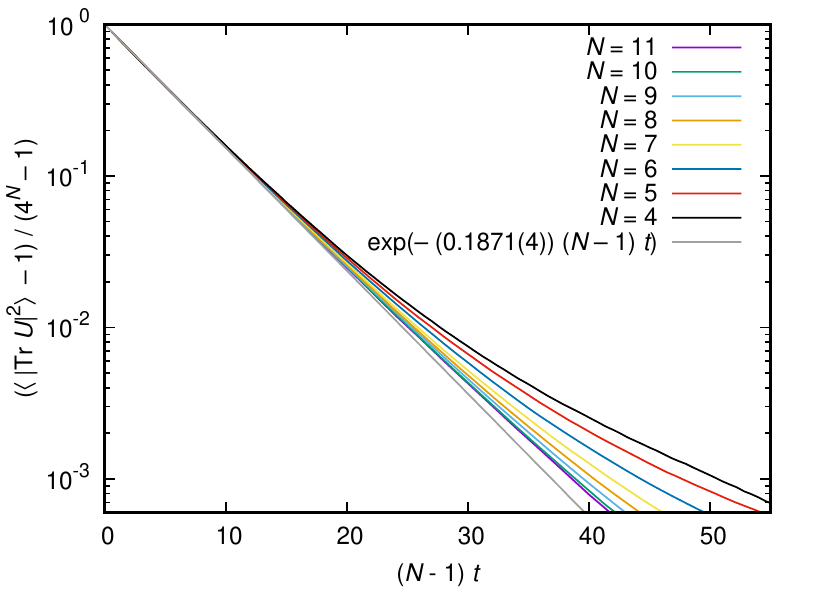}
\includegraphics[width=0.47\textwidth]{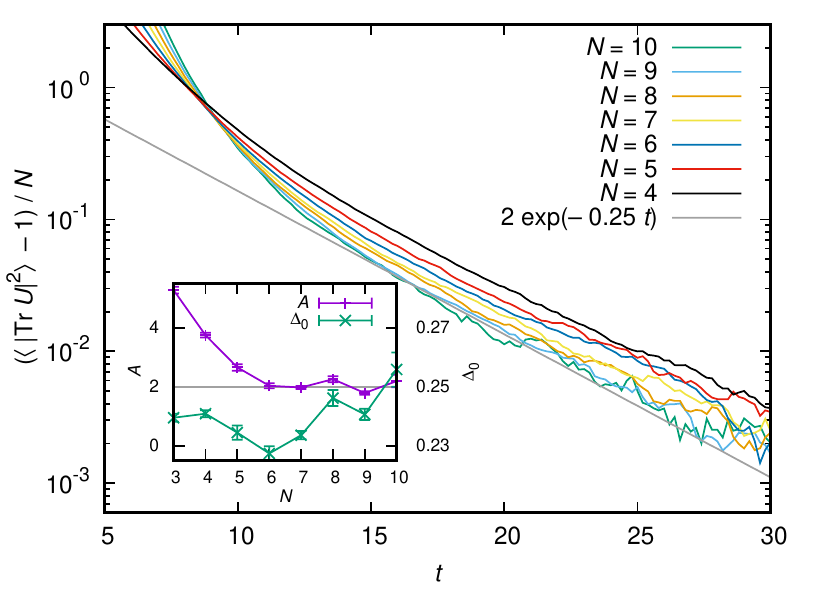}
      \caption{[Left] Plot of $(\langle\vert\Tr U\vert^2\rangle-1) / (4^N-1)$ against $(N-1)t$ for the local RCQ chain with $dt = 0.2$ for $N = 11, 10, 9, 8, 7, 6, 5, 4$ qubits. 
[Right] Plot of $(\langle\vert\Tr U\vert^2\rangle-1)/N$ for the local RCQ chain with $dt = 0.2$ for $N = 10, 9, 8, 7, 6, 5, 4$ qubits. 
}\label{fig:SFF-LQSG-circuit}
\end{figure}

\begin{figure}
  \centering
    \includegraphics[width=0.95\textwidth]{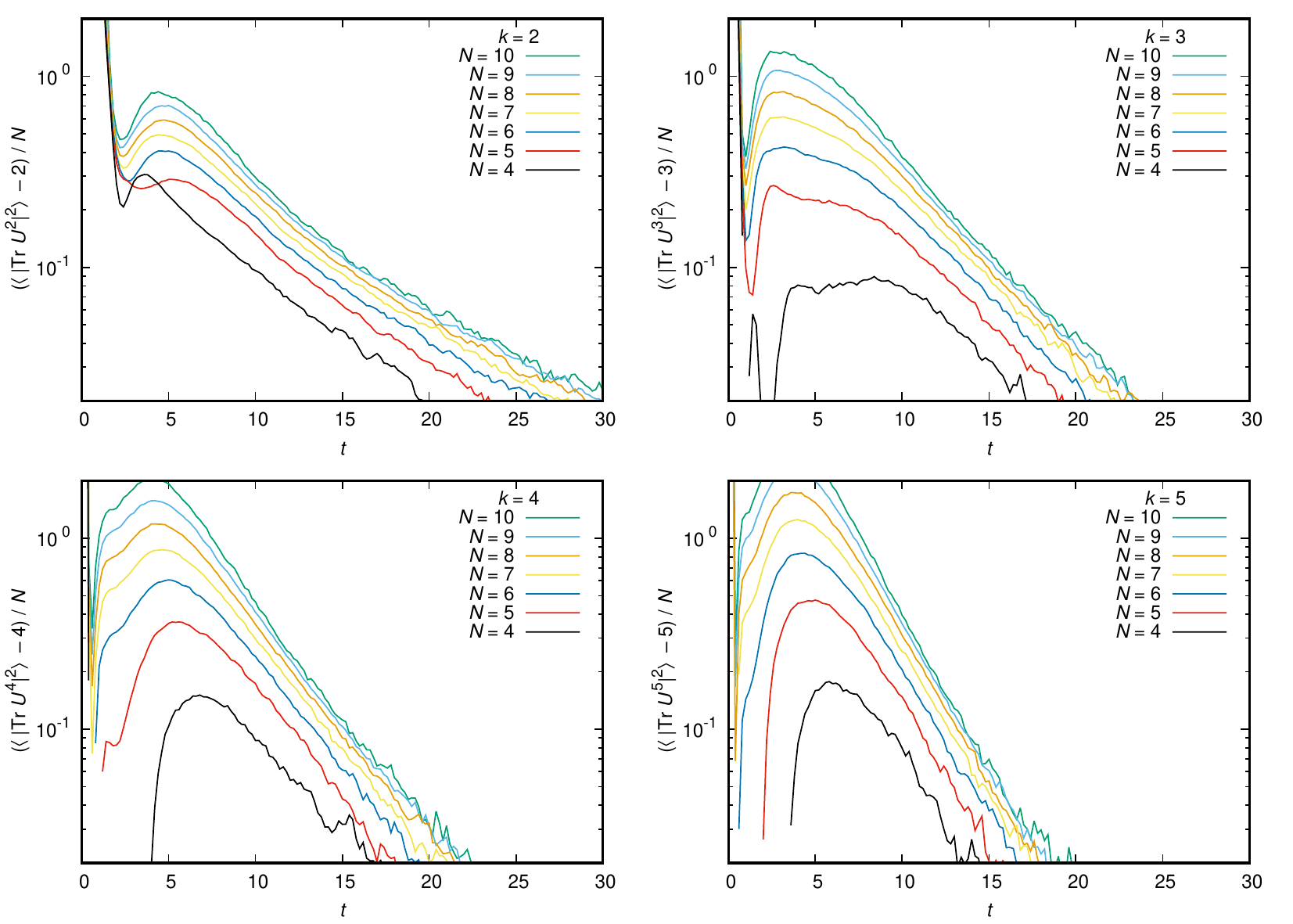}
\caption{Plot of $(\langle\vert\Tr U^k\vert^2\rangle-k)/N^2$ against $t$ for the local RCQ chain with $dt = 0.2$ for $N = 10, 9, 8, 7, 6, 5, 4$ qubits, $k=2,3,4,5$.
An exponential fit of the $N=9$ curve is also shown.
\label{fig:SFF-LQSG-circuit-various-k}
}
\end{figure}

\newpage
\subsection{XXZ Random Circuit and Diffusion}\label{sec:spin-conserving-circuit}
In this section, we will calculate  $u(k,t)$ defined in \eqref{utk} for an XXZ type random circuit  introduced in \cite{Khemani:2017nda, rakovszky2017} which has a conserved quantity and so helps
clarify the relationship between $t_{\rm Haar}$ and $t_{\rm diff}$.  
The quantum circuit is composed of a brickwork array of two-qubit gates as in  previous examples (see Fig.~\ref{fig:local-random-cicuit}).  
In order to motivate this  circuit, 
let us consider a Brownian circuit made from 
the Hamiltonian \eqref{Hamiltonian-XXZ} (which we repeat here) with time varying couplings: 
\begin{align}\nonumber
H =  \sum_{i=1}^{N} \Big( \frac{1}{4} \vec{\sigma}_i \cdot  \vec{\sigma}_{i+1} +  \frac{\omega_i}{2} \sigma^3_{i} \Big).
\end{align}

The total spin along the $z$-direction, $S_z^{\rm tot} = \frac{1}{2}\sum_i \sigma_i^3$, commutes with the Hamiltonian regardless of the value of the random magnetic field, 
so Brownian circuit time evolution commutes with $S_z^{\rm tot}$, and hence the total spin is conserved.   The random circuit under consideration is the discrete analog of this Brownian circuit.

Because the part of the Hamiltonian relevant for the interaction between $i$-th and $j$-th sites, 
$\Big( \frac{1}{4} \vec{\sigma}_i \cdot  \vec{\sigma}_{j} +  \frac{\omega_i}{2} \sigma^3_{i}+  \frac{\omega_{j}}{2} \sigma^3_{j} \Big)$, 
commutes with $\sigma_i^3+\sigma_{j}^3$, the unitary gate which describe the infinitesimal time evolution 
takes the following form, 
\begin{eqnarray}
G_{i,j}= 
\begin{pmatrix}
    U^{(1)}_{11}       & 0& 0 & 0 \\
    0    & U^{(2)}_{11}& U^{(2)}_{12} & 0  \\
    0    & U^{(2)}_{21}& U^{(2)}_{22} & 0  \\ 
    0    & 0 & 0  &U^{(3)}_{11}  \\
\end{pmatrix},  
\end{eqnarray}
where $U^{(1)} $ and  $U^{(3)} $ are independent random elements of U$(1)$
and $U^{(2)}$  is a Haar random unitary element in U$(2)$. The indices $i$ and $j$ indicates the locations of the qubits.
Here $U^{(1)} $ and  $U^{(3)}$ act on the spin $+1$ sector ($\ket{0,0}$) and the spin $-1$ sector ($\ket{1,1}$), respectively, 
and $U^{(2)}$ acts on the spin $0$ sector ($\ket{0,1}$ and $\ket{1,0}$).

Because of the symmetry we use the  Pauli basis $\{I, \sigma^{+}, \sigma^{-}, Z\}$, where 
\begin{eqnarray}
I = \ket{0}\bra{0}+\ket{1}\bra{1}, 
\qquad
Z = \ket{0}\bra{0}-\ket{1}\bra{1},
\end{eqnarray}
\begin{eqnarray}
\sigma^{+} = \frac{1}{2} (X+i Y) = \ket{0}\bra{1},
 \quad 
 \sigma^{-} = \frac{1}{2} (X - i Y) = \ket{1}\bra{0}~. 
\end{eqnarray}
In this notation, the sum of the spins at the $i$-th and $j$-th sites is 
\begin{eqnarray}
\frac{\sigma_i^3+\sigma_{j}^3}{2}
=
\frac{Z_iI_j+I_iZ_j}{2}~. 
\end{eqnarray}
The unitary gate can be expressed as 
\begin{eqnarray}
G_{i, j} 
&=& 
\frac{U^{(1)}_{11}}{4}(I_i+Z_i)(I_j+Z_j)
\nonumber\\
& &
+
\frac{U^{(2)}_{11}}{4}(I_i+Z_i)(I_j-Z_j)
+
U^{(2)}_{12}\sigma^+_i\sigma^-_j
+
U^{(2)}_{21}\sigma^-_i\sigma^+_j
+
\frac{U^{(2)}_{22}}{4}(I_i-Z_i)(I_j+Z_j)
\nonumber\\
& &
+
\frac{U^{(3)}_{11}}{4}(I_i-Z_i)(I_j-Z_j)~.
\end{eqnarray}
We now consider the adjoint action of $G_{i, j}$ on pairs of Paulis in the string.  This has the following properties \cite{Khemani:2017nda}:
(for simplicity, we suppress $i,j$ indices): 
\begin{enumerate}
\item 
$II$, $ZZ$ and $(IZ+ZI)/2$ are invariant. 

\item 
$(IZ-ZI)/2$, $\sigma^+\sigma^-$ and $\sigma^-\sigma^+$ will mix among each other with equal probability.  After Haar averaging,  these terms vanish.  

\item 
$\sigma^{+} \sigma^{+}$  and $\sigma^{-} \sigma^{-} $ get random phases $U^{(1)}_{11}U^{(3)}_{11}{}^\ast$ and $U^{(1)}_{11}{}^\ast U^{(3)}_{11}$, respectively. 
Because of these phases, the averages become zero.  

\item The raise-by-one operators $\sigma^{+} I$, $\sigma^{+} Z$, $I \sigma^{+} $ and $Z \sigma^{+} $ will transition between each other with equal probability. 
The averages of these linear combinations are zero.  
The same will happen to the lower-by-one operators. 

\end{enumerate}

The important point here is that any chain other than $II$, $ZZ$ and $(IZ+ZI)/2$ are completely randomized
 after one application of the unitary gate, so that they average to zero immediately.   The transitions between the remaining $I, Z$ chains form a Markov process. Note that unlike the $U(4)$ gate the evolution is a Markov chain already at $k=1$.
This Markov chain can be understood more easily if we use another basis 
\begin{eqnarray}
P=\ket{0}\bra{0}=\frac{I+Z}{2}, 
\qquad
M=\ket{1}\bra{1}=\frac{I-Z}{2}.  
\label{def-M-P}
\end{eqnarray}
In this basis, $PP$ and $MM$ are invariant, while $PM$ and $MP$ are mapped to $\frac{1}{2}PM+\frac{1}{2}MP$.   The Markov property is now evident.
\subsubsection{ $\left\langle\left| {\rm Tr} \ U(t) \right|^2\right\rangle$ in the XXZ Random Circuit}
Let us use the setup above to estimate the $k=1$ moment,  
 \begin{eqnarray}
 \left\langle|\Tr U(t)|^2\right\rangle = \sum_{i, j = 1}^{2^N} \left\langle U_{ii} U^*_{jj} \right\rangle. 
\end{eqnarray}
We first consider the `local' model defined above, which has  nearest-neighbor interactions via the gates $G_{i,i,+1}$. 

As in the case of RCQ, we can write the spectral form factor as 
\begin{eqnarray}
\left\langle U_{ii} U^*_{jj} \right\rangle 
=
\left\langle\Tr \left[ U_t |i\rangle\langle j|U_t^\dagger|j\rangle\langle i|\right] \right\rangle
\end{eqnarray}
where as before $ |i\rangle\langle j|$ can be expressed as a sum over strings of  $\sigma$-matrices. 
Let us use the $M,P,\sigma^\pm$ basis. When $i\neq j$, at least one $\sigma^\pm$ appears in the chain. 
Then, as we have seen, as soon as $\sigma^\pm$ is hit by a random gate, $U_t \left(\sigma\sigma\cdots\right) U_t^\dagger$ averages to zero. 
Hence contributions from $i\neq j$ terms disappear in one time step. 
Therefore we only need to consider  $i=j$. 

First let us consider the case with a single $M$, 
\begin{eqnarray}
\left\langle U_{ii} U^*_{ii} \right\rangle = \left\langle\Tr \left[ U_t \left(PP \cdots PMP \cdots P\right) U_t^\dagger  \left(PP \cdots PMP \cdots P\right)\right]\right\rangle.  
\label{eq:Markov-chain-XXZ}
\end{eqnarray}
We need to see how $ U_t \left(PP \cdots PMP \cdots P\right) U_t^\dagger$ evolves. 
After averaging $MP$ and $PM$ evolve to $\frac{PM+MP}{2}$, while $PP$ is invariant. 
Hence, when the random gate hits $MP$, the chain  remains $MPP\cdots P$ with probability $1/2$, and changes to $PMP\cdots P$ with probability $1/2$. 
So $M$ performs a random walk in the string.  Because there is no preferred direction there is no net drift of the walker.   We analyze this system in detail below but for now make a few qualitative remarks.  

The probability distribution of $M$ locations spreads diffusively so $\left\langle U_{ii} U^*_{ii} \right\rangle$, given by the overlap with the initial state, decays like  $\left\langle U_{ii} U^*_{ii} \right\rangle \sim 1/\sqrt{t}$ at early time.  At late time it is uniformly spread on the string 
and eventually $\left\langle U_{ii} U^*_{ii} \right\rangle$ converges to $1/N$. 
For general $i$ there are $q$ $M$'s and $N-q$ $P$'s. Then the diffusive decay at early time is $\left(1/\sqrt{t}\right)^q$ (when $q \ll N$). The late-time value is 
$1/{N \choose q}$. 

In order to give a precise analysis we identify $M$ and $P$ with two dimensional vectors,
\begin{align}
\ket{M}  = 
\left(
\begin{array}{c}
1\\
0
\end{array}
\right)
\qquad  \text{and} \qquad \ket{P} = 
\left(
\begin{array}{c}
0\\
1
\end{array}
\right). 
\end{align} 
In other words we identify $|M\rangle$ and $|P\rangle$ to be two states of a qubit. 
Let us denote by $P_{i,j}$ a unitary operator that acts on the vector space of $N$-qubits and swaps the states in $i, j$ positions: 
\begin{eqnarray}
P_{i, j} \left(\ket{P}_i \otimes  \ket{M}_j\right) 
&=&
 \ket{M}_i \otimes \ket{P}_j, 
\nonumber\\
P_{i, j} \left(\ket{M}_i \otimes  \ket{P}_j\right) 
&=&
 \ket{P}_i \otimes \ket{M}_j, 
\nonumber\\
P_{i, j} \left(\ket{M}_i \otimes  \ket{M}_j\right) 
&=&
  \ket{M}_i \otimes \ket{M}_j, 
\nonumber\\
P_{i, j} \left(\ket{P}_i \otimes  \ket{P}_j\right) 
&=&
 \ket{P}_i \otimes \ket{P}_j. 
 \label{def_Pij}
\end{eqnarray}
A single step of our process can be identified with the action of 
\begin{align}
\widetilde Q_{\rm loc}  =  \left[ \bigotimes_{i \in {\rm even}}\frac{1}{2} \Big(P_{i, i+1} +I\Big) \right] \cdot\left[ \bigotimes_{i \in {\rm odd}}\frac{1}{2} \Big(P_{i, i+1} +I\Big) \right]  ~.
\end{align}
Instead of analyzing this Markov process, we will approximate it by $N$ repetitions of a  process where in each step a random site $i$ is selected and the gate is applied between $i$ and $i+1$. This should approximate the original parallelized circuit at long times and in particular we expect it to give the same $N$ scaling of the gap.\footnote{We confirm numerically in the next section that the analytic results for the Markov dynamics of $ Q_{\rm loc}$ are fully consistent with the numerical results on the original parallelized circuit $\widetilde Q_{\rm loc}$. }  The Markov operator for this process is
\begin{align}
 Q_{\rm loc}  =  \left[ \frac{1}{N}\sum_{i=1}^N\frac{1}{2} \Big(P_{i, i+1} +I\Big) \right]^{\otimes N}  ~.
\end{align}

The number $r$ of $M$'s in a string is conserved by this process.  We will use the subscript $r$ to denote the restricted action of $Q_{\rm loc, r}$ in this subspace.  This process has already been studied extensively. It is related to an interchange model on cards studied by Diaconis and Ng \cite{Ng1996}.\footnote{We thank Persi Diaconis for bringing this work to our attention.}  They point out that this model can be mapped to a Heisenberg spin chain 
\begin{align}
 Q_{\rm loc}  =  \left[\frac{3}{4} I - \frac{1}{2N}H_{\rm XXX}\right]^{\otimes N}~, \text{where} \quad H_{\rm XXX} = -\frac{1}{2}\sum_{i=1}^N \vec\sigma_i\cdot\vec\sigma_{i+1} ~.
\end{align}
This connection becomes clear by noting that
\begin{align}
P_{i, j} = \frac{1}{2}\left(I+\vec\sigma_i\cdot\vec\sigma_j\right)~. 
\end{align}

It is known that the interchange model for local and 2-local cases defines a reversible and irreducible Markov chain. This process has a uniform stationary distribution on the $ {N \choose r}$ different strings.   The largest eigenvalue of $Q_{\rm loc, r}$ corresponding to the stationary distribution will be $\lambda_{0,r} =1$.  and all the eigenvalues, as we will discuss,  will be between $0$ and $1$.  

Now we go through a standard Markov chain analysis.
Suppose we have an initial string with $r$ $M$'s, $\ket{v_r}=\ket{MPPM\cdots MP}$. 
Then, we can express 
\begin{eqnarray}
 \left\langle\Tr \left[ U_t \left(MPPM\cdots MP\right) U_t^\dagger  \left(MPPM\cdots MP\right)\right]\right\rangle 
 = \bra{v_r} Q_{\rm loc}^t \ket{v_r} 
\end{eqnarray}
Denote by $\mathcal H_r$ the subspace of the Hilbert space spanned by vectors that have exactly $r$ elements of $M$. 
Let us denote by $\lambda_{a,r}$ the eigenvalues of $Q_{{\rm loc},r}$ for $a=0,1,.., {N \choose r}-1$ in decreasing order; 
$1=\lambda_{0,r} > \lambda_{1,r} \geq \lambda_{2,r} \geq \cdots \geq \lambda_{{N \choose r}-1,r}>0$. 

The stationary state $\ket{\lambda_{0,r}}$ is the eigenvector with the eigenvalue  $\lambda_{0,r}=1$ and has uniform weight on each string. 

Now, we can calculate $u(k=1, t)$ for each sector. 
Recalling the notation \eqref{def-M-P} $M=|1\rangle\langle 1|$ and $P=|0\rangle\langle 0|$, the number $r$ of $M$'s is the number of $1$'s
in the binary representation of the state $|i\rangle$ appearing in \eqref{eq:Markov-chain-XXZ}. 
In terms of the XXZ model, $0$ and $1$ are spin up ($+1/2$) and down ($-1/2$), so this state $|i\rangle$ has 
total spin  $S_{z} = \frac{1}{2} (N-r) - \frac{1}{2}r = (N/2)-r$. Therefore, 
\begin{align}
 \left\langle|\Tr U(t)|^2\right\rangle_{S_z=N/2-r} 
 &=
\sum_{i\in \{S_z=N/2-r\}}  \left\langle U_{ii}U_{ii}^\ast\right\rangle  =
 \sum_{v_r}  \bra{v_r} Q_{{\rm loc},r}^{\hspace{.1em}t} \ket{v_r} \\
 &= \Tr ~ Q_{{\rm loc},r}^{\hspace{.1em}t} = \sum_{a=0}^{{N \choose r} -1} \lambda_{a, r}^t \label{evaluesum}
\end{align}
where the  $\ket{v_r}$ that are summed over are all the permutations of the vector $\ket{MPPM\cdots MP}$ with $r$ $M$'s. 

We now define a gap parameter  $\Delta_r$ by $e^{-\Delta_r} = \lambda_{1,r}$.  At large $N$ this is just the gap in the $H_{\rm XXX}$ spin chain.\footnote{Up to an order one constant.}  We can give a simple lower bound on $u(1,t)$ by just keeping the first two terms of \eqref{evaluesum}
\begin{eqnarray}\label{lowerbound}
u(1, t) > 1+ e^{- \Delta_r t} ~.
\end{eqnarray}
So the time to get $\epsilon$ close to the Haar value\footnote{$t_\epsilon$ is a precise version of $t_{\rm Haar}$.} is lower bounded by
\begin{eqnarray}
t_\epsilon > \frac{1}{\Delta_r}   \log\frac{1}{\epsilon}   ~. 
\end{eqnarray}

The gap is determined by the properties of the XXX spin chain which is one of the canonical examples of an integrable system.    Reference \cite{Ng1996} quotes the following results.\footnote{We have not exhaustively studied the literature on this model.  It may well be the case that more analytic results are known.}   For $r =1,2$ the gap is the same and is given by $\Delta_1 = \Delta_2 = \frac{c}{N^2}$  where $c$ is an order one constant.  Further it is conjectured that the gap is the same for all $r$.  Our numerics (see the following section) support this.   Using this result we have
\begin{eqnarray}
t_\epsilon > \frac{N^2}{c} \log\frac{1}{\epsilon}  ~. 
\end{eqnarray}
As we  discussed earlier the $N^2$ dependence is a signal of diffusion.

A sharp upper bound is more subtle to obtain. The terms ignored in \eqref{lowerbound} contribute an $N$ dependent offset to the time required to reach the Haar value.  We define $t_{\rm offset}$ by
\begin{eqnarray}
t_\epsilon = t_{\rm offset} + \frac{N^2}{c} \log\frac{1}{\epsilon}  ~. 
\end{eqnarray}

We can begin to estimate $t_{\rm offset}$ by approximating the XXX chain spectrum.
In the large $N$ finite $r$ limit the spectrum  is dominated by $r$ free magnons, each with nonrelativistic spectrum
\begin{eqnarray}\label{dispersion}
E(p) \sim  p^2/2 ~.
\end{eqnarray}
Here $p$ is a continuum momentum, where we assume $\frac{1}{N} \ll p \ll 1$.

Assuming $r \ll N$ and $t \ll N^2$ we can approximate
\begin{equation}
u(1,t) \approx \frac{1}{r!} \left( N \int dp e^{-p^2 t/2} \right)^r \sim \frac{1}{r!}\Big(\frac{N}{\sqrt{t}}\Big)^r ~.
\end{equation}
This becomes order one at a time $t_{\rm offset} \sim \frac{N^2}{r^2}$ .  Here we see that the diffusional dynamics is related to the nonrelativistic dispersion relation \eqref{dispersion} of the XXX chain.  

But this is not the slowest dynamics in this system.  It is known that there are bound states of $b$ magnons for all $b \le r$.  These have ``mass" of order $b$ and hence a spectrum (see \cite{Karabach1997})
\begin{eqnarray}\label{dispersion-b}
E(p) \sim  p^2/(2b) ~.
\end{eqnarray}
These bound states diffuse more slowly because of their greater mass.  Consider the case when all $r$ magnons are bound up into bound states of size $b$, producing $r/b$ ``particles".  This makes the following contribution to $u(1,t)$:
\begin{eqnarray}\label{bdep} 
u(1,t) \supset \frac{1}{(r/b)!} \left( N \int dp e^{-p^2 t/(2b)} \right)^{r/b} \sim \frac{1}{(r/b)!}\Big(\frac{N}{\sqrt{t/b}}\Big)^{r/b}
\end{eqnarray}
This interesting function of time  becomes order one at a time $t_{\rm offset} \sim N^2 b^3/r^2$.   This suggests we can make $t_{\rm offset} \sim N^2$ by choosing $b \sim r^{2/3}$. In fact it seems we can make  $t_{\rm offset}$ parametrically longer, $\sim N^2 r$, by choosing the largest allowed value $b \sim r$.   For $r \sim N$ this would give $t_{\rm offset} \sim N^3$.   However, when $t \sim N^2 $ the discreteness of the spectrum becomes important and these approximations are no longer valid.    Instead we turn to numerics (discussed below) which indicate that $t_{\rm offset}$ is independent of $r$ for large enough $r$ and hence $t_{\rm offset} \sim N^2$.\footnote {We cannot rule out a weak $r$ dependence in $t_{\rm offset}$ from these numerics so for example an $N^2 \log( N/C)$ behavior with a large $C$ coefficient is possible.}   Our best estimate becomes
\begin{eqnarray}
t_\epsilon = \tilde{c} N^2 + \frac{N^2}{c} \log\frac{1}{\epsilon}  ~. 
\end{eqnarray}

Again we point out that this $N^2$ dependence is parametrically slower than the scrambling time for this system which is $N$ \cite{Khemani:2017nda, rakovszky2017}, corresponding to ballistic propagation.
 
We now turn to  the 2-local XXZ random circuit.   
The analysis is analogous to the local system and we end up with a Markov chain.  We first consider a process where at each step a gate acts on one randomly selected pair of qubits.  One parallel step of the full process  corresponds to $N$ repetitions of this.
The Markov operator for this process is
\begin{align}
Q_{\rm 2\mathchar`-loc}  =  \left[\frac{1}{{N \choose 2}}\sum_{i<j}^N \frac{1}{2} \Big(P_{i, j} +I\Big) \right]^{\otimes N} ~.
\end{align}
The spectrum of this operator differs in important ways from the geometrically local case giving a different $N$ scaling for $t_\epsilon$.  

The spectrum of $Q_{\rm 2\mathchar`-loc, r}$ is related to the gap of the Heisenberg model on a complete graph. More specifically, we express 
\begin{align}
Q_{\rm 2\mathchar`-loc}  =    \left[ \frac{3}{4} I - \frac{1}{2N}\widetilde{H}_{\rm XXX}\right]^{\otimes N}~, \text{where} \quad \widetilde{H}_{\rm XXX} =  -\frac{N}{{N \choose 2}}\frac{1}{2} \sum_{i<j}^N\vec\sigma_i\cdot\vec\sigma_j 
 ~.
\end{align}

Then, we use a simple identity 
\begin{align}
 \frac{1}{2}\sum_{i<j}^N\vec\sigma_i\cdot\vec\sigma_j  =  \left( \frac{1}{2}\sum_{i=1}^N\vec\sigma_i \right)^2 -\frac{1}{4} \sum_{i=1}^N\vec\sigma_i^2 =  \left( \frac{1}{2}\sum_{i=1}^N\vec\sigma_i \right)^2 -\frac{3N}{4} = \vec S^2 -\frac{3N}{4}
 \end{align}
 where we use $\vec S^2$ to denote the total spin squared operator.
So we find
\begin{eqnarray}
\widetilde{H}_{\rm XXX} = \frac{N}{{N \choose 2}} \Big( \frac{3N}{4}  I - {\vec S}^2 \Big)
\end{eqnarray}
 
   Recall that $S_z = \frac{1}{2}\sum_{i}^N \sigma_i^3$ and it commutes with $\vec S^2$. In the basis where both are diagonal  $\vec S^2$ has eigenvalues $s(s+1)$, where $s=0, 1,\cdots, N/2$ for even $N$ and $S_z$ has eigenvalues $s_z=-s, -s+1, \cdots, +s$.  Note that $r = \frac{N}{2} - s_z$. The eigenvalues of $\widetilde{H}_{\rm XXX}$ are given by 
 \begin{align}
 E_{(s, s_z)} = \frac{3N^2}{4}\frac{1}{{N \choose 2}} - \frac{N}{{N \choose 2}}s(s+1)
 \label{eq:2-local-XXZ-eig}
 \end{align}
Note that the eigenvalue is independent of $s_z$  and only depends on $s$.   In a given $r$ sector, that is for a given value of $s_z$, all values of $s$ such that $\frac{N}{2} \ge s \ge s_z$ contribute.

At large $N$ the important eigenvalues $Q_{\rm 2\mathchar`-loc}$ are given by $\lambda_{a} =e^{-E_a}$.  So we can write
\begin{eqnarray}
u(1,t) = \sum_a e^{-E_a t}
\end{eqnarray}
and the gap is given by $E_1 -E_0$.  For all $r$ sectors $E_0$ corresponds to $s = \frac{N}{2}$ and $E_1$ to $s = \frac{N-2}{2}$.  So at large $N$ the gap is $N$ independent.
\begin{eqnarray}
\Delta_r = E_1 -E_0 \rightarrow 2 ~.
\end{eqnarray}

Diffusion on this completely connected graph takes place in order one time.    The degeneracy of this first excited state is $N$ \cite{Mendonca2013} and so a lower bound for $u(1, t)$ is 
\begin{eqnarray}
u(1,t) > 1 + N \cdot e^{-2t}
\end{eqnarray}

giving a Haar time of
\begin{eqnarray}\label{haartwolocal}
t_\epsilon > \frac{1}{2}\log N +\frac{1}{2} \log{\epsilon}
\end{eqnarray}

In fact this is an accurate estimate.   The small gaps grow linearly in level, $E_n - E_0 = 2n$ and the degeneracies grow like ${N \choose n} - { N \choose n-1}$. \footnote{Note that this is the degeneracy of the energy levels for each sector of fixed spin in $Z$ direction (which we label by $s_z$ or $r$).} So the full formula is well approximated by 
\begin{eqnarray}
u(1,t) \approx (1 + e^{-2t})^N \cdot (1-e^{-2t}) +e^{-2t(N+1)}~.
\end{eqnarray}
The Haar time is again given by \eqref{haartwolocal}.  Note again that this logarithmic dependence is due to a different mechanism than the logarithmic scrambling time.

One can study $u(2,t)$ for this model using the techniques discussed in Appendix \ref{app:k=2RCQcalc}. We do not perform a full analysis of this case, but note that  the ``unequal chain" part of equation~\eqref{eq:U1jj11j_terms_sep} (the last term) will converge slowly because of the same diffusive effects described above.   This indicates that the time scale to converge to the Haar value for $u(2, t)$ will be parametrically the same as in the $u(1, t)$ analysis. 

\subsubsection{Numerical Results for XXZ Markov Chain}

We conclude this section by showing some numerical results for local XXZ random circuits. For the local XXZ Markov chain the upper left panel of Fig.~\ref{fig:XXZ_Markov_chain} illustrates that for a fixed value of $N$ the low lying eigenvalues are independent of the value of $r$, indicating that bound state effects of the kind described in \eqref{bdep} are not important  for the final approach to Haar. We can see that the offset is roughly $\sim N^2$ by looking at the behavior of $u(1,t)$   in the lower panel.  We plot the $u(1,t)$ for $r = N/2$ (i.e.,  $S_z=0$) sector for different values of $N$. Note that when the time is rescaled by $N^2$ the results align completely. This suggests that the  best numerical fit to the time to get $\epsilon$ close to  Haar is given by 
\begin{eqnarray}
t_\epsilon \sim c_1 N^2 \left[c_2+ \log\frac{1}{\epsilon}  \right]. 
\end{eqnarray}
where $c_1$ and $c_2$ are constants independent of $N$. However, we would like to emphasize that we cannot rule out a possibility of a weak $N$ dependence in  the coefficient $c_2$.  
\begin{figure}
\begin{center}
\scalebox{0.42}{
\includegraphics{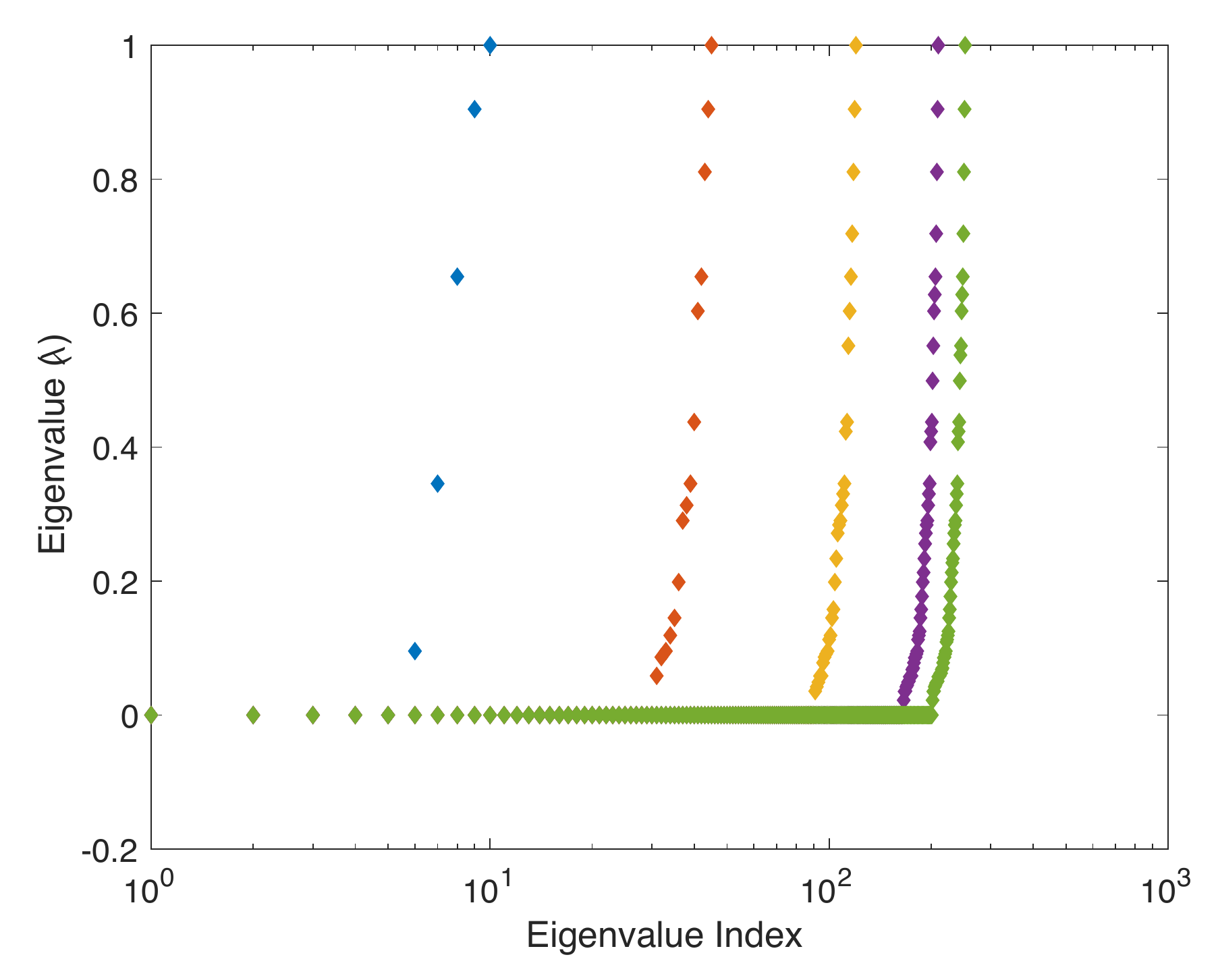}}
\scalebox{0.42}{
\includegraphics{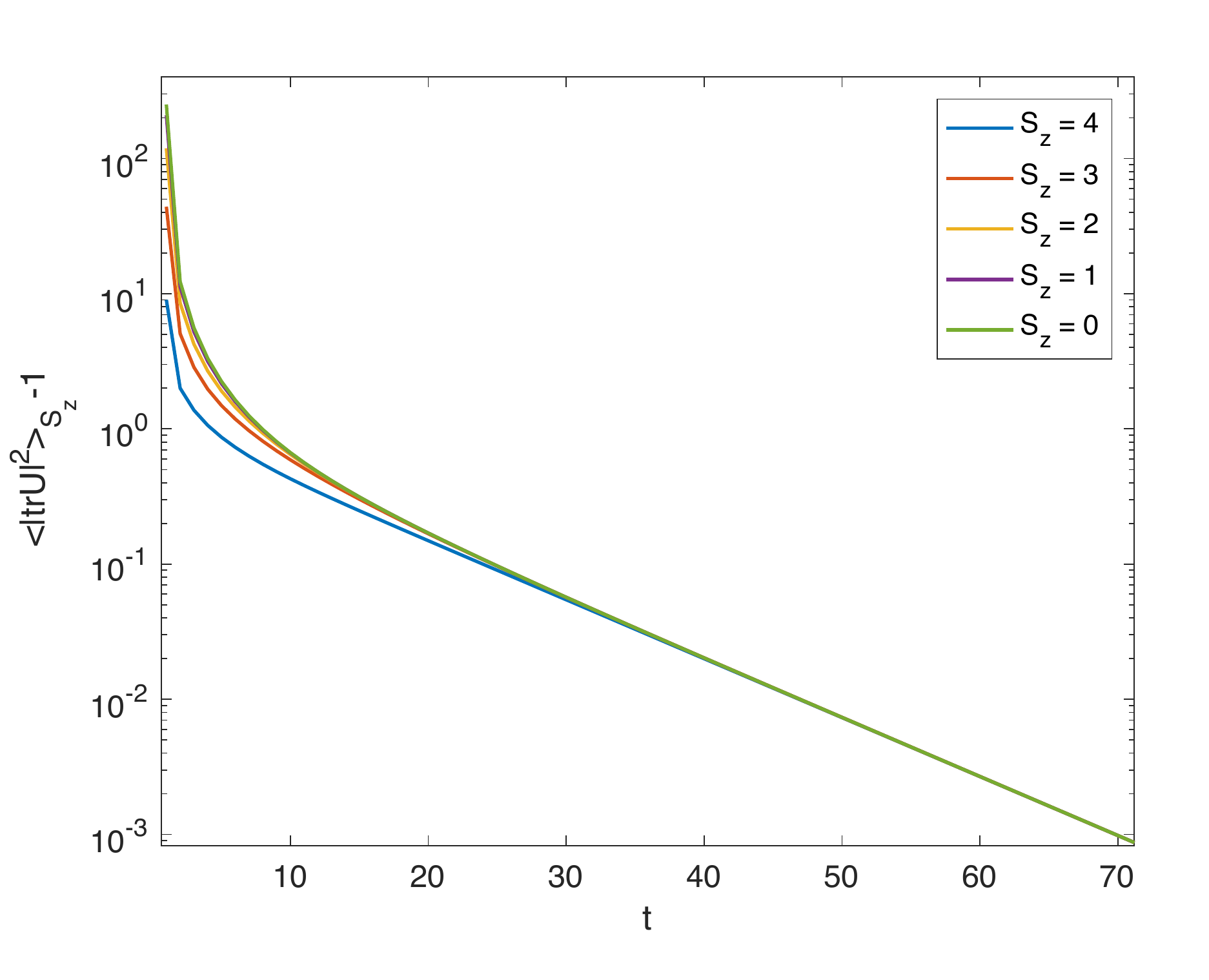}}
\scalebox{0.5}{
\includegraphics{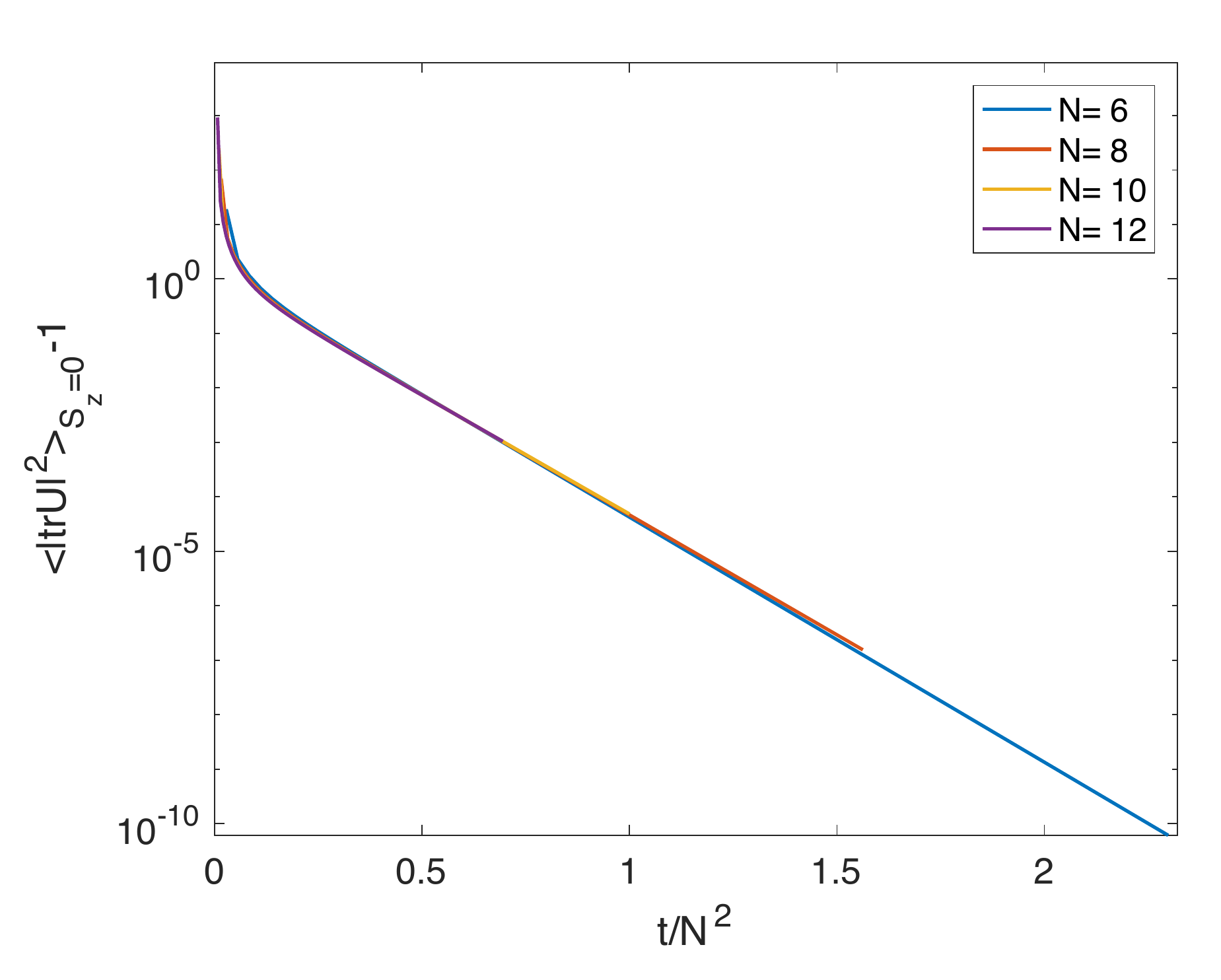}}
\end{center}
\caption{
Numerical results for local XXZ random circuit. 
[Top-Left] The eigenvalues of $Q_{{\rm loc},r}$ for $r=0,1,2,3,4$ for $N=8$. The second largest eigenvalues take the same value. 
[Top-Right] The $u(1, t)$ for $r=0,1,2,3,4$ at $N=8$. They all converge to the same exponential decay in late times. 
[Bottom] The spectral form factors for several values of $N$ in the largest sector, where $r=N/2$ (or $S_z = 0$). 
}\label{fig:XXZ_Markov_chain}
\end{figure}
\\

\newpage
\section{Connection to Correlation Functions}\label{sec:hamestimate}
In this section we discuss the connection between the ramp time and the relaxation time scale of correlation functions of simple operators in random quantum circuits.   This may well be useful for determining the ramp time for Hamiltonian systems but there is an obstruction to a simple connection which we discuss below. 

  We use the approach of \cite{Roberts:2016hpo, Cotler:2017jue} and rewrite the spectral form factor as a sum over correlation functions. Rewriting the  operator trace in terms of all  Pauli operators $P_j$ \cite{Roberts:2016hpo} one has for any  operator $\widehat O$
\begin{eqnarray}
I  \cdot   \Tr ( \widehat O )= \frac{1}{L} \sum_{j=1}^{L^2}  P_j \widehat O P_j
\end{eqnarray}
Introducing another operator $\widehat Q$ and taking the trace on both sides we get 
\vspace{-0.3cm}
\begin{eqnarray}\label{trQtrUrewrite}
 \Tr(\widehat Q) \cdot \Tr ( \widehat O )= \frac{1}{L} \sum_{j=1}^{L^2}   \Tr \Big( \widehat Q P_j \widehat O P_j \Big) 
\end{eqnarray}

This is a general relation that can be applied to random quantum circuits as well as Hamiltonian dynamics. Let $\widehat Q^\dagger =  \widehat O = U(t)$ and average over the $U(t)$ of random quantum circuit realizations.  We can then rewrite  the infinite temperature spectral form factor 
 \vspace{-0.3cm}
 \begin{eqnarray}\label{correxp} 
 \langle \Tr \ U^\dagger(t) \cdot \Tr \ U(t) \rangle= 1+  \sum_{j=2}^{L^2}  \frac{1}{L} \Big\langle \Tr \Big( U^\dagger(t) P_j U(t) P_j \Big) \Big\rangle.
\end{eqnarray}

To understand the approach to the ramp we must understand the decay of the two point correlators of all $L^2-1$ non-identity Pauli strings ($\langle P_j(t) P_j \rangle $). In Section \ref{localandk-localRCQrandomcircuit} we showed that this sum is dominated by simple Pauli strings and the leading contribution to come from the Pauli strings with a single non-identity element -- there are $3N$ such strings. Each of these simple two point functions will decay exponentially $\sim e^{-mt}$ for some gap parameter $m$. Therefore, the spectral form factor can be approximated by 
 \begin{eqnarray}\label{leadingcorr}
 \langle \Tr \ U^\dagger(t) \cdot \Tr \ U(t) \rangle  \sim 1 + 3N\cdot e^{-mt}
 \end{eqnarray}
demonstrating a $\log N$ scaling for the ramp time with a prefactor of order $1/m$.  So the time scale is related to the decay of two point functions, not the Lyapunov exponent.  

We expect the same to be true for Hamiltonian evolution, where $U(t)  = e^{-iHt}$ and the average is over an ensemble of Hamiltonians. In a previous version of this paper, we gave a heuristic argument estimating the ramp time of Hamiltonian systems by assuming, as above,  that the slowest decay was that of such simple operators.  However that argument had an error.\footnote{We thank Alex Altland,  Dmitry Bagrets and Alexei Kitaev for pointing this problem out to us. }    Two point functions of operators that couple to the Hamiltonian  have  subleading terms that do not decay in time because of energy conservation. \footnote{An example of such a two point function in SYK is $\langle \psi_{1}\psi_{2} \psi_{3}\psi_{4}(t)  \psi_{1}\psi_{2} \psi_{3}\psi_{4}(0) \rangle_\beta$ that has a $1/N^3$ suppressed term that doesn't decay in time, whereas leading term decays exponentially.}  They remain constant for all times. These terms are small, but are much larger than the ramp time value and so need to be controlled to make a reliable estimate.   In fact we believe that these terms almost exactly cancel at the ramp time.   A reliable calculation of the ramp time has been done for Hamiltonian  SYK in \cite{saad2018} using the collective field formalism.  The result there is consistent with the intuition described here.

\section{Closing Remarks}\label{closingremarks}

In this paper we have studied the onset of random matrix behavior in strongly chaotic many-body systems.   But it remains to give a convincing argument for random matrix behavior over the full range of energy and time scales present.  For example it is an important task to give a general argument for presence of the full ramp/plateau structure in the spectral form factor.  One approach that has been suggested in \cite{SerbynMoore2016, chalker1996}  is to extend the ideas of Dyson Brownian motion used with great effect in \cite{chalker1996} to many-body systems.  Further exploring this approach is an interesting subject for future research.  One important issue will be to understand the role of the comparatively large  ${\cal O}(1/N)$ (rather than ${\cal O}(1/L)$) fluctuations in eigenvalue density that are characteristic of many-body systems.

Another interesting approach has been developed recently in \cite{AltlandBagrets2017}.  They apply ideas of single-body random hopping problems to the exponentially large Hilbert space of the many-body system.    Their approach does not get the value of the ramp time we argue for but we think this technique  may well be useful to describe the later time ramp and plateau. 

  This approach is closely connected to the observation in Appendix F of \cite{Cotler:2016fpe} that the connected contributions to $\langle \Tr H^k \Tr H^k \rangle$ where $H$ is the SYK Hamiltonian have the  perturbative $SYK$ value for $k$ fixed as $N$ gets large, but at $k^* \approx N \log N$ this quantity assumes the value it would have if $H$ were a random matrix in the second quantized Hilbert space.   Naively converting this to a time scale (remembering that the width of the spectrum is $\sim \sqrt{N}$) one finds $t \sim \sqrt{N }\log N$, the scale found in \cite{AltlandBagrets2017}.   

The result of the calculation in \cite{AltlandBagrets2017} is a formula rather similar to \eqref{correxp}, a sum over an exponentially large number of more and more rapidly decaying terms.    The leading correction is of the form $Ne^{-\mu t}$ much like \eqref{leadingcorr}.  But here $\mu \sim 1/\sqrt{N}$, corresponding to a parametrically slower decay rate than in the SYK version of \eqref{leadingcorr}.   It is this slow decay that causes the large estimate $t_{\rm ramp} \sim \sqrt{N} \log N$.  There are no perturbative modes in SYK with such a slow decay rate\footnote{There are power-law decays in correlators due to the sharp edge in the density of states \cite{Bagrets:2016cdf}. But for quantities focussed in the bulk of the band these decays should be exponentially suppressed because of the small matrix elements between bulk states and edge states.}  so their significance is a puzzle  to which we hope to return.\footnote{The methods being used in \cite{saad2018} should be helpful.}

One of the motivations for thinking about these   strongly chaotic, scrambling, systems is their connection to the physics of quantum black holes.   One important thing to understand is which phenomena are important in both small and large  AdS black holes and which  only occur in large ones.    The time scale for evaporation of a small black hole is order $S$ (where $S$ is the black hole entropy).  The ramp time we have found for the $k$-local systems appropriate to the black hole horizon is order $\log S$, parametrically smaller than the evaporation time.  These effects will be exponentially subdominant in simple observables at these short times so studying them will be difficult, but they are present and are a signature of the nonperturbative quantum dynamics of these systems.

Perhaps the most important question from the gravitational point of view is the bulk dual in the sense of AdS/CFT of these random matrix phenomena \cite{Cotler:2016fpe}. In SYK a  related question is the explanation of these phenomena in terms of the $G, \Sigma$ collective fields.    We have made progress on this problem which we hope to report on in the near future \cite{saad2018}. 

\section*{Acknowledgments}

The authors thank Alex Altland, Dmitry Bagrets, Fernando Brandao, Persi Diaconis, Antonio Garc\'ia-Garc\'ia, Patrick Hayden, David Huse, Vedika Khemani, Alexei Kitaev, Sepehr Nezami, Dan Roberts, Phil Saad, Douglas Stanford, Alex Streicher, Lenny Susskind and Beni Yoshida for valuable discussions.  We thank Dan Roberts for sharing his MATLAB code with us, and David Huse and Vedika Khemani for sharing a draft of their paper before publication.
Discussions during the YITP workshop (YITP-W-17-01) on ``Quantum Gravity, String Theory and Holography" were useful in completing this work.

Parts of this work were presented at Strings 2017 (June 2017) and at the KITP workshop ``Frontiers of Quantum Information Physics" (October 2017). 

HG and SHS were supported in part by NSF grant PHY-1720397.
MH is  supported  in  part  by  the  Grant-in-Aid of the Japanese Ministry of Education, Sciences and Technology, Sports and Culture (MEXT) for Scientific Research (No. 17K14285).
MT acknowledges JSPS KAKENHI Grants No.~JP15H05855, No. JP15K21717, and No.~JP17K17822.

In accordance with institutional policy the data
used in the preparation of this paper is available to other scientists on request.

\appendix
\section{Additional Numerical Results}
\subsection{Number Variance in SYK}\label{sec:NR-NV}

Garc\'ia-Garc\'ia and Verbaarschot studied the number variance in the SYK model; 
see Fig.~9 and Fig.~10 in \cite{Garcia-Garcia:2016mno}. 
They observed a deviation from the RMT result at  $K \sim N^2$ \cite{Garcia-Garcia:2018ruf}.    In  Sec. \ref{numbervarsec} we have explained that this very small value of $K$ is due to the large fluctuations in the variance of the energy distribution.  As a preliminary check we have implemented a slightly improved unfolding procedure:
\begin{enumerate}
\item Shift the origin of the energy spectrum so that $\langle E\rangle=0$ for each sample.
\item Then we rescale the eigenvalues for each sample so that the variance for each sample is the same as $\langle \Tr H^2 \rangle$, the average variance. 
\item Then we unfold each sample using the density profile determined by $\langle \rho(E) \rangle$.
\item Then we compute $\Sigma^2(K)$. 
\end{enumerate}
Steps (1) and (2) are not implemented in \cite{Garcia-Garcia:2016mno}.
The number variance calculated in this manner is shown in 
Fig.~\ref{fig:SYK-number-variance}, 
for $N=32$ and $N=34$. The agreement with RMT persists to larger $K$, roughly a factor of two larger, than in \cite{Garcia-Garcia:2016mno}.  Thus we see that the number variance is sensitive to the unfolding procedure.  

\begin{figure}
\includegraphics{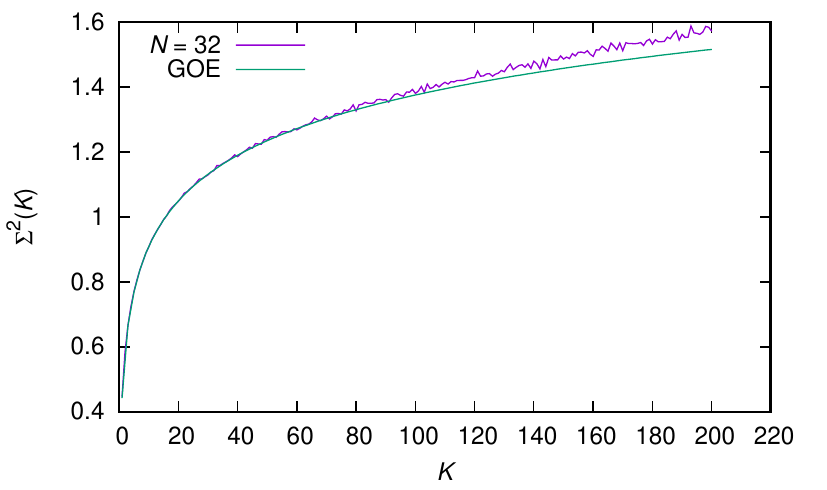}
\includegraphics{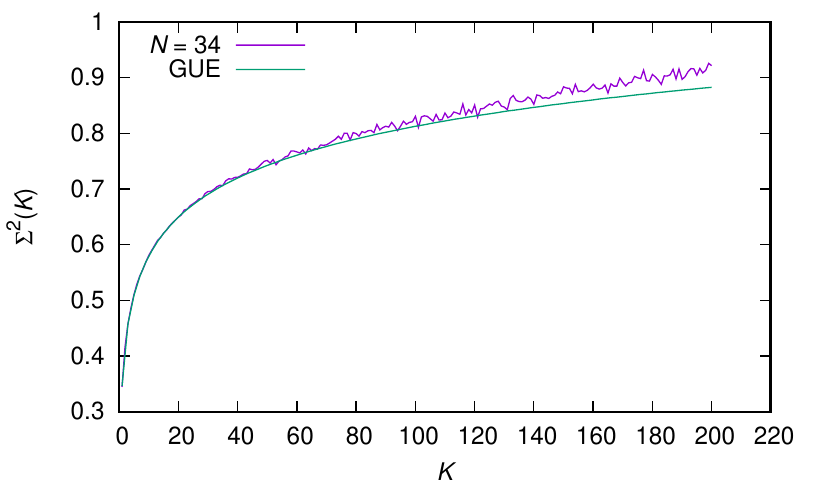}
\caption{$\Sigma^2(K)$ for the SYK model, $N=32$ (250 samples) and $N=34$ (110 samples),
obtained using the unfolded spectra after shifting and rescaling.
The curves for corresponding random matrix ensembles, GOE and GUE, are also plotted.}\label{fig:SYK-number-variance}
\end{figure}  

As discussed in the text we expect that RMT behavior extends to an exponentially large value of $K$,  much larger than shown even by these improved results.  Our unfolding procedure is still rather crude.  As a first step one could try to fully unfold the spectrum sample by sample.  But it is unclear to us whether this is even in principle sufficient.  At a minimum an exponentially large number of samples would be required to get a sufficiently accurate value of the variance.

For this reason the number variance seems not to be the best quantity to probe long range spectral rigidity in many-body systems. 
As discussed in the text the spectral form factor is less sensitive to such errors. 

\subsection{Estimating $t_{\mathrm{min}}$ in the SYK Model}\label{sec:t_min}
In this appendix we give our detailed results for determining $t_\mathrm{min}$ in the SYK model numerically.
The algorithm used to select an optimal $\alpha$ is as follows:  In the second panel of Figure~\ref{fig:Y-original-SYK-spectrum} we have varied $\alpha$ with step size 0.5, and observed that 
some $\alpha$ between $2.5 < \alpha < 3.0$ gives the smallest value of $t_\mathrm{min}$; 
at $\alpha\leq 2.5$ the oscillation due to the sharp edges survives; 
the oscillation disappears at $\alpha= 3.0$, and after then the min moves toward later time.  
By closer inspection we conclude that $\alpha$ for the smallest $t_\mathrm{min}$ is $2.9\pm 0.1$.
In Fig.~\ref{fig:Z-Y-comparison}, we plotted the near-dip region of $g(t)$ and $h(\alpha=2.5, t)$.  
$h(\alpha=2.5, t)$ is rather smooth near the dip and $h(\alpha\geq 3.0, t)$ has a single minimum.
Hence we can estimate the min time by choosing the value of $t$ in the data which gives a single minimum of $h(\alpha, t)$.
The neighboring data points can give the edges of the error bar for $\alpha$.    We should emphasize that this procedure is somewhat arbitrary.

\begin{table}[h]
\caption{The value of smallest $t_\mathrm{min}$, defined as the value of $t$ when $|Y(\alpha,\beta=0,t)|^2/|Y(\alpha,\beta=0,t=0)|^2$ has a single minimum as a function of $t$ for the smallest $\alpha$. The value of such $\alpha$ is also given.
The number of samples is
$240000\times2^{(14-N)/2}$ for $N=10,12,14$,
$600000\times2^{(18-N)/2}$ for $N=16,18$,
$120000\times2^{(22-N)/2}$ for $N=20,22$,
$48000$ for $N=24$, $10000$ for $N=26$, $3000$ for $N=28$, $1000$ for $N=30$, $1500$ for $N=32$ and $110$ for $N=34$.
 }
\label{tbl:tmin}
\begin{center}
\begin{tabular}{|c|c|c|}
\hline
$N$ &  smallest $t_\mathrm{min}$ & $\alpha$ for smallest $t_\mathrm{min}$\\
\hline
\hline
10 & $ 9.1 \pm 0.1$ & $4.1 \pm 0.1$ \\\hline
12 & $ 8.7 \pm 0.1$ & $4.2 \pm 0.1$ \\\hline
14 & $10.1 \pm 0.1$ & $4.9 \pm 0.1$ \\\hline
16 & $12.9 \pm 0.1$ & $4.7 \pm 0.1$ \\\hline
18 & $12.2 \pm 0.1$ & $3.9 \pm 0.1$ \\\hline
20 & $11.5 \pm 0.1$ & $3.7 \pm 0.1$ \\\hline
22 & $11.2 \pm 0.1$ & $3.5 \pm 0.1$ \\\hline
24 & $11.7 \pm 0.1$ & $3.5 \pm 0.1$ \\\hline
26 & $14.5 \pm 0.1$ & $3.3 \pm 0.1$ \\\hline
28 & $13.4 \pm 0.1$ & $3.0 \pm 0.1$ \\\hline
30 & $13.0 \pm 0.1$ & $2.9 \pm 0.1$ \\\hline
32 & $12.5 \pm 0.1$ & $2.9 \pm 0.1$ \\\hline
34 & $12.1 \pm 0.1$ & $2.8 \pm 0.1$ \\\hline
\end{tabular}
\end{center}
\end{table}

It is hard to determine the $N$-dependence of $t_\mathrm{ramp}$ from Table~\ref{tbl:tmin}.   The fact that $t_\mathrm{min}$ and hence an upper bound for $t_{\rm ramp}$ is nonmonotonic,  a physically unlikely result, indicates the uncertainty in the procedure.    The best we can say is that $t_{\rm ramp}$ is certainly far smaller than the dip time $t_{\rm dip}$ and increases at most slowly with $N$.   Assuming order one coefficients a power law  faster than linear or an exponential behavior (as suggested in \cite{Garcia-Garcia:2016mno}) is disfavored.

\subsection{Eigenvalue Behavior for 2-local RCQ}\label{sec:2-local-RCQ}
In order to test the spectral rigidity of the energy spectrum, 
we performed the unfolding with steps 1 -- 4 explained in Sec.~\ref{sec:NR-NV}.
The shifted-and-rescaled energy spectrum (obtained by performing steps 1, 2 and 3), 
before the unfolding, is shown in the right panel of Fig.~\ref{Fig:QSG-var-dos-1}. 
Unlike the SYK model, the edge of the spectrum is not sharp. 
Due to this, the slope of SFF decays much faster, as we will see shortly. 
The nearest-level separation obtained from the unfolded spectrum is plotted in the left panel of Fig.~\ref{Fig:QSG-var-dos-2}. 
A good agreement with GUE ensemble at large $N$ can be seen. 
The number variance $\Sigma^2(K)$ is plotted in the right panel of Fig.~\ref{Fig:QSG-var-dos-2}. 
At $N=16$, an agreement with RMT can be seen only at $K\lesssim 10$.

\begin{figure}
\includegraphics{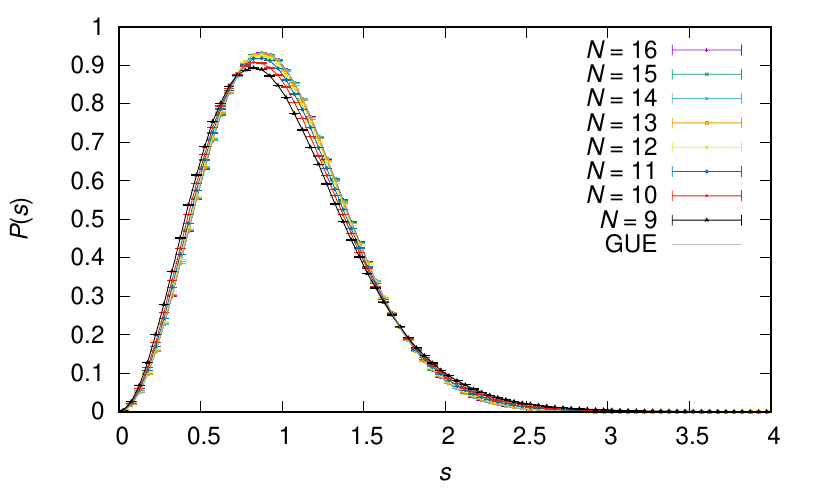}
\includegraphics{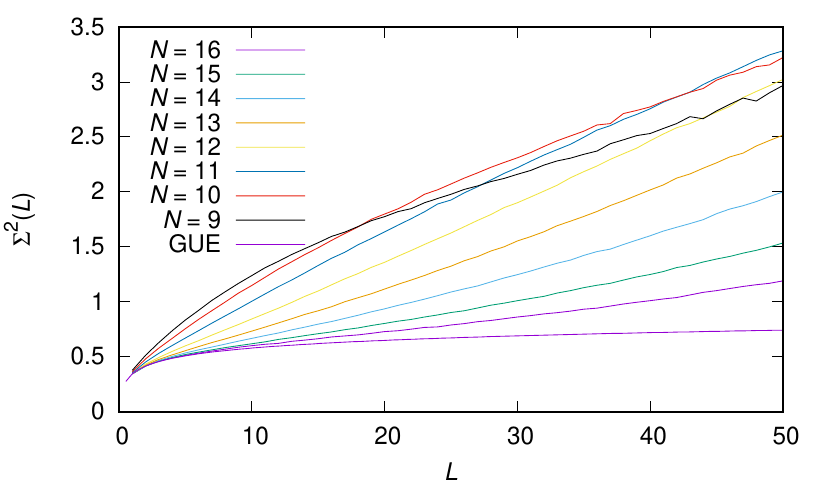}
\caption{Unfolded level separation distribution $P(s)$ for the RCQ model for $N=9,10,\ldots,16$ compared against the Pad\'e approximant for GUE,
and the number variance $\Sigma^2(K)$. $L$ on the right panel corresponds to $K$ in the text.
}
\label{Fig:QSG-var-dos-2}
\end{figure}

\section{Analytic Results for Random Quantum Circuit}\label{appendix-analytics-spin-circuit}
In this appendix we collect several analytic results which complement the discussion in Sec.~\ref{sec:Random-and-Brownian-circuits}.

\subsection{$k$-th Moment of the Haar Measure}\label{appendix-analytics-Haar}
If we take the average with respect to the Haar measure on U$(L)$, $u(k, t)=\left\langle |\Tr U^k(t)|^2\right\rangle$ is equal to $k$ for $k \leq L$ and is equal to $L$ for $k>L$. 
For $k=1$ and $k=2$, it can be seen by contracting the indices in (see e.~g.~Ref.~\cite{Nahum:2017yvy})
\begin{eqnarray}
\left\langle U_{i_1 i_2} U^*_{j_1 j_2}\right\rangle_{\rm Haar}
= 
\frac{1}{L}\delta_{i_1j_1}\delta_{i_2j_2} 
\end{eqnarray}
and
\begin{eqnarray}\label{twoutwoustar}
\left\langle U_{i_1 i_2} U^*_{j_1 j_2}  U_{i_3 i_4}  U^*_{j_3 j_4}\right\rangle_{\rm Haar}
= 
\frac{1}{L^2-1} \Big[ \delta_{i_1 j_1}\delta_{i_2 j_2}  \delta_{i_3 j_3} \delta_{i_4 j_4} +  \delta_{i_1 j_3}\delta_{i_2 j_4}  \delta_{i_3 j_1} \delta_{i_4 j_2}  \\ \nonumber
-\frac{1}{L} (\delta_{i_2 j_2}\delta_{i_4 j_4}  \delta_{i_1 j_3} \delta_{i_3 j_1}  + \delta_{i_1 j_1}\delta_{i_3 j_3}  \delta_{i_2 j_4} \delta_{i_4 j_2} )  \Big], 
\end{eqnarray}
as
\begin{eqnarray}
\left\langle |\Tr U(t)|^2  \right\rangle_{\rm Haar}
=
\sum_{i_1,i_2,j_1,j_2}
\delta_{i_1i_2}\delta_{j_1j_2}\cdot
\frac{1}{L}\delta_{i_1j_1}\delta_{i_2j_2} 
=
1
\end{eqnarray}
and
\begin{eqnarray}
\left\langle |\Tr U^2(t)|^2  \right\rangle_{\rm Haar}
&=&  \sum_{i_1, i_2, j_1, j_2 = 1}^L \frac{1}{L^2-1} \Big[\delta_{i_1 j_1} \delta_{i_2 j_2} + \delta_{i_1 j_2} \delta_{i_2 j_1} - \frac{1}{L}  2\delta_{i_1 i_2 j_1 j_2}\Big] \\
&=& \sum_{i, j =1}^L \frac{2}{L^2 -1}  - \frac{1}{L} \sum_{i=1}^L \frac{2}{L^2-1}  \\
&=& 2 \frac{L^2}{L^2-1} - \frac{2}{L^2 -1} = 2.
\end{eqnarray}
A proof for generic $k$ is discussed in Section \ref{sec:sff-gue-cue} (also see \cite{1999chao.dyn..6024H}).

\subsection{Calculating $\left\langle| \Tr \ U^2(t)|^2\right\rangle$ in RCQ Random Circuit} \label{app:k=2RCQcalc}
In this section we study
 \begin{eqnarray}
u(2,t) = \left\langle|\Tr U^2(t)|^2 \right\rangle
=
 \sum_{i_1, i_2, j_1, j_2 = 1}^L \left\langle U_{i_1i_2} U_{i_2 i_1} U^*_{j_1 j_2} U^*_{j_2 j_1} \right\rangle. 
\end{eqnarray}
for the RCQ random circuit using the techniques of \cite{Harrow2009}. The generalization to the Brownian circuit is straightforward. 
Here we specialize to the U(4) gateset. 
We need to estimate two kinds of contributions:
\begin{itemize}
\item
$\left\langle U_{ii} U_{ii} U^*_{j j} U^*_{j j} \right\rangle$. 
These terms start out at 1 and decay to their Haar value which from \eqref{twoutwoustar} has a  large $L$ value of  $ 2/L^2$ for $i=j$. It is zero for $i\neq j$.

\item
$\left\langle U_{ij} U_{ji} U^*_{ij} U^*_{ji}\right\rangle$ with $i\neq j$. 
These terms start at zero and reach their Haar value $\sim 1/L^2$.

\end{itemize}

\subsubsection*{Estimating $\left\langle U_{ij} U_{ji} U^*_{ij} U^*_{ji}\right\rangle$ ($i\neq j$)}
Without loss of generality we can set $i=1$ and consider 
$\left\langle U_{1j} U_{j1} U^*_{1j} U^*_{j1}\right\rangle$ ($j\neq 1$). 
These terms start out at zero and reach their Haar value $\sim 1/L^2$. In the $0,1$-basis, the $\ket{1}$ state is expressed by $|0^N\rangle$. 
In the same way as the discussion below equation \eqref{U11U11*}, we take the state $j$ to be of the form $|1^{q}0^{N-q}\rangle$, again without loss of generality. 
The argument is a simple generalization of the one for $\left\langle U_{ii} U^*_{j j} \right\rangle$, 
presented below \eqref{U11U11*}. 
We start from the initial state on two copies of the system, \begin{eqnarray}
\hat{\rho}^{(2)}(0) = \left(\ket{1^q 0^{N-q}} \bra{1^q 0^{N-q}}\right) \otimes \left( \ket{0^N} \bra{0^N}\right). 
\end{eqnarray}
Then, we can express 
\begin{eqnarray}
U_{1j} U_{j1} U^*_{1j} U^*_{j1}  
= 
\Tr \Big[  \hat{\rho}^{(2)}(t) \tilde{\hat{\rho}}^{(2)}(0) \Big],  
\end{eqnarray}
where
\begin{eqnarray}
\tilde{\hat{\rho}}^{(2)}(0) =  \left(\ket{0^N} \bra{0^N}\right) \otimes \left(\ket{1^q 0^{N-q}} \bra{1^q 0^{N-q}}\right). 
\end{eqnarray}  

We will use $\tilde \gamma_0(p_1, p_2)$ to denote the Pauli coefficients for $\tilde{\hat{\rho}}^{(2)}(0)$. Recalling that 
\begin{eqnarray}
\ket{0}\bra{0} = \frac{1}{2}(I+Z)  \quad \text{and} \quad \ket{1}\bra{1} = \frac{1}{2}(I-Z),  
\end{eqnarray}
we can see that $\tilde{\hat{\rho}}^{(2)}(0)$ consists of the Pauli strings made only of $I$ and $Z$. 
Therefore, 
\begin{eqnarray}
\left\langle U_{1j} U_{j1} U^*_{1j} U^*_{j1}  \right\rangle= \sum_{p'_1, p'_2} \gamma_t(p'_1, p'_2) \tilde \gamma_0(p'_1, p'_2), 
\label{eq:U1jj11j_terms}
\end{eqnarray}
where $p_1'$ and $p_2'$ run through the Pauli strings made only of $I$ and $Z$. 
That $\left\langle U_{1j} U_{j1} U^*_{1j} U^*_{j1}  \right\rangle$ is zero at $t=0$ means 
$\gamma_t(p'_1, p'_2) \tilde \gamma_0(p'_1, p'_2)$ cancel with each other at early time (due to sign differences). 
 In order to see the time evolution of the second and third terms, we need to understand how the unitary gates selected from $\Gamma ={\rm U}(4)$
act on two copies of Pauli strings $\sigma_{p'_1}$ and $\sigma_{p'_2}$. 
We write them as $\sigma_{p'_1}=\sigma_{p'_{1,1}}\otimes\sigma_{p'_{1,2}}\otimes\cdots\otimes\sigma_{p'_{1,N}}$ 
and $\sigma_{p'_1}=\sigma_{p'_{2,1}}\otimes\sigma_{p'_{2,2}}\otimes\cdots\otimes\sigma_{p'_{2,N}}$, respectively. 
Because each gate acts on Pauli matrices at two sites $i$ and $j$, let us focus on these two sites:
$\sigma_{p'_{1,i}}\otimes\sigma_{p'_{1,j}}$ and $\sigma_{p'_{2,i}}\otimes\sigma_{p'_{2,j}}$.  
The gate acts on $(\sigma_{p'_{1,i}}\otimes\sigma_{p'_{1,j}})\otimes(\sigma_{p'_{2,i}}\otimes\sigma_{p'_{2,j}})$ as 
\begin{eqnarray} 
& &
\left\langle
\left\{G_{i,j}(\sigma_{p'_{1,i}}\otimes\sigma_{p'_{1,j}})\left(G_{i,j}\right)^\dagger\right\}
\otimes
\left\{G_{i,j}(\sigma_{p'_{2,i}}\otimes\sigma_{p'_{2,j}})\left(G_{i,j}\right)^\dagger\right\}
\right\rangle_{G\in\Gamma}
\nonumber\\
& &
\qquad
=
\sum_{p''}c_{p',p''}\left(\sigma_{p''_{1,i}}\otimes\sigma_{p''_{1,j}}\right)\otimes\left(\sigma_{p''_{2,i}}\otimes\sigma_{p''_{2,j}}\right).  
\end{eqnarray} 
Here $G_{i,j}\in\Gamma = $U$(4)$ is a unitary gate and the Haar average over the gate set $\Gamma$ has been taken. 
The coefficients $c_{p',q'}$ are \cite{Harrow2009}
\begin{eqnarray}\label{markovamplitudes}
c_{p',p''}
=
\begin{cases}
    0,&  \left(p'_{1,i},p'_{1,j}\right) \neq \left(p'_{2,i},p'_{2,j}\right) \ \text{or}\ \left(p''_{1,i},p''_{1,j}\right) \neq \left(p''_{2,i},p''_{2,j}\right),  \\
    1 ,& \left(p'_{1,i},p'_{1,j}\right) = \left(p'_{2,i},p'_{2,j}\right) = \left(p''_{1,i},p''_{1,j}\right) = \left(p''_{2,i},p''_{2,j}\right) = (0,0)\\
    \frac{1}{15},  & \left(p'_{1,i},p'_{1,j}\right) = \left(p'_{2,i},p'_{2,j}\right)  \neq (0,0)  \ \text{and}\ \left(p''_{1,i},p''_{1,j}\right) = \left(p''_{2,i},p''_{2,j}\right) \neq (0,0).
\end{cases} 
\end{eqnarray} 
In words, unequal strings are mapped to zero in one step.  When strings are equal and $II$ they remain unchanged.  When strings are equal and not $II$ they are uniformly sprayed among the 15 non $II$ configurations.  
(When the gate set is a subset of $U$(4) the first, unequal case, is given by $1 - \Delta$ where $\Delta$ is the gap of the gate set.  These unequal configurations die away exponentially, like the $k=1$ discussion in the text. The last coefficient is also modified with no qualitative effect.)

By using this expression, we can estimate the time dependence both for local and 2-local circuits. 
We first focus on the local circuit.  

Note that we can write the sum in Eq.~\eqref{eq:U1jj11j_terms} as
\begin{eqnarray}
\left\langle U_{1j} U_{j1} U^*_{1j} U^*_{j1}  \right\rangle = 4^{-N} + 2^{-N} \sum_{p' \neq0} \gamma_t(p', p') +2^{-N} \sum_{p'_1 \neq p'_2} \gamma_t(p'_1, p'_2)~. 
\label{eq:U1jj11j_terms_sep}
\end{eqnarray}
We can bound the second and third terms separately. For the $U(4)$ gateset the third term in the sum is set to zero after a single step of the random circuit. One can check that the terms in the second term have the same sign, so we can rewrite the second term as
\begin{eqnarray}\label{gammaoverlap}
(*2) = 4^{-N} \sum_{p' \neq 0} 2^N \cdot|\gamma_{t}(p',p')| 
\end{eqnarray}
where $p'$ is a sum of all $2^N-1$ Pauli strings of $I$'s and $Z$'s. For each fixed $p'$, we have a Markov chain that starts at position $p'$ (note that $2^N \cdot|\gamma_{0}(p',p')| =1 $)and converges to a uniform distribution on $4^N-1$ strings.
\begin{eqnarray}\label{uniformvalue}
\lim_{t \to \infty} 2^N \cdot |\gamma_{t}(p',p')| = \frac{1}{ 4^{N} -1}
\end{eqnarray}
This implies that $\lim_{t \to \infty} (*2)  \approx 4^{-N} \cdot 2^{-N}$, which is exponentially suppressed relative to the Haar value ($= 4^{-N}$).  So we need only focus on the early time decay of this quantity.    In particular to establish $\epsilon$ closeness to Haar for fixed $\epsilon$ we do not need to consider the final decay governed by the gap of the full Markov chain.

Because of the large number of configurations, and in particular the 15 possible two Pauli strings not equal to $II$, the return probability in this chain is small and can be ignored (up to fractional errors of $1/15$).  So we can think of the configuration space as a large tree. The slowest decaying $N$ qubit strings are those with lots of $II$ substrings and hence very few $Z$'s.  We can ignore the rare cases where the $Z$'s are adjacent.   Then the overlap \eqref{gammaoverlap} decays with a rate proportional to the number of $Z$'s.  The estimate becomes analogous to that presented in Section \ref{localandk-localRCQrandomcircuit}.

As in Section \ref{localandk-localRCQrandomcircuit} we will find a time of order $\log \frac{N}{\epsilon}$ for the sum of the slowest terms to be small enough to get $\epsilon$ close to the Haar value.    At this time each of the slowest chains, those with a few $Z$'s,  have only spread to polynomially many other chains.   So the gap of the full Markov process should not come into play.

 The initial decay rate of the Pauli string with $d$ $Z$'s will be given by $e^{-\tilde \Delta d}$ per step of the process where we write the probability for a two qubit string to remain in a given non $II$ configuration as $e^{-\tilde{\Delta}} = \frac{1}{15}$.    Therefore we have
\begin{eqnarray}
(*2) &=& 4^{-N} \sum_{p' \neq 0} 2^N \cdot|\gamma_{t}(p',p')| \approx 4^{-N} \sum_{d=1}^N {N \choose d} e^{-\tilde \Delta d t} \\
&=& 4^{-N} \Big((1+ e^{-\tilde \Delta t })^N-1\Big)
\end{eqnarray}
Putting everything together and assuming $N$ is sufficiently large, we conclude
\begin{eqnarray}
\left\langle U_{1j} U_{j1} U^*_{1j} U^*_{j1}\right\rangle  \approx 4^{-N}(1 + Ne^{-\tilde \Delta t})
\qquad (j\neq 1). 
\end{eqnarray}
That is the strings with just one $Z$ dominate this decay.

\subsubsection*{Estimating $\left\langle U_{ii} U_{ii} U^*_{j j} U^*_{j j} \right\rangle$}
We consider $i=1$, without loss of generality.  Again in the $0,1$-basis, $\ket{1}$ is expressed by $|0^N\rangle$. 
In the same way as the discussion below \eqref{U11U11*}, we take the state $\ket{j}$ to be of the form $|1^{q}0^{N-q}\rangle$.

The argument is a simple generalization of the one for $\left\langle U_{ii} U^*_{j j} \right\rangle$, 
presented below \eqref{U11U11*}. 
We start from the initial state on two copies of the system, 
\begin{eqnarray}
\hat{O}^{(2)}(0) = \left(\ket{0^N} \bra{1^q 0^{N-q}}\right) \otimes \left( \ket{0^N} \bra{1^q 0^{N-q}}\right).  
\end{eqnarray}
The time evolution of $\hat{\rho}$ is defined by 
\begin{eqnarray}
\hat{O}^{(2)}(t) = \left(U(t)\ket{0^N} \bra{1^q 0^{N-q}}U^\dagger(t)\right) \otimes \left( U(t)\ket{0^N} \bra{1^q 0^{N-q}}U^\dagger(t)\right).  
\end{eqnarray}
Then, by construction, 
\begin{eqnarray}
 U_{11} U^*_{jj} U_{11} U^*_{jj}   =\Tr \Big[ \hat{O}^{(2)}(t) \hat{O}^{(2)\dagger}(0) \Big]. 
\end{eqnarray}
We expand $\hat{O}^{(2)}$ by Pauli strings as 
\begin{eqnarray}
\hat{O}^{(2)}(t)  = 2^{-N} \sum_{p_1, p_2} \gamma_t(p_1, p_2) \sigma_{p_1} \otimes \sigma_{p_2}.   
\end{eqnarray}  
The coefficients at $t=0$ are given by 
\begin{eqnarray}
\gamma_0(p'_1, p'_2) = 2^{-N} (+i)^{f(\vec \alpha(p'_1)) +f(\vec \alpha(p'_2))}, 
\end{eqnarray} 
when 
 $p'_1$ and $p'_2$ are Pauli strings whose first $q$ elements are $X$ or $Y$ and the remaining $N-q$ elements are $I$ or $Z$. 
For other $p_1$ and $p_2$, $\gamma_0(p_1, p_2)=0$. 
We can now write
\begin{eqnarray}
U_{11} U^*_{jj} U_{11} U^*_{jj}  
= 
\sum_{p'_1, p'_2} \gamma_t(p'_1, p'_2)  \gamma^*_0(p'_1, p'_2) 
\end{eqnarray}
We can split the sum into three parts 
\begin{eqnarray}
\langle U_{11} U^*_{jj} U_{11} U^*_{jj}\rangle
=
4^{-N}\delta_{1j} + \sum_{p'\neq 0} \gamma_{t}(p',p')\gamma^*_{0}(p',p')  +  \sum_{p'_1 \neq p'_2}  \gamma_t(p'_1, p'_2) \gamma^*_0(p'_1, p'_2).
\label{eq:U11jj11jj_term}
\end{eqnarray}

The strings we have chosen have $q$ $X$ or $Y$s.  We denote the number of  $Z$'s by $d$, where $d$ can runs from zero to $n-q$. When $j \neq 1$ or equivalently $(q\geq 1)$ we note that for every string with an $X$ at a given position there is an identical string with a $Y$ at that position. Within the second term, contributions of string with $X$ and $Y$ in the same position will differ by a sign, due to the fact that there are two copies of the system and each one will contribute an $i$ for each copy of $Y$.  So the contributions of all such strings cancel exactly. This implies that when $j \neq 1$ the second term in \eqref{eq:U11jj11jj_term} will be set to zero in one step.  Again for the $U(4)$ gateset, the third term in the sum is also set to zero after a single step of the random circuit.

 In the remaining case $(j=1)$, we will only have strings of $I$ and $Z$ and the evolution will be analogous to the previous section.  In fact we not need to repeat this estimate because there is only one such term, $i=j=1$, which makes an exponentially small contribution to the final value of $u(2, t)$.    
 
\subsubsection*{Summing up all the contributions}
We can rewrite the sum as (recall $L=2^N$ and assume $t>1$) 
\begin{eqnarray}
u(2,t) = \left\langle |\Tr U^2(t)|^2 \right\rangle
&= &
L \sum_{j = 1}^L \left\langle U_{11} U_{11} U^*_{jj} U^*_{jj} \right\rangle 
+ 
2L \sum_{j = 2}^{L} \left\langle  U_{1j} U_{j1} U^*_{1j} U^*_{j1} \right\rangle. 
\nonumber\\
&\lesssim&  
\frac{1}{L} (1 + N\cdot e^{-\tilde \Delta t})  + 2 (1+N\cdot e^{-\tilde \Delta t}) 
\nonumber\\
&\simeq &
2 + 2 N\cdot e^{-\tilde \Delta t}. 
\label{eq:trU2_sum}
\end{eqnarray}
The factor of $2$  comes from the $i, j$ symmetry in the sum.
Note that this result becomes  $\epsilon$ close to the Haar value in times $\frac{1}{\tilde \Delta} \log(N/\epsilon)$.

For the 2-local system the above bound goes through without modification.   There is one subtlety however.   The results of Harrow-Low \cite{Harrow2009} show that full Markov chain convergence occurs in order $\log N$ time, indicating that the entire Pauli string configuration space is covered in that time.  So the tree approximation may no longer be valid.   But by this time our quantity is exponentially close to the Haar value \eqref{uniformvalue} and so there should be no increase in the estimate.

\subsubsection*{Alternate Strategy}

 Another method for calculation the $u(2,t)$ is to use the rewrite presented in equation \eqref{trQtrUrewrite},
 \begin{eqnarray}
 u(2, t)= \Big\langle \Tr U_t^2  \cdot \Tr (U_t^\dagger)^2 \Big\rangle = 1 +\sum_{j=2}^{L^2}  \frac{1}{L}  \Big\langle \Tr \Big( U_t^2 P_j (U_t^\dagger)^2 P_j \Big) \Big\rangle 
\end{eqnarray}
 where the sum is over all the non-identity Pauli strings. In the second term of the above formula, we translate the average over product of circuits to an average on two copies of the system followed by a swap operation $\mathcal F$. 
 \begin{eqnarray}
  \Big\langle \Tr \Big( U_t^2 P_j (U_t^\dagger)^2 P_j \Big) \Big\rangle  =  \Big\langle \Tr  \Big[ \Big ( (U_t P_j U_t^\dagger)\otimes (U_t^\dagger P_j U_t)  \Big)  \mathcal F  \Big] \Big\rangle 
 \end{eqnarray}
We have not analyzed this systematically, but observe that averaging can be done independently for each gate and so a new, perhaps more complicated, chain will result. The question then is to understand how coefficients of the Pauli strings decay in this chain for each application of a gate.  If we ignore backtracking and approximate the chain evolution by a tree, as above, the question reduces to computing the decay rate for single gate from $U(4)$. 
 
The average of the U(4) gateset on a non-identity Pauli string $AB$ ($A, B = I, X, Y, Z$) dotted with the same string can be computed and is,
 \begin{eqnarray}
\int_{U(4)} dU \  \Tr \Big(U^2 A B (U^\dagger)^2 AB\Big) = \frac{1}{15}~.
\end{eqnarray}
This means that in the local and 2-local circuit the strings with  $d$ non-identity elements will decay like $e^{- d \tilde{\Delta} t}$, where again $e^{-\tilde{\Delta}} = 1/15$. This indicates that the decay is dominated by the slowest decaying terms which have $d=1$. The total number of such terms is $3N$, giving the $\log N$ result described above. 

There is a small subtlety here.  There are $d=2$ terms with the two Paulis adjacent to each other.  These decay like a $d=1$ term.   However there are only $O(N)$ terms of this kind so this will just produce an addition to the numerical factor in front of $N$.   This is a random circuit version of the factorization issue discussed in Section \ref{sec:hamestimate}.  Therefore, the Haar time scaling, $\sim \log \frac{N}{\epsilon}$, is unaffected. 
 
\section{Random Band Matrix}\label{sec:band-matrix}

\subsection{Hamiltonian Evolution}
We consider an $L\times L$ `band matrix' $M$ with width $w$.  
Rather than introducing a sharp cutoff, we introduce a Gaussian suppression factor, 
\begin{eqnarray}
M_{ij}=M_{ij}^{(0)} \cdot\frac{e^{-\frac{(i-j)^2}{2w^2}}}{\sqrt{w}}, 
\end{eqnarray}
where $M_{ij}^{(0)}$ is the standard GUE ensemble.  

By definition, banded random matrix becomes usual RMT at $w\gtrsim L$. 
It is not chaotic when the band is too narrow, namely at $w\lesssim \sqrt{L}$. 
In the following we consider large $w\ge\sqrt{L}$. 

\begin{figure}[!ht]
  \centering
    \includegraphics[width=0.6\textwidth, clip]{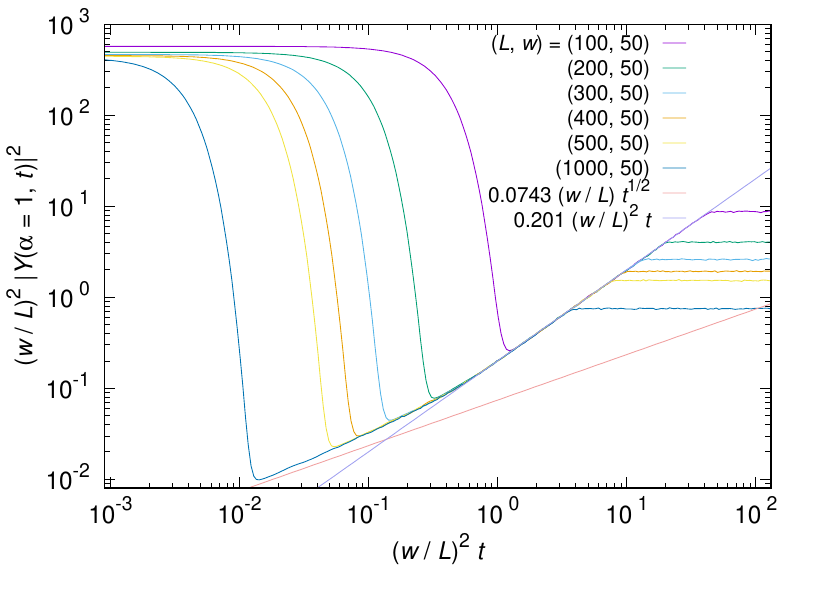}
          \caption{ $(w/L)^2|Y(\alpha=1,t)|^2$ vs $(w/L)^2\cdot t$, $L=100, 200, 300, 400, 500, 1000$, $w=50$, $\alpha=1$.
$10000$ samples have been taken.
     }\label{fig:Y-band-rescaled}
\end{figure}

By using $Y$ with appropriate value of $\alpha$, we can see the emergence of the ramp. 
As shown in Figs.~\ref{fig:Y-band-rescaled} and \ref{fig:Y-band-rescaled-alpha0}, 
if we plot $(w/L)^2|Y^2|$ against $(w/L)^2\cdot t$, 
we can see a convergence to a smooth curve in a wide range, 
across the onset of the ramp. 
Therefore, regardless of the detail of the procedure to define $t_\mathrm{ramp}$, it must scale as $t_\mathrm{ramp}\sim (L/w)^2$. 
This scaling was derived in \cite{ErdosKnowles2015}, where the authors showed that the energy scale at which sine kernel description of the two-point function breaks down is the same as the Thouless energy, which is the inverse of the time-scale for a local perturbation to diffuse through the entire system, namely $t_{\rm diff}$. 
Note that the earlier growth of $(w/L)^2|Y^2|$ is slower, $\sim t^{1/2}$. 
Our numerical plots are consistent with this scaling, as can be seen in Fig.~\ref{fig:Y-band-rescaled-alpha0}.

\begin{figure}[t]
  \centering
\includegraphics[width=0.6\textwidth]{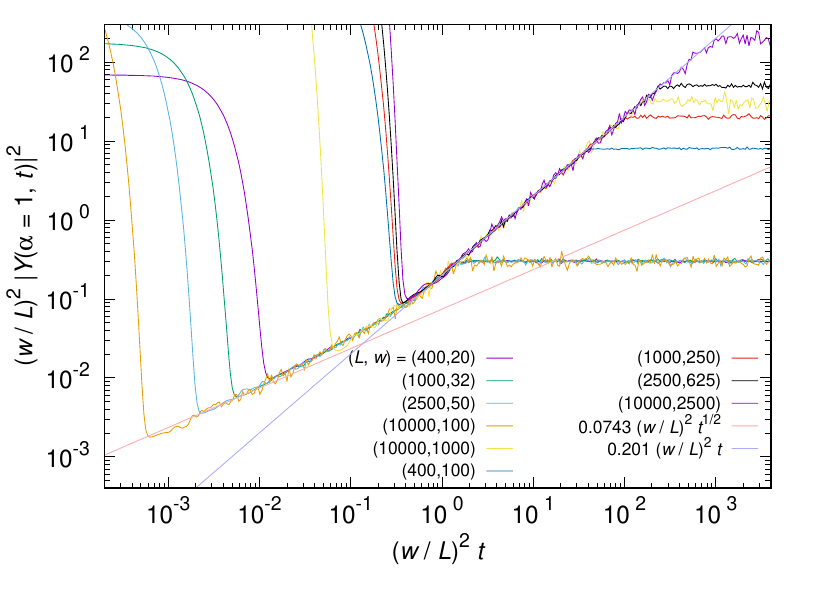}
      \caption{The value of $(w/L)^2|Y(\alpha=1, t)|^2$ plotted against $(w/L)^2\cdot t$ for various $(L, w)$ satisfying $\sqrt{L}\leq w \leq L/4$, $\alpha=1$.
	  The number of samples with each value of $w$ is $10^6 / L = 2500, 1000, 400, 100$ for $L=400, 1000, 2500, 10000$ cases. For $w \ll L/4$ the initial growth indicates that $|Y(\alpha=1, \tau)|^2 \sim (L/w)\sqrt{t}$, while for all $(L, w)$ plotted, the growth for larger $t$ before the plateau agrees with $|Y(\alpha=1, t)|^2 \sim t$.
     }\label{fig:Y-band-rescaled-alpha0}
\end{figure}

It would be instructive to mention the difference between the banded matrices and local 
or $k$-local Hamiltonians.  
The band matrix mimics a Hamiltonian acting on $L$-state Hilbert space, 
which connects $w$ nearby states. This looks `local' if we identify the states with 
a single particle hopping on $L$ lattice sites, 
but it is useful to note that this is rather different from a many-body local Hamiltonian, 
say a local spin system. In the latter, in a natural local basis, the Hamiltonian is sparse 
but the nonzero entries are not aligned near the diagonal as in the banded matrix.
  
In local Hamiltonians with $N$ spins, each row and column has $L=2^N$ entries, 
among which there are $O(N)$ nonzero elements. This is much sparser than 
a banded matrix with $w\sim\sqrt{L}$, but it can already be chaotic. 
The time scales are also very different; the ramp time for the local Hamiltonian, 
$t_{\rm ramp}\sim N^2\sim (\log L)^2$ (see Sec.~\ref{sec:local-Hamiltonian}), is much shorter than $(L/w)^2$ when $w\sim\sqrt{L}$.  

\subsection{Brownian Circuit}\label{sec:Brownian-circuit-band-matrix}
The Brownian circuit of random band matrices can describe diffusion, because the number of particles --- which is one --- is conserved. 
For this reason, as we will see, we can confirm $t_{\rm Haar}\sim (L/w)^2\sim t_{\rm diff}$. 

The numerical procedure is the same: we calculate 
$U(t)=\prod_{k=1}^{n}e^{-iH_kdt}$, 
where $t=n\cdot dt$ and $H_k$ are $L\times L$ random band matrices with a width $w$. 
From this we calculate the $\langle|Tr U|^2\rangle$. 
\begin{figure}[h]
\includegraphics{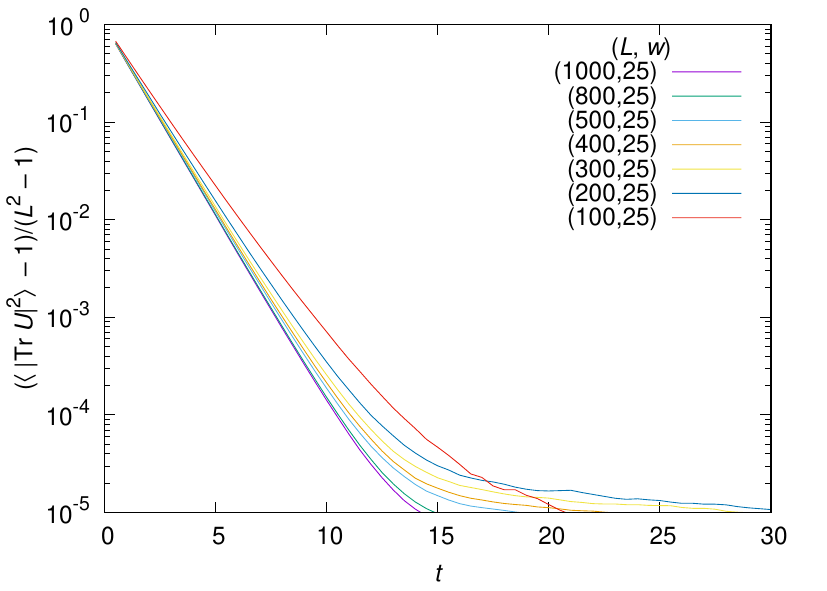}
\includegraphics{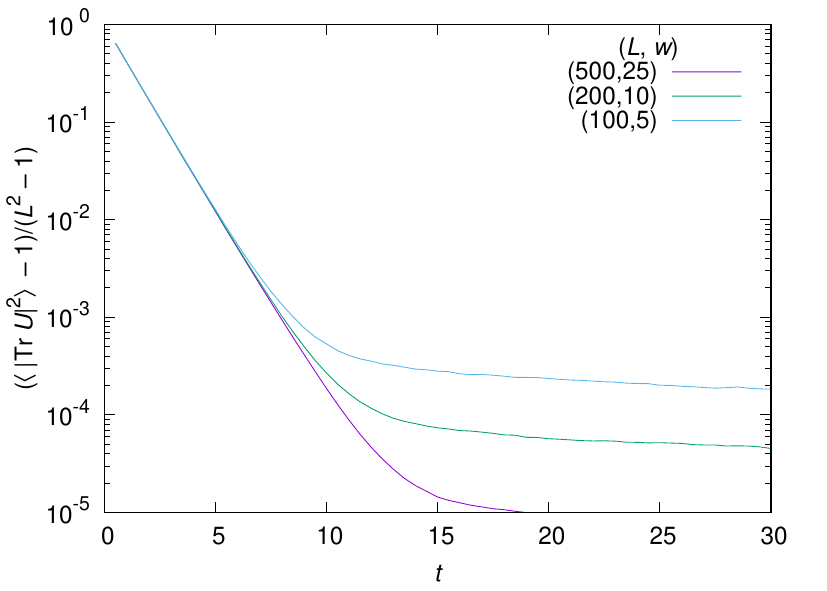}
\caption{Plot of $(\langle\vert\Tr U\vert^2\rangle-1)/(L^2-1)$ against $t$ for $w=25$ and various values of $L$ (left)
and for $L/w = 20$ with $L=500, 200, 100$ (right). $dt = 0.5$ is fixed.
}\label{fig:TrUsq-L2m1}
\end{figure}

First let us see that the specific choice of $dt$ is not important. 
When $dt$ is sufficiently small, if we take $tdt=n(dt)^2$ to be the horizontal axis, the $dt$-dependence is gone; see 
Fig.~\ref{fig:Brownian-band-matrix-dt-dependence}. 
This scaling can be understood if the time evolution is described as a random walk 
with step-size $wdt$. After $n$ steps, the typical distance from the starting point should be 
$\sqrt{n}wdt$, if the random walk picture is true. 
Then for fixed $L$ and $w$ the $dt$-dependence should disappear when 
the horizontal axis is $n(dt)^2$. 
Below we fix $dt$ to be $0.5$. 

\begin{figure}
\begin{center}\scalebox{1.2}{
\rotatebox{0}{
\includegraphics[width=0.47\textwidth]{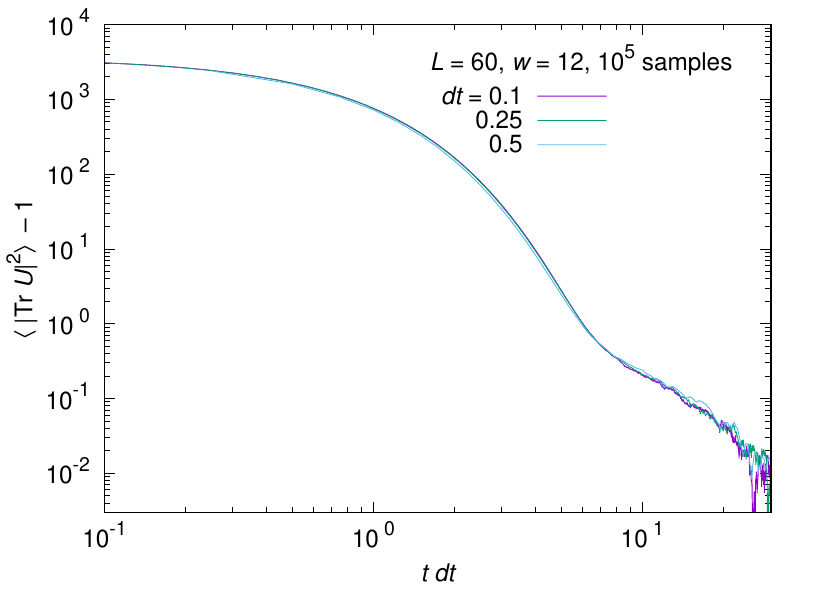}}}
\end{center}
\caption{
Plot of $\langle\vert\Tr U\vert^2\rangle-1$ against $t~dt$ for $dt = 0.1, 0.25, 0.5$ for Brownian circuits of random band matrices with $L=60$ and $w=12$.
$10^5$ samples have been taken for each choice of $dt$.
}\label{fig:Brownian-band-matrix-dt-dependence}
\end{figure}

In Fig.~\ref{fig:TrUsq-L2m1} we have plotted $\langle\vert\Tr U\vert^2\rangle-1$ against $(t(w/L)^2)/(L^2-1)$
for fixed $w=25$ and various $L$, and for a fixed $L/w$.
In both cases we can see an exponential decay. 
The exponent is a function of $L/w$, as we can see from the right panel of Fig.~\ref{fig:TrUsq-L2m1}. 

As we can see from Fig.~\ref{fig:TrUsq-w25-late}, we can show the exponential decay at late time as well,
$\langle |{\rm Tr}U|^2\rangle-1\sim e^{-2t(w/L)^2}$. Note that the exponent is different from the early time.  
If we define $t_{\rm Haar}$ as the time $\langle |{\rm Tr}U|^2\rangle-1$ reaches to a certain value, say 0.1, 
then this scaling leads to $t_{\rm Haar}\sim (L/w)^2$.

\begin{figure}[h!]
\begin{center}
\includegraphics[width=0.57\textwidth]{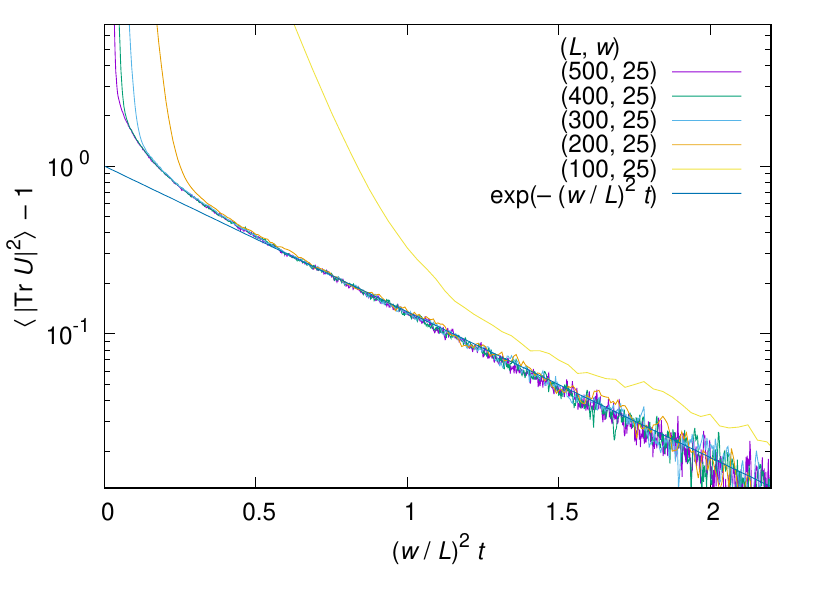}
\end{center}
\caption{Log plot of $\langle\vert\Tr U\vert^2\rangle-1$ against $t(w/L)^2$
for $w=25$ and various values of $L$. $dt = 0.5$ is fixed. $10^5$ samples have been used for $L=300, 200, 100$.
}\label{fig:TrUsq-w25-late}
\end{figure}

It is possible to understand this scaling from the random walk picture. 
When the average distance from the starting point, which is $\sqrt{t}w$, is of order $L$, 
it is natural to expect that the transfer matrix $U$ is almost a random unitary. 
Therefore $\sqrt{t_{\rm Haar}}w\sim L$, or equivalently, $t_{\rm Haar}\sim (L/w)^2$.

\end{document}